# Fundamental Limits of Communications in Interference Networks

## Part II: Information Flow in Degraded Networks

Reza K. Farsani[1]

Email: reza_khosravi@alum.sharif.ir

***Abstract:*** **In this second part of our multi-part papers, the information flow in *degraded interference networks* is studied. A full characterization of the sum-rate capacity for the degraded networks with any possible configuration is established. Here, *network with any possible configuration* is referred to as a general single-hop communication scenario with any number of transmitters, any number of receivers and any possible distribution of messages among transmitters and receivers. It is shown that a *successive decoding scheme* is sum-rate optimal for these networks. Also, it is proved that the transmission of only a certain subset of messages is sufficient to achieve the sum-rate capacity in such networks. Algorithms are presented to determine this subset of messages explicitly. According to these algorithms, the optimal strategy to achieve the sum-rate capacity in degraded networks is that the transmitters try to send information for the stronger receivers and, if possible, avoid sending the messages with respect to the weaker receivers. This is indeed a fundamental characteristic of the information flow in degraded networks. The algorithms are easily understood using our graphical illustrations for the achievability schemes based on directed graphs. The sum-rate expression for the degraded networks is then used to derive a unified outer bound on the sum-rate capacity of arbitrary non-degraded networks. Several variations of the degraded networks are identified for which the derived outer bound is sum-rate optimal. Specifically, noisy interference regimes are derived for certain classes of multi-user/multi-message interference networks. Also, for the first time, network scenarios are identified where the incorporation of both successive decoding and treating interference as noise achieves their sum-rate capacity. Finally, by taking insight from our results for degraded networks, we establish a unified outer bound on the entire capacity region of the general interference networks. These outer bounds for a broad range of network scenarios are tighter than the existing cut-set bound.**

[1] Reza K. Farsani was with the department of electrical engineering, Sharif University of Technology. He is by now with the school of cognitive sciences, Institute for Research in Fundamental Sciences (IPM), Tehran, Iran.

**To:**

**THE MAHDI**

# Contents

# I. INTRODUCTION

In Part I of our multi-part papers [1], we discussed in details the problem of characterizing capacity limits for the general single-hop communication networks. We referred to such scenarios as the *interference networks*. We provided a detailed review of the existing literature which deals with capacity bounds for such networks. We also introduced the Multiple Access Channel (MAC), the Broadcast Channel (BC), the Classical Interference Channel (CIC), and the Cognitive Radio Channel (CRC) as the main building blocks for all the interference networks. Our attempt in Part I [1] was regarded to highlight similarities in the derivations of capacity bounds for these basic structures. Now, we intend to launch the study of multi-message network topologies of arbitrary large sizes. Specifically, in Part II, we shall investigate the behavior of information flow in degraded interference networks.

As discussed in the introduction of Part I [1], two types of degradedness are commonly recognized: the *physical* and the *stochastical* degradedness. The physically degraded interference networks are those for which the receivers can be arranged in a successive order from stronger to weaker such that the signal with respect to each receiver is statistically independent of the input signals conditioned on the signal of a stronger receiver. The stochastically degraded networks are those for which there exists an equivalent physically degraded network, where the equivalency of two networks implies that the marginal distributions of the transition probability function for both networks are identical. For interference networks, since the two equivalent networks have the same capacity region, it is not required to make a distinction between the two types of degradedness; hence, we generally refer to these two as "degraded interference networks".

To study the information flow in large multi-message networks, first we need to construct an appropriate language for the description of capacity bounds for such scenarios. The interference networks are essentially composed of two infrastructures: the transmitters and the receivers. Before analyzing for both cases, it is essential to discuss each one separately. Hence, we discuss in details the MAC with Common Messages (MACCM) and the BC with Common Messages (BCCM) in Section II. These two scenarios represent all the transmission and the reception aspects of the interference networks. As discussed earlier, the capacity region of the MAC with arbitrary arrangement of messages among transmitters was established in [9], however, by the approach of [9], describing the capacity region for the network with a large number of transmitters and messages is difficult as indicated in [9, p. 1059]. In Subsection II.A, we will revisit this problem. We will show that based on our graphical illustrations, one can easily characterize the capacity region for the MACCM with any arbitrary configuration. For this purpose, we expand the messages on certain directed graphs named the "***MACCM Graph***". This graph enables us to easily describe the encoding and decoding schemes to achieve the capacity for the network. Moreover, the developed graphs are very helpful to determine the sum-rate capacity for an arbitrary multi-message MACCM. Clearly, in order to achieve the sum-rate capacity for the MACCM with a given message set, the transmission of only a certain subset of messages is sufficient. Using the MACCM graphs, we provide an efficient algorithm to determine these messages. We then consider the BCCM in Subsection II.B. For this network, the capacity region is unknown. We first explain the Marton's coding scheme for the two-user channel using our graphical illustrations. Next, we will consider a multi-receiver degraded BCCM with an arbitrary set of messages which are required to be sent to the receivers. For this scenario, we shall prove that it is sufficient to transmit only one message (which should be carefully picked) and ignore the others to achieve the sum-rate capacity. This remarkable point regarding the BCCMs constitutes one of the keys to explicitly derive the sum-rate capacity for any arbitrary degraded interference network in Section III.

In Section III, we shall prove the main theorem of Part II where we provide a full characterization for the sum-rate capacity of the degraded interference networks with any arbitrary topologies. First, we establish an outer bound for the sum-rate capacity of the general degraded interference network. We then provide a coding scheme to achieve this outer bound which yields the exact sum-rate capacity. We prove that for all degraded networks, a successive decoding scheme is sum-rate optimal. In this decoding strategy, each receiver decodes its messages as well as all the messages corresponding to the receivers weaker than itself in a successive order from weaker receivers to the stronger ones till its own messages. We shall show that to achieve the sum-rate capacity for degraded networks, only a carefully picked subset of messages is required to be considered for the transmission scheme. To this end, inspired by the graphical illustrations developed for the MACCMs, we will present an order to expand the messages with respect to a given arbitrary network over certain directed graphs; we name such illustrations as the "***MACCM Plan of Messages***". The MACCM plan of messages clearly depicts both the multiple access and broadcast natures included in a given network. These graphical tools indeed play a central role in describing the behavior of information flow in large multi-message networks as discussed in various parts of our multi-part papers. Using the MACCM plan of messages, we will explicitly determine those messages which are required to be considered for the transmission scheme to achieve the sum-rate capacity for degraded networks. Also, we provide examples to clarify our results.

We next make use of our sum-rate capacity result for the degraded networks to establish unified outer bounds for the general non-degraded interference networks. In Subsection IV.A, we derive a unified outer bound on the sum-rate capacity of general networks.

By using of the derived outer bounds, we obtain the sum-rate capacity for several variations of degraded networks. Specifically, we introduce new network scenarios such as "*Generalized Z-Interference Networks*" and "*Many-to-One Interference Networks*" and identify noisy interference regimes for them. In this regime, treating interference as noise is sum-rate optimal. Also, for the first time, we identify network topologies where the incorporation of both successive decoding and treating interference as noise schemes achieves their sum-rate capacity.

Finally, in Subsection IV.B, we propose a unified outer bound on the entire capacity region of the general interference networks. These outer bounds for almost all network scenarios are tighter than the existing trivial cut-set bound [10].

In general, our results provide a deep understanding regarding the nature of information flow in general single-hop communication networks, specifically those with degraded structures. These results are also very important from the viewpoint of practical applications because they do not depend on the network topology and one can apply them to a broad range of practical communication systems. Such general results are indeed rare in network information theory [11].

Note that the notations and definitions in Part II of our multi-part papers are exactly similar to those we used in Part I [1].

# II. On Multiple Access and Broadcast Channels with Common Messages

The interference networks are essentially composed of two infrastructures: the transmitters and the receivers. All characteristics of these networks are directly concerned to either the transmission or the reception aspects. Before analyzing for both aspects, it is required to discuss each one separately. Whenever we consider the problem from the view point of "transmission", we have a MAC scenario; on the other hand, whenever we look at it from the view point of "reception", we have a BC scenario. Therefore, in what follows, we deal with the multi-transmitter MAC and the multi-receiver BC with arbitrary distribution of messages among transmitters/receivers.

In the rest of the paper, when referring to a coding scheme for a given network, we mean a random code with vanishing average error probability. The superscript $n$ that appears in the description of the codewords indicates a random code of length $n$, where $n \in \mathbb{N}$. The codewrods of the code are all generated i.i.d. according to a certain probability distribution law. The probability distribution law, the relation among the generated codewords and the decoding procedure will be exactly specified for each given coding scheme.

## II.A) The MAC with Common Messages (MACCM)

Let us first consider the Multiple Access Channels with Common Messages (MACCMs). In this scenario, several transmitters send independent messages to a receiver. In the general case, each non-empty subset of transmitters transmits an individual message. It was shown by Slepian and Wolf [9] that superposition coding achieves the capacity region of the MACCMs. Consider the two-user MACCM shown in Fig. 1.

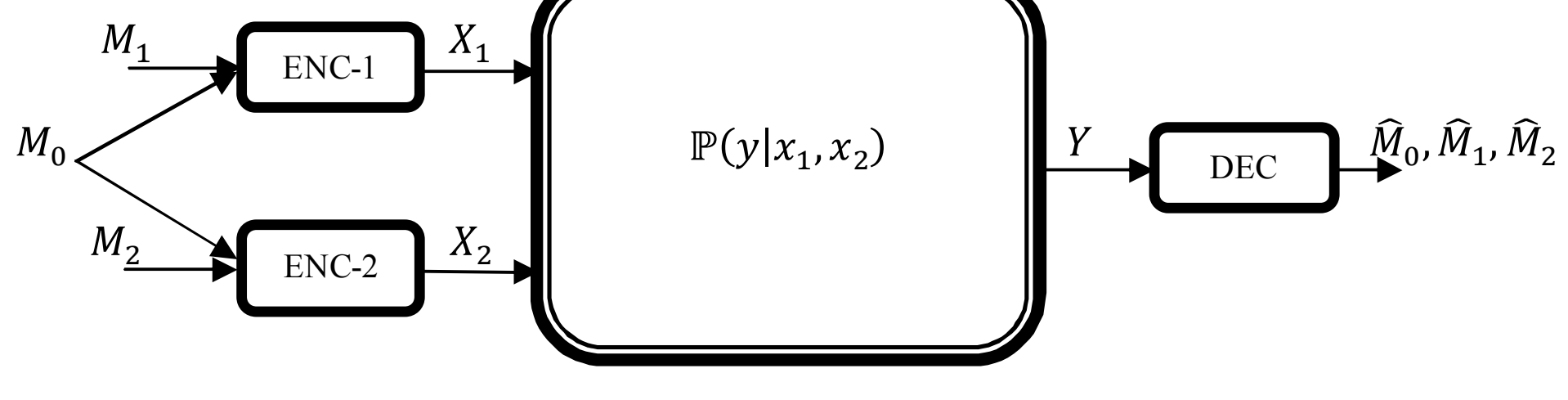

Figure 1. The two-user MACCM.

The two-user MACCM is obtained from the general interference network defined in Section II.B of Part I [1] (see also Fig. 14 in Section III) by setting $K_1 = 2, K_2 = 1, \mathbb{M} = \mathbb{M}_Y = \{M_0, M_1, M_2\}, \mathbb{M}_{X_1} = \{M_0, M_1\}$, and $\mathbb{M}_{X_2} = \{M_0, M_2\}$. The capacity region of this channel is given by:

$$\bigcup_{P_{W_0}P_{X_1|W_0}P_{X_2|W_0}} \left\{ \begin{array}{l} (R_0, R_1, R_2) \in \mathbb{R}_+^3 : \\ R_1 \leq I(X_1; Y|X_2, W_0) \\ R_2 \leq I(X_2; Y|X_1, W_0) \\ R_1 + R_2 \leq I(X_1, X_2; Y|W_0) \\ R_0 + R_1 + R_2 \leq I(X_1, X_2; Y) \end{array} \right\} \tag{1}$$

This result can also be extended to the case of MACCM with arbitrary number of transmitters. Nevertheless, by the approach of [9], characterizing the capacity region for the MACCM with a large number of transmitters is difficult as indicated in [9, p.1059]. To overcome this difficulty, we revisit the coding theorem leading to this capacity result both from the achievability scheme and the converse viewpoints. Indeed, we intend to present a new approach to describe the capacity region of the MACCM. As we will see later, this new approach is a powerful technique to understand the behavior of information flow in large multi-message networks.

As a brief discussion regarding the coding scheme which yields the capacity result (1), the common message $M_0$ is encoded by a codeword $W_0^n$ generated based on $P_{W_0}$. Then, for each of the private messages, a codeword is generated superimposing on the common message codeword $W_0^n$: the private message $M_i$ is encoded using a codeword $X_i^n$ generated based on $P_{X_i|W_0}$, $i = 1,2$. The $i^{th}$ transmitter ($i = 1,2$) sends $X_i^n(M_i, M_0)$ over the channel. The decoder decodes the messages using a jointly typical decoder. We use a graphical representation for this coding scheme as depicted in Fig. 2.

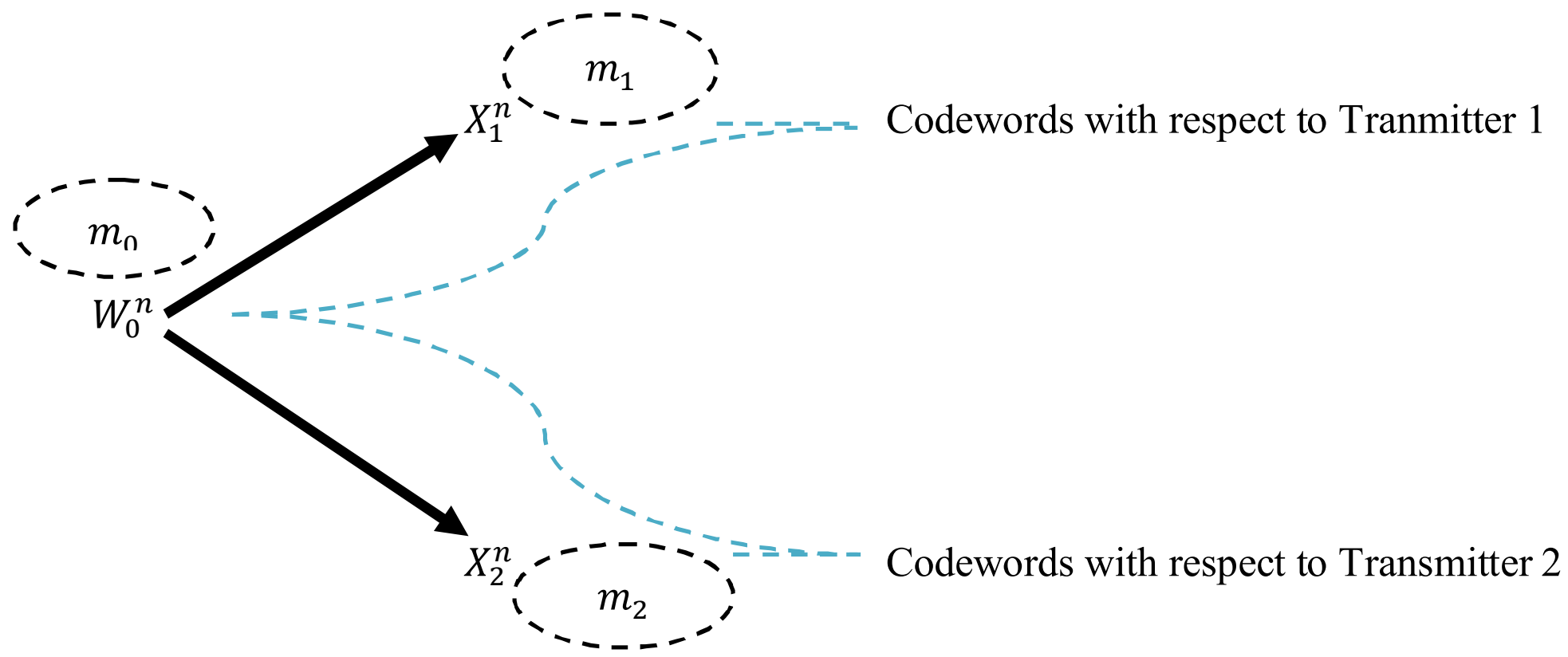

Figure 2. The encoding graph for the two-user MACCM: Every two codewords connected by an arrow form a superposition structure: The codeword at the beginning of the arrow is the cloud center and the one at the end of the arrow is the satellite. The ellipses by the codewords show what is conveyed by the corresponding codeword in addition to the ones conveyed by its cloud centers.

In this illustration, we use a directed graph (digraph) to represent the superposition structures among the generated codewords: Every two codewords connected by an arrow (directed age) build a superposition structure where the codeword at the beginning of the arrow is the cloud center and the one at the end of the arrow is the satellite. We refer to such digraph as the *encoding graph*. The ellipses by the codewords show what is conveyed by the corresponding codeword in addition to the ones conveyed by its cloud centers.

Now, let us prove the converse theorem for the capacity region of the two-user MACCM. Consider a length-$n$ code for the two-user MACCM with vanishing average error probability. Let $R_0, R_1, R_2$ be the communication rates with respect to the messages $M_0, M_1, M_2$. Using the Fano's inequality, we can write:

$$\begin{aligned} nR_1 \leq I(M_1; Y^n) + n\epsilon_{1,n} &\overset{(a)}{\leq} I(M_1; Y^n|M_0, M_2) + n\epsilon_{1,n} \\ &= \sum_{t=1}^{n} I(M_1; Y_t|M_0, M_2, Y^{t-1}) + n\epsilon_{1,n} \\ &\leq \sum_{t=1}^{n} I(M_1, Y^{t-1}; Y_t|M_0, M_2) + n\epsilon_{1,n} \end{aligned}$$

$$\stackrel{(b)}{=} \sum_{t=1}^{n} I(M_1; Y_t|M_0, M_2) + n\epsilon_{1,n}$$

(2)

where inequality (a) holds because $M_1$ is independent of $M_0, M_2$ and equality (b) holds because $X_{1,t}, X_{2,t}$ are deterministic function of $(M_0, M_1, M_2)$ and $Y^{t-1}, M_0, M_1, M_2 \rightarrow X_{1,t}, X_{2,t} \rightarrow Y_t$ forms a Markov chain. Similarly, one can derive:

$$\begin{aligned} nR_2 &\leq \textstyle\sum_{t=1}^{n} I(M_2; Y_t|M_0, M_1) + n\epsilon_{2,n} \\ n(R_1 + R_2) &\leq \textstyle\sum_{t=1}^{n} I(M_1, M_2; Y_t|M_0) + n(\epsilon_{1,n} + \epsilon_{2,n}) \\ n(R_0 + R_1 + R_2) &\leq \textstyle\sum_{t=1}^{n} I(M_0, M_1, M_2; Y_t) + n\epsilon_{3,n} \end{aligned}$$

(3)

By applying a standard time-sharing argument, we derive the following capacity outer bound for two-user MACCM:

$$\bigcup_{\substack{P_Q P_{M_0} P_{M_1} P_{M_2} \\ X_i = f_i(M_0, M_i, Q)}} \left\{ \begin{array}{l} (R_0, R_1, R_2) \in \mathbb{R}_+^3 : \\ R_1 \leq I(M_1; Y|M_0, M_2, Q) \\ R_2 \leq I(M_2; Y|M_0, M_1, Q) \\ R_1 + R_2 \leq I(M_1, M_2; Y|M_0, Q) \\ R_0 + R_1 + R_2 \leq I(M_0, M_1, M_2; Y|Q) \end{array} \right\}$$

(4)

One may restrict the PDFs $P_{M_0}, P_{M_1}, P_{M_2}$ in (4) to be uniform. Note that the parameters appeared in (4) are as follows:

1. The Random Variable (RV) representing the receiver signal, i.e., $Y$.
2. The RVs representing the messages, $M_0, M_1, M_2$.
3. The parameter $Q$ which is the time-sharing RV.

Surprisingly, the rate regions (1) and (4) are equivalent. These two characterizations are two sides of the same coin. The rate region (1) can be easily transformed to (4) by using the functional representation lemma [11]; on the other hand, by redefining $W_0 \cong (M_0, Q)$ and using the fact that $X_i$ is a deterministic function of $(M_0, M_i, Q)$, $i = 1,2$, and also the Markov chain $M_0, M_1, M_2, Q \rightarrow X_1, X_2 \rightarrow Y$, the rate region (4) is translated to (1). Roughly speaking, in characterization (4) the cover of the transmitters random variables $X_1, X_2$ and also the auxiliary random variable $W_0$ is removed from the messages $M_1, M_2, M_0$, respectively. In fact, the characterization (4) suggests an alternative coding scheme to achieve the capacity region of the MACCM. Clearly, one can encode the messages using a superposition coding scheme where the cloud center and the satellite codewords are generated independently. Let us describe this coding scheme. The common message $M_0$ is encoded using a codeword $M_0^n$ generated based on $P_{M_0}$. For each of the private messages, a codeword is generated superimposing on the common message codeword $M_0^n$; however, here the satellite codewords are generated statistically independent of the cloud center. Precisely, the private message $M_i, i = 1,2$, is encoded by a codeword $M_i^n$ superimposed on $M_0^n$, nonetheless, it is generated according to $P_{M_i}$ (not according to $P_{M_i|M_0}$)[2]. A time-sharing codeword $Q^n$ is also generated and revealed to all parties. The $i^{th}$ transmitter ($i = 1,2$) sends the codeword $X_i^n = f_i(M_0^n, M_i^n, Q^n)$ over the channel where $f_i(.)$ is an arbitrary componentwise deterministic function. The decoder decodes all the codewords using a jointly typical decoder. A simple analysis of this scheme leads to the rate region (4). Based on the equivalence between the characterizations (1) and (4), we can arrange the messages in a digraph as shown in Fig. 3 to describe the capacity region of the two-user MACCM.

[2] Note that, however, the codeword $M_i^n, i = 1,2$, is generated independent of $M_0^n$, it still conveys both the private and the common messages, i.e., the pair $(M_i, M_0)$. This is because $M_i^n$ is superimposed on $M_0^n$.

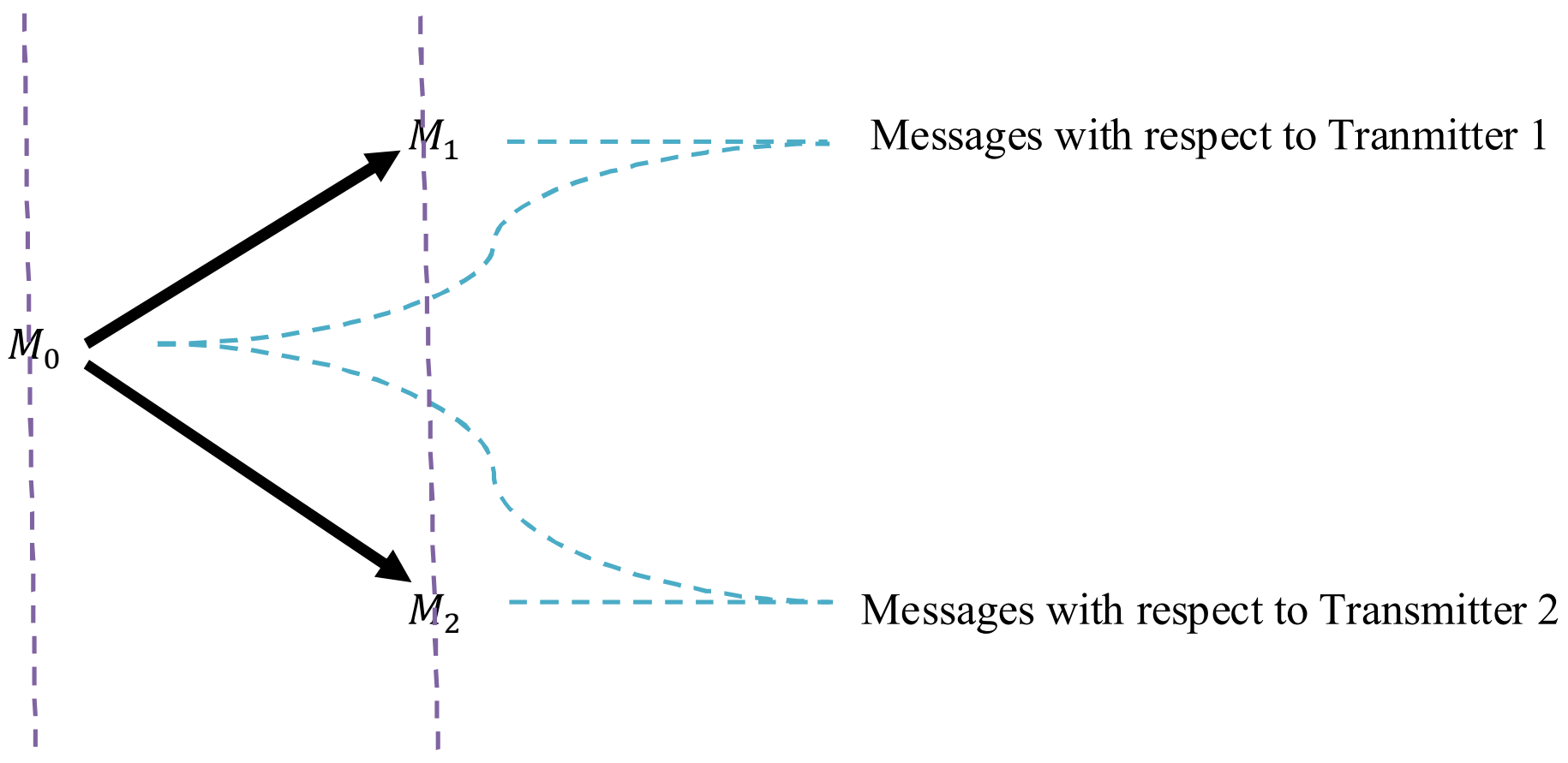

Figure 3. The message graph for the two-user MACCM: Every two messages connected by an arrow form a superposition structure: The message at the beginning of the arrow is the cloud center and the one at the end of the arrow is the satellite.

To configure this digraph, the messages are arranged in two columns (the number of transmitters); in the first column the private messages (which are transmitted only by one transmitter) are situated and in the second column the common message (which is transmitted by both transmitters). The message in Column 2 is connected by two arrows to the messages in Column 1. We refer to this digraph as the *massage graph*.

In fact, to derive the message graph each codeword in the encoding graph is replaced by the message which is conveyed by that codeword but not by its cloud centers. On the other hand, the encoding graph can be also derived from the message graph. For this purpose, each message in the message graph is replaced by a codeword. Then, in the resultant scheme, it is assumed that every two codewords which are connected by an arrow form a superposition structure. Moreover, each codeword conveys its corresponding message as well as all the messages conveyed by its cloud centers. We refer to this conversion procedure as the *Message Graph/ Encoding Graph (MG/EG) translation*. Therefore, the encoding and the message graph are also two sides of the same coin; the message graph represents the capacity region in the form of (4) while the encoding graph represents the capacity region in the form of (1).

Now based on the developed representation of the capacity achieving scheme for the two-user MACCM, we present a simple and efficient language to describe the capacity region of MACCM of arbitrary large size. As we will see later, the same approach is extensively used in our multi-part papers to describe the behavior of information flow in large interference networks.

Let us consider a $K_1$-transmitter MACCM. In this network scenario, $K_1$ transmitters send independent messages to a single receiver. The network model has been depicted in Fig. 4.

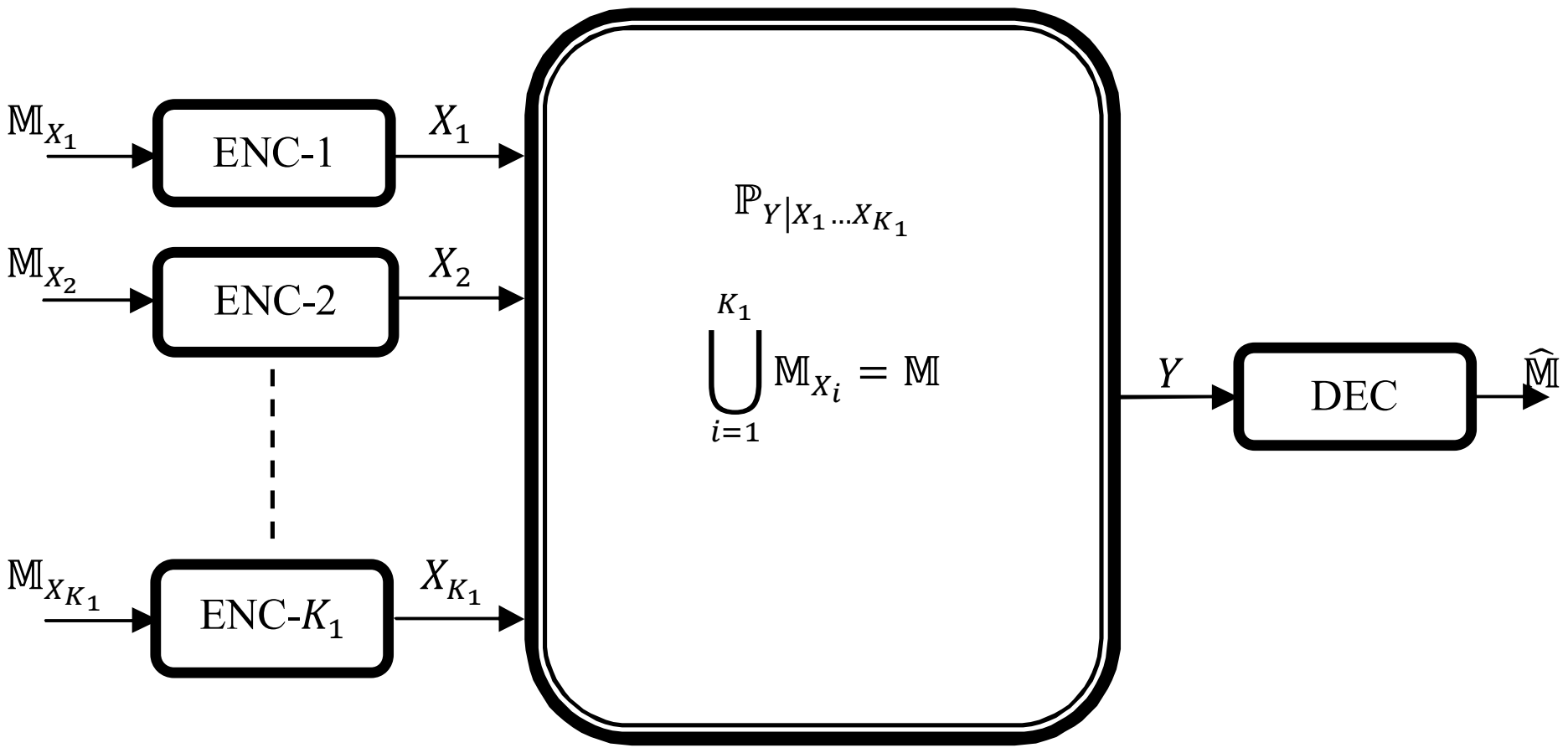

Figure 4. The general Multiple Access Channel with Common Messages (MACCM).

In the general case each nonempty subset of transmitters transmits an individual message; thus, $2^{K_1}-1$ messages are designated for transmission. We refer to this setup as the MACCM with *complete set of messages*. The MACCM is derived from the general interference network defined in Subsection II.B of Part I [1] (see also Fig. 14 in Section III) by setting $K_2=1$ and $\mathbb{M}=\mathbb{M}_Y$.

We intend to derive an explicit characterization of the capacity region of the $K_1$-transmitter MACCM with arbitrary distribution of messages. To this end, we first consider the general case in which each subset of transmitters sends an individual message to the receiver. As mentioned, in this case we have $2^{K_1}-1$ messages. Therefore, we can label each message by a nonempty subset of $\{1,\dots,K_1\}$ to determine which subset of transmitters transmits this message. For example, $M_{\{1,2,3\}}$ is a message which is cooperatively sent over the channel by transmitters 1, 2 and 3. Then, we expand these messages in a message graph similar to the one we described for the two-user case as depicted in Fig. 5.

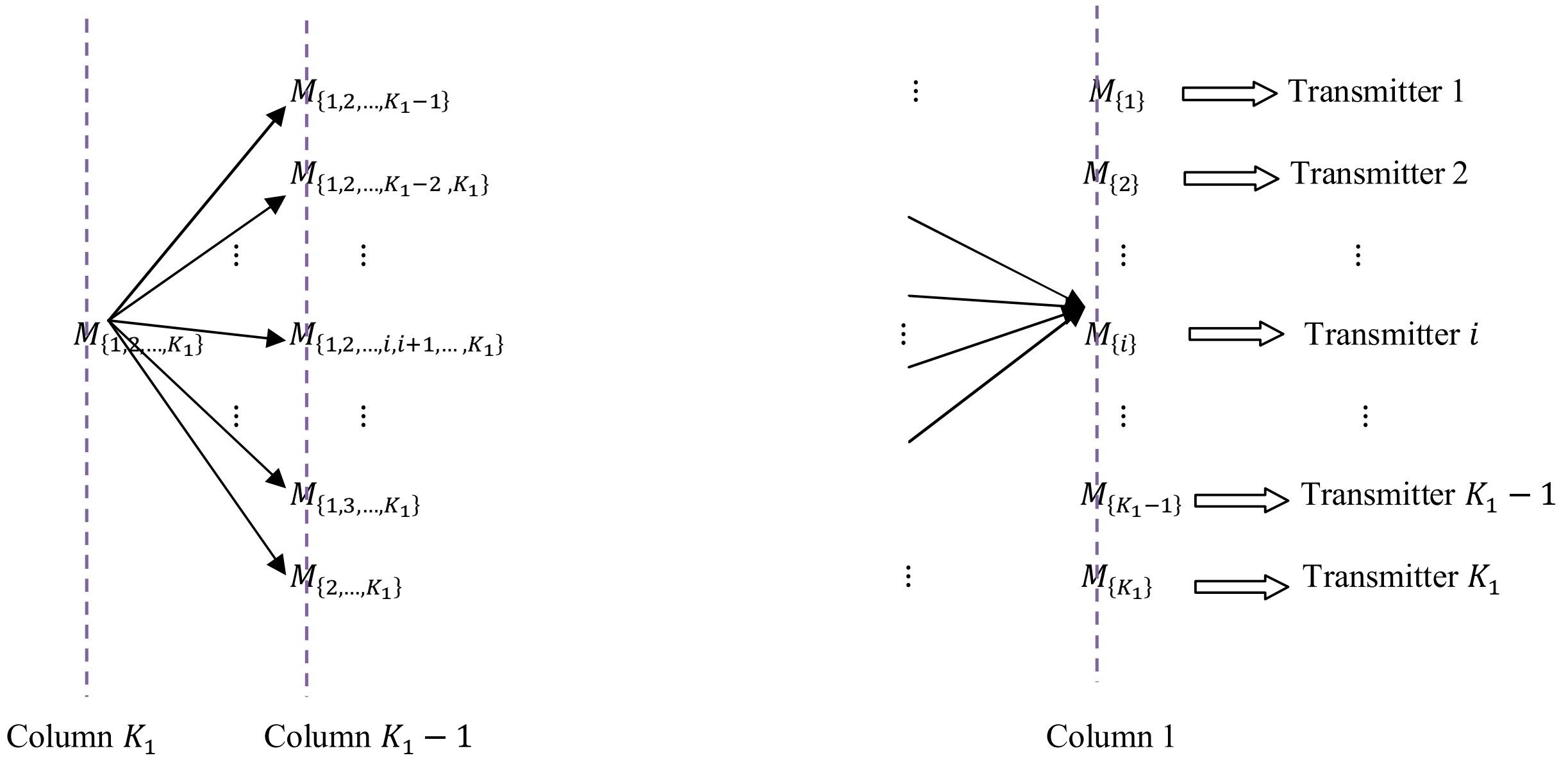

Figure 5. The message graph for the general MACCM.

This message graph is constructed as follows. The messages are positioned on $K_1$ columns such that in the $i^{th}$ column, $i=1,\dots,K_1$, one can find messages which have $i$ elements in their label sets. Then, the message $M_{\Delta_1}$ in column $i$, $i=2,\dots,K_1$, is connected to the message $M_{\Delta_2}$ in column $i-1$ provided that $\Delta_2$ is a subset of $\Delta_1$, i.e., $\Delta_2\subseteq\Delta_1$. For example, for the case of four transmitters the message graph (for the MACCM with complete set of messages) is configured as in Fig. 6.

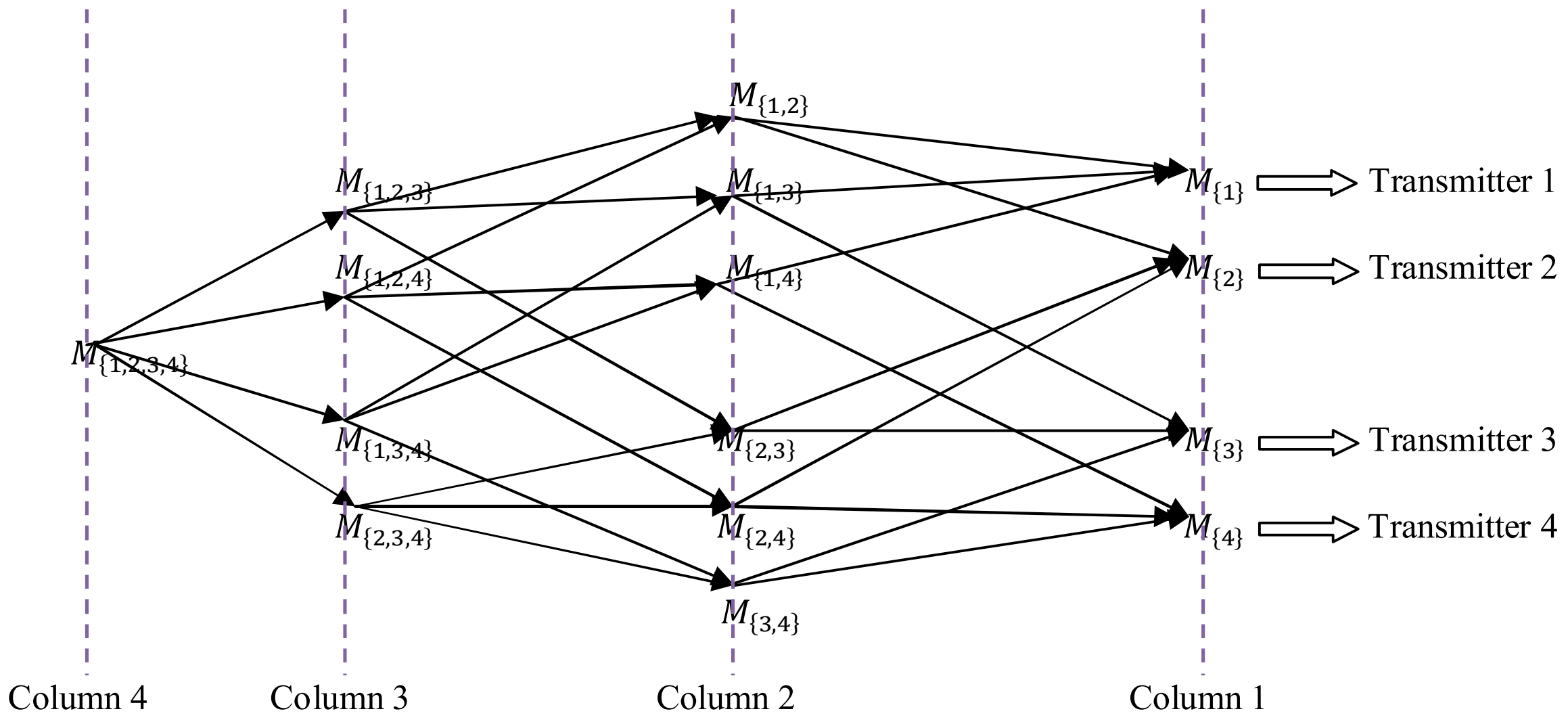

Figure 6. The message graph for a 4-transmitter MACCM with complete messages..

The message graph indeed has interesting properties, e.g., there exists a directed path connecting $M_{\Delta_1}$ to $M_{\Delta_2}$ if and only if $\Delta_2 \subseteq \Delta_1$. Let us now consider a network with any arbitrary arrangement of messages among transmitters (not necessarily complete set of messages as in Fig. 5). Assume the message set $\mathbb{M}$, where:

$$\mathbb{M} \subseteq \{M_\Delta : \Delta \subseteq \{1, \dots, K_1\}\} \tag{5}$$

are intended to be sent over the channel. To derive the message graph with respect to this scenario, in the message graph for MACCM with complete messages, first those messages which are not in $\mathbb{M}$ as well as the directed edges crossing at their places are removed. Then, we follow the resultant sub-graph from column $K_1$ towards column 1; a message $M_{\Delta_1}$ is connected to $M_{\Delta_2}$ by a directed edge if and only if $\Delta_2 \subseteq \Delta_1$ and there is no other message $M_{\Delta_3}$ with $\Delta_2 \subseteq \Delta_3 \subseteq \Delta_1$ and also there is no directed path connecting $M_{\Delta_1}$ to $M_{\Delta_2}$. For example, for the 4-transmitter MACCM the message graph with respect to the following message set:

$$\mathbb{M} = \{M_{\{1,2,3,4\}}, M_{\{2,3,4\}}, M_{\{1,2\}}, M_{\{1,4\}}, M_{\{3,4\}}, M_{\{1\}}, M_{\{2\}}, M_{\{3\}}\} \tag{6}$$

is given in Fig. 7. The red edges indicate that these edges have been added to the graph after removing non-desired messages and the edges crossing at their positions. Also, note that the dashed arrows in Fig. 7 are not really parts of the message graph; these arrows indeed depict the respective messages of Transmitter 4.

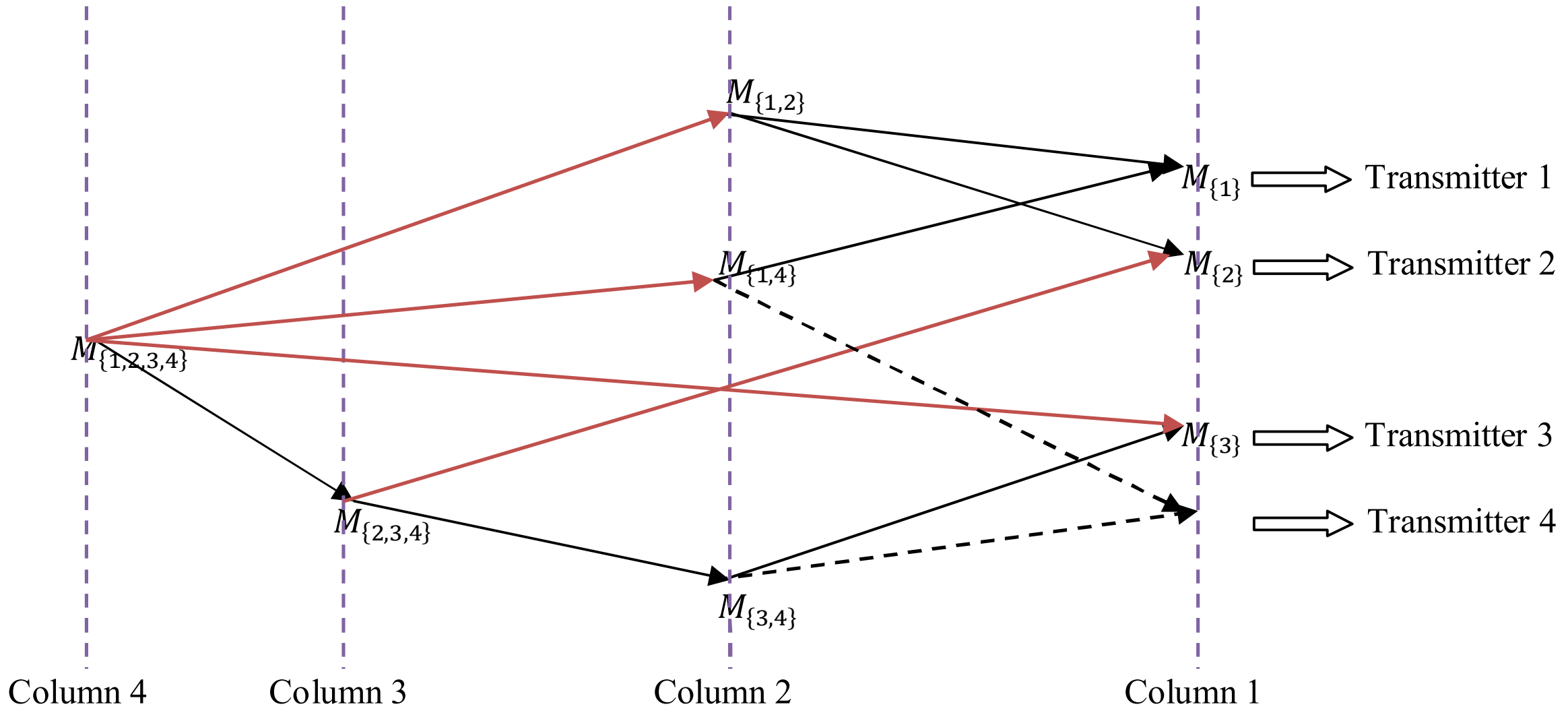

Figure 7. The message graph with respect to the messages (6). The blue edges these edges have been added to the graph after removing non-desired messages and the edges crossing at their places.

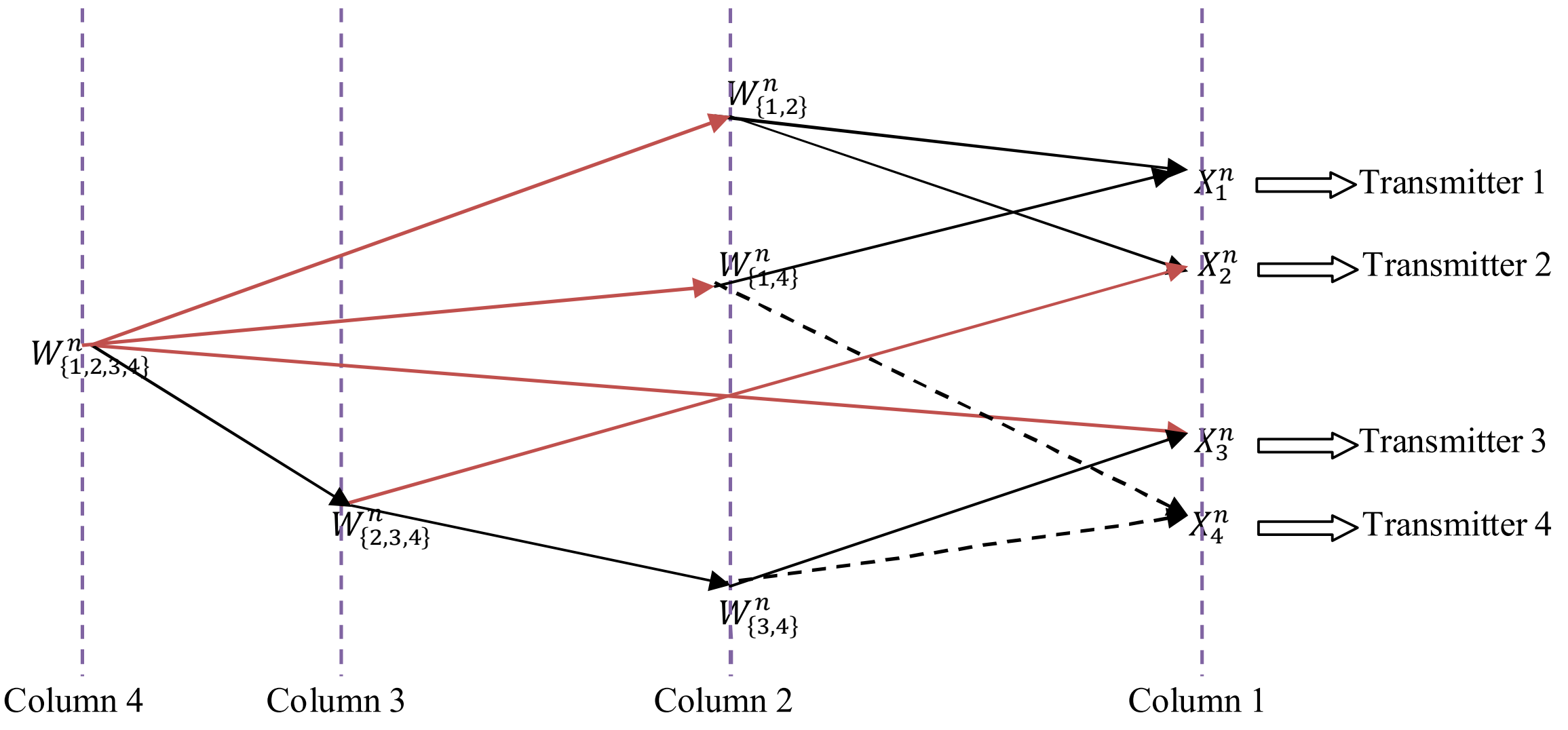

Figure 8. The encoding graph corresponding to the message graph in Fig. 7.

One can readily translate the message graph to the encoding graph. To this end, similar to the two-user case, each message in the message graph is replaced by a codeword. The message $M_\Delta$ is replaced by the codeword[3] $W_\Delta^n$, for $\Delta \subseteq \{1, \dots, K_1\}$. It is supposed that the message $M_{\{i\}}$ in Column 1 is replaced by the codeword $X_i^n$, $i = 1, \dots, K_1$; in other words, $W_{\{i\}} \equiv X_i$. Moreover, in the encoding graph for MACCM, we need to situate a codeword $X_i^n$ in the place of $M_{\{i\}}$, $i = 1, \dots, K_1$, even if this message does not appear in the message graph, i.e. it does not belong to $\mathbb{M}$. Then, in the resulting scheme each two codewords which are connected by an arrow form a superposition structure. Moreover, each codeword conveys the message which is replaced as well as all the messages conveyed by the cloud centers of that codeword. Fig. 8 represents the encoding graph with respect to the messages (6).

The codewords in the encoding graph are generated based on the following criterion.

***MACCM Encoding Graph Probability Law:*** *The codeword $W_\Delta^n$ is generated i.i.d. based on the PDF $P_{W_\Delta|\{W_\Gamma : \Delta \subsetneq \Gamma\},Q}$, $\forall \Delta \subseteq \{1, \dots, K_1\}$; also, $W_{\{i\}} \equiv X_i$, $i = 1, \dots, K_1$, is a deterministic function of $(\{W_\Gamma : \{i\} \subsetneq \Gamma\}, Q)$ provided that $M_{\{i\}}$ does not belong to $\mathbb{M}$. Note that $\mathbb{M}$ is the set of messages designated for transmission. Also, if the message $M_{\{1,\dots,K_1\}}$ belongs to $\mathbb{M}$, then there is no need for the time-sharing parameter $Q$. We call this the probability law of MACCM encoding graph.*

Note that, $\{W_\Gamma^n : \Delta \subsetneq \Gamma\}$ constitutes the set of all cloud centers of the codeword $W_\Delta^n$ in the encoding graph; in other words, each codeword in this set is connected to the codeword $W_\Delta^n$ through a directed path. For example, for the encoding graph of Fig. 8, which corresponds to the messages (6), the respective probability law is given below:

$$P_{W_{\{1,2,3,4\}}} P_{W_{\{2,3,4\}}|W_{\{1,2,3,4\}}} P_{W_{\{1,2\}}|W_{\{1,2,3,4\}}} P_{W_{\{1,4\}}|W_{\{1,2,3,4\}}} P_{W_{\{3,4\}}|W_{\{2,3,4\}},W_{\{1,2,3,4\}}}$$
$$\times P_{X_1|W_{\{1,2\}},W_{\{1,4\}},W_{\{1,2,3,4\}}} P_{X_2|W_{\{1,2\}},W_{\{2,3,4\}},W_{\{1,2,3,4\}}} P_{X_3|W_{\{3,4\}},W_{\{2,3,4\}},W_{\{1,2,3,4\}}} P_{X_4|W_{\{1,4\}},W_{\{3,4\}},W_{\{2,3,4\}},W_{\{1,2,3,4\}}} \quad (7)$$

where $P_{X_4|W_{\{1,4\}},W_{\{3,4\}},W_{\{2,3,4\}},W_{\{1,2,3,4\}}} \in \{0,1\}$ because the message $M_4$ does not belong to the messages $\mathbb{M}$ designated for transmission.

***Definition 1: MACCM Graph***

*A digraph is said to be a MACCM graph if it can be considered as a massage graph with respect to a MACCM with a given distribution of messages.*

[3] To avoid ambiguity, we use the letter "$W$" to denote the codewords' symbols for the encoding graph. Nevertheless, one can still use the letter "$M$". In this case, to derive the encoding graph, each message $M_\Delta$ in the message graph is replaced by the codeword $M_\Delta^n$. In other words, given a message graph, we can directly perceive the corresponding encoding scheme.

Here, we mention that the MACCM graphs are digraphs without any directed cycles; nevertheless, this does not mean that every acyclic digraph can be a MACCM graph. For example, in the MACCM graphs if two nodes (corresponding to some messages) are connected by a directed path of length two or more, then they are no longer connected by a directed edge. As we will see subsequently, the MACCM graphs are very useful tools, not only to describe a given encoding scheme for a certain channel model but also to design coding strategies with satisfactory performance for interference networks of arbitrary large size. In the following, using the MACCM graph we easily describe the capacity region of MACCM with any distribution of messages.

***Definition 2: Cloud Center and Satellite Messages;***

*Consider a MACCM graph with respect to a given message set $\mathbb{M}$. A message $M_{\Delta_1}$ is said to be a satellite for the message $M_{\Delta_2}$ provided that $\Delta_1 \subseteq \Delta_2$. Also, in this case, $M_{\Delta_2}$ is said to be a cloud center for $M_{\Delta_1}$.*

***Definition 3: Right-Sided Message Sets***

*Consider a MACCM with the associated message set $\mathbb{M}$ where $\mathbb{M} \subseteq \{M_\Delta : \Delta \subseteq \{1,\dots,K_1\}\}$. A subset $\Omega$ of $\mathbb{M}$ is said to be right-sided provided that:*

$$if\ \ M_\Delta \in \Omega \ \ \Rightarrow \forall\, \bar{\bar{\Delta}} \subseteq \Delta :\ M_{\bar{\bar{\Delta}}} \in \Omega \tag{8}$$

From the viewpoint of MACCM message graph, $\Omega$ is right-sided provided that if $M_\Delta$ belongs to $\Omega$, then all messages in the right hand side of $M_\Delta$, which are connected to it by a directed path, also belong to $\Omega$.

Using the developed graphical illustrations, we can now readily describe the capacity region of the $K_1$-transmitter MACCM with any arbitrary distribution of messages.

***Proposition 1)*** *Consider the $K_1$-transmitter MACCM in Fig. 4 with the associated message set $\mathbb{M} = \left\{M_{\Delta_1},\dots,M_{\Delta_{\|\mathbb{M}\|}}\right\}$ where $\Delta_l \subseteq \{1,\dots,K_1\}$, $l = 1,\dots,\|\mathbb{M}\|$. The capacity region denoted by $\mathcal{C}_{\mathbb{M}}^{MACCM}$ is given below:*

$$\mathcal{C}_{\mathbb{M}}^{MACCM} \triangleq \bigcup_{\mathcal{P}_{\mathbb{M}}^{MACCM}} \left\{ \begin{array}{l} \left(R_{\Delta_1},\dots,R_{\Delta_{\|\mathbb{M}\|}}\right) \in \mathbb{R}_+^{\|\mathbb{M}\|} : \\ \forall\, \Omega \subseteq \mathbb{M} :\ \ \Omega\ is\ right-sided \\ \displaystyle\sum_{\substack{l:\\ M_{\Delta_l} \in \Omega}} R_{\Delta_l} \leq I(\Omega; Y|\mathbb{M}-\Omega, Q) \end{array} \right\} \tag{9}$$

*where $\mathcal{P}_{\mathbb{M}}^{MACCM}$ denotes the set of all joint PDFs $P_{Q\mathbb{M}X_1\dots X_{K_1}}$ satisfying:*

$$P_{Q\mathbb{M}X_1\dots X_{K_1}} = P_Q \prod_{l=1}^{\|\mathbb{M}\|} P_{M_{\Delta_l}} \prod_{i=1}^{K_1} P_{X_i|\mathbb{M}_{X_i},Q} \tag{10}$$

*Moreover, $P_{X_i|\mathbb{M}_{X_i}Q} \in \{0,1\}, i = 1,\dots,K_1$, i.e., $X_i$ is a deterministic function of $\left(\mathbb{M}_{X_i}, Q\right)$.*

It should be noted that, similar to the rate region (4), the random variables appearing in expression (9) are composed of the receiver signal, i.e., $Y$, the messages $M_{\Delta_1},\dots,M_{\Delta_{\|\mathbb{M}\|}}$, and the time-sharing RV, i.e., $Q$. This representation of the capacity region can be easily interpreted using the encoding graph. First note that we can easily re-express the capacity region using the encoding graph. It is sufficient to exchange each message in the expression (9) with the RV from which the respective codeword of that message in the encoding graph is constructed. Also, the time-sharing parameter $Q$ remains unchanged; however, if the message $M_{\{1,2,\dots,K_1\}}$ exists among the designated messages for transmission, i.e., $\mathbb{M}$, it is possible to combine $Q$ with $W_{\{1,2,\dots,K_1\}}$ and represent them with a unique

RV similar to the two-user case in (1). The joint PDFs associated to the resultant rate region are also given according to the MACCM encoding graph probability law. These joint PDFs are factorized as follows:

$$\prod_{i=K_1:-1:1} \prod_{\substack{\Delta: \\ M_\Delta \in \mathbb{M}, \\ \|\Delta\|=i}} P_{W_\Delta | \{W_\Gamma : \Delta \subsetneq \Gamma\}, Q} \tag{11}$$

The notation $i = K_1: -1: 1$ in the factorization (11) indicates that the counter $i$ is decreasing by one in each round. In fact, to derive this factorization, we parse the MACCM graph from column $K_1$ towards column 1 and write the probability law based on which each codeword is generated (there is no preference in treating two codewords positioned in the same column). Also, we recall that we have assumed by default that $W_{\{i\}} = X_i, i = 1, \dots, K_1$.

Note that from the view point of encoding graph, a right-sided set of codewords contains also all their satellites, (the right-sided sets are closed with respect to "being satellite"). Now consider the following constraint of the description (9):

$$\sum_{\substack{l: \\ M_{\Delta_l} \in \Omega}} R_{\Delta_l} \leq I(\Omega; Y | \mathbb{M} - \Omega, Q)$$

When this expression is translated into the encoding graph representation, it is the necessary condition for vanishing the error probability of incorrect decoding of the codewords with respect to $\Omega$. Therefore, the capacity expression (9) indeed reflects a fact regarding the decoding of two codewords which form a superposition structure: incorrect decoding of the cloud center leads to incorrect decoding of the satellite codeword.

***Example 1;***

Consider a 4-transmitter MACCM in which the following messages are designated for transmission:

$$\mathbb{M} = \{M_{\{1,2,3\}}, M_{\{2,4\}}, M_{\{1,2\}}, M_{\{3\}}\}$$

In Fig. 9, we have depicted the message graph and encoding graph (conjugated with each other) for this scenario.

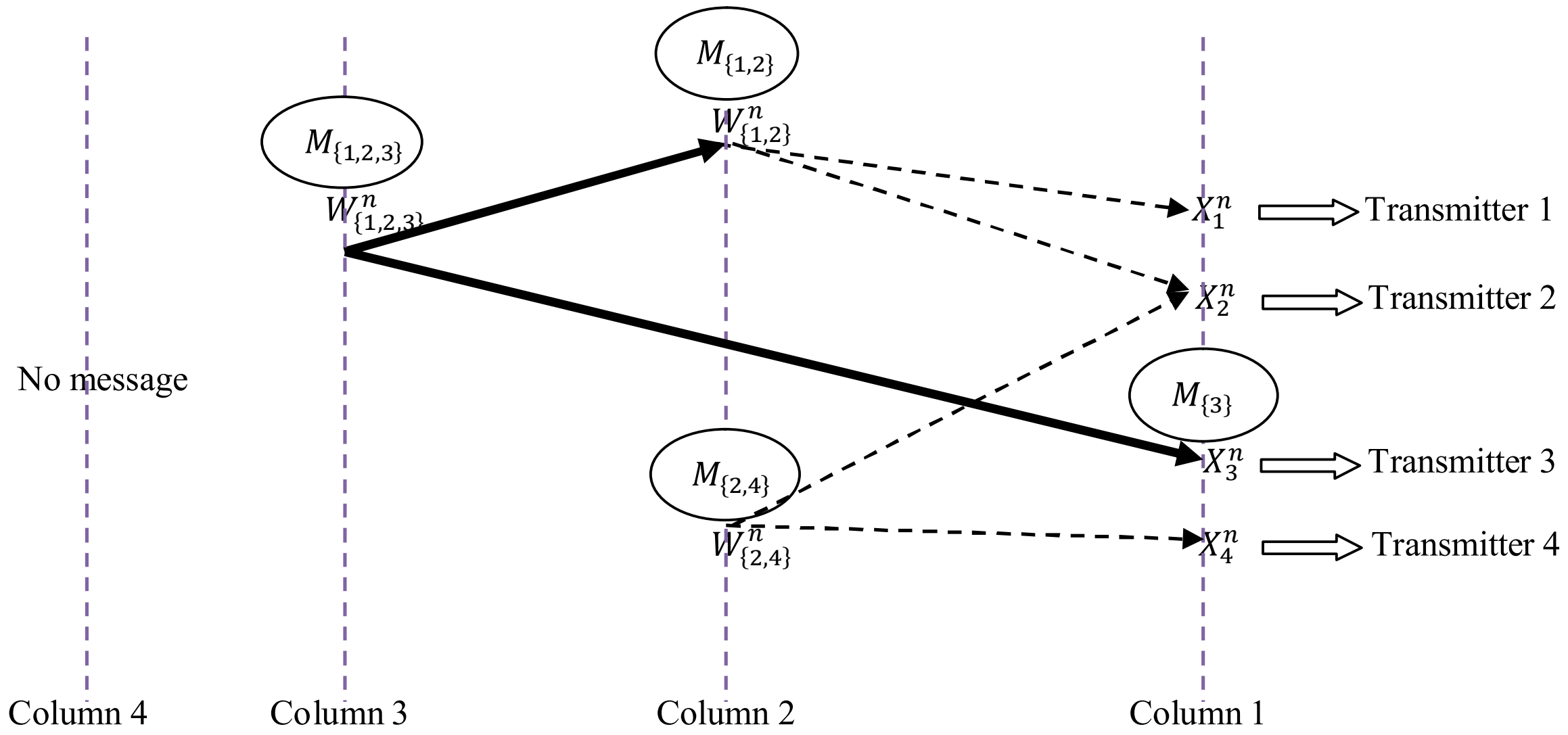

Figure 9. The message and the encoding graphs corresponding to the MACCM scenario of Example 1.

From Proposition 1, the capacity region for this scenario is given by:

$$\bigcup_{\mathcal{P}^{EX\ IV.1}} \left\{ \begin{array}{l} \left(R_{\{1,2,3\}}, R_{\{1,2\}}, R_{\{2,4\}}, R_{\{3\}}\right) \in \mathbb{R}_+^4 : \\ , \\ R_{\{3\}} \leq I\left(M_{\{3\}}; Y \middle| M_{\{2,4\}}, M_{\{1,2\}}, M_{\{1,2,3\}}, Q\right) \\ R_{\{1,2\}} \leq I\left(M_{\{1,2\}}; Y \middle| M_{\{2,4\}}, M_{\{3\}}, M_{\{1,2,3\}}, Q\right) \\ R_{\{2,4\}} \leq I\left(M_{\{2,4\}}; Y \middle| M_{\{1,2\}}, M_{\{3\}}, M_{\{1,2,3\}}, Q\right) \\ R_{\{1,2\}} + R_{\{3\}} \leq I\left(M_{\{1,2\}}, M_{\{3\}}; Y \middle| M_{\{2,4\}}, M_{\{1,2,3\}}, Q\right) \\ R_{\{2,4\}} + R_{\{3\}} \leq I\left(M_{\{2,4\}}, M_{\{3\}}; Y \middle| M_{\{1,2\}}, M_{\{1,2,3\}}, Q\right) \\ R_{\{1,2\}} + R_{\{2,4\}} \leq I\left(M_{\{2,4\}}, M_{\{1,2\}}; Y \middle| M_{\{3\}}, M_{\{1,2,3\}}, Q\right) \\ R_{\{1,2\}} + R_{\{2,4\}} + R_{\{3\}} \leq I\left(M_{\{2,4\}}, M_{\{1,2\}}, M_{\{3\}}; Y \middle| M_{\{1,2,3\}}, Q\right) \\ R_{\{1,2,3\}} + R_{\{1,2\}} + R_{\{3\}} \leq I\left(M_{\{1,2,3\}}, M_{\{1,2\}}, M_{\{3\}}; Y \middle| M_{\{2,4\}}, Q\right) \\ R_{\{1,2,3\}} + R_{\{1,2\}} + R_{\{3\}} + R_{\{2,4\}} \leq I\left(M_{\{1,2,3\}}, M_{\{1,2\}}, M_{\{3\}}, M_{\{2,4\}}; Y \middle| Q\right) \end{array} \right\} \tag{12}$$

where $\mathcal{P}^{EX\ IV.1}$ denotes the set of all joint PDFs given below:

$$P_Q \times P_{M_{\{3\}}} \times P_{M_{\{1,2\}}} \times P_{M_{\{2,4\}}} \times P_{M_{\{1,2,3\}}} \times P_{X_1|M_{\{1,2\}},M_{\{1,2,3\}},Q} \times P_{X_2|M_{\{1,2\}},M_{\{2,4\}},M_{\{1,2,3\}},Q} \times P_{X_3|M_{\{3\}},M_{\{1,2,3\}},Q} \times P_{X_4|M_{\{2,4\}},Q} \tag{13}$$

Using the MG/EG translation, one can simply re-express the capacity region based on the encoding graph to derive a characterization similar to (1).

Let us now deal with the problem of determining the sum-rate capacity of the MACCM with a certain distribution of messages. First consider the two-user channel. Using the capacity expression in (1), we can readily derive the sum-rate capacity of this channel as follows:

$$\max_{P_{W_0}P_{X_1|W_0}P_{X_2|W_0}} I(X_1, X_2; Y) \tag{14}$$

In fact, one can easily verify that the following point:

$$(R_0, R_1, R_2) = \left(I(W_0; Y), I(X_1; Y|W_0), I(X_2; Y|X_1, W_0)\right)_{P_{W_0}P_{X_1|W_0}P_{X_2|W_0}} \tag{15}$$

belongs to the capacity region. This point can also be achieved using a successive decoding strategy at the decoder: the receiver decodes the common message $M_0$ conveyed by $W_0^n$ first, then decodes the private message $M_1$ conveyed by $X_1^n$, and finally decodes the private message $M_2$ conveyed by $X_2^n$. Moreover, by considering the expression (4), we can also re-write the sum-rate capacity as follows:

$$\max_{\substack{P_Q P_{M_0} P_{M_1} P_{M_2} \\ X_i = f_i(M_0, M_i, Q)}} I(M_0, M_1, M_2; Y|Q) \tag{16}$$

In fact, the following point belongs to the capacity region:

$$(R_0, R_1, R_2) = \left(I(M_0; Y|Q), I(M_1; Y|M_0, Q), I(M_2; Y|M_1, M_0, Q)\right)_{\substack{P_Q P_{M_0} P_{M_1} P_{M_2} \\ X_i = f_i(M_0, M_i, Q)}} \tag{17}$$

Now, consider the maximization in (14). It is readily derived that $W_0 \equiv X_1$ (or $W_0 \equiv X_2$) is the solution to this optimization problem. Therefore, the sum-rate capacity is given by:

$$\max_{P_{X_1 X_2}} I(X_1, X_2; Y) \tag{18}$$

An immediate consequence of this observation is that the sum-rate capacity for the two-user MACCM is achieved at the corner point of the capacity region which is the maximum achievable common rate $R_0$. In other words, a scheme to achieve the rate (18) is that both transmitters cooperatively transmit only the common message and withdraw transmission of private messages. Consequently, using (16) we can re-express the sum-rate capacity as follows:

$$\max_{\substack{P_Q P_{M_0} \\ X_i=f_i(M_0,Q)}} I(M_0;Y|Q) \tag{19}$$

It should be noted that (19) is the sum-rate capacity only when there exists the common message. But for the case where no common message is available in the network, the sum-rate capacity is given by:

$$\max_{P_Q P_{X_1|Q} P_{X_2|Q}} I(X_1,X_2;Y|Q) \equiv \max_{\substack{P_Q P_{M_1} P_{M_2} \\ X_i=f_i(M_i,Q)}} I(M_1,M_2;Y|Q) \tag{20}$$

To achieve the sum-rate (20), each transmitter transmits separately its respective private message and the decoder jointly decodes both messages.

Next, consider the $K_1$-transmitter MACCM in which the messages $\mathbb{M}$ are intended to be transmitted, where $\mathbb{M} \subseteq \mathbb{M}_{2^{K_1}-1} = \{M_\Delta : \Delta \subseteq \{1,\dots,K_1\}\}$.

***Proposition 2)*** *The sum-rate capacity of the $K_1$-transmitter MACCM with the messages $\mathbb{M}$ is given by:*

$$\mathcal{C}_{\mathbb{M}-sum}^{MACCM} = \max_{\mathcal{P}_{\mathbb{M}-sum}^{MACCM}} I(\mathbb{M};Y|Q) \tag{21}$$

*where $\mathcal{P}_{\mathbb{M}-sum}^{MACCM}$ denotes the set of all joint PDFs $P_{Q\mathbb{M}X_1\dots X_{K_1}}$ that can be factorized as (10).*

*Proof of Proposition 2)* We expand the messages into the MACCM graph. Then, we follow this graph from column $K_1$ towards column 1, each column in up to down direction (this is an arbitrary choice), and enumerate/re-label the messages as $\{M_1,\dots,M_{\|\mathbb{M}\|}\}$. For example, in the MACCM graph in Fig. 9 the messages are re-labeled as:

$$M_{\{1,2,3\}} \cong M_1, \quad M_{\{1,2\}} = M_2, \quad M_{\{2,4\}} = M_3, \quad M_{\{3\}} = M_4 \tag{22}$$

Now consider the following point:

$$\Big(I(M_1;Y|Q), I(M_2;Y|M_1,Q), \dots, I\big(M_{\|\mathbb{M}\|};Y\big|M_{\|\mathbb{M}\|-1}, M_{\|\mathbb{M}\|-2},\dots,M_1,Q\big)\Big) \in \mathbb{R}_+^{\|\mathbb{M}\|} \tag{23}$$

which is computed by $P_Q \prod_{M_\Delta \in \mathbb{M}} P_{M_\Delta} \prod_{i=1}^{K_1} P_{X_i|\mathbb{M}_{X_i}Q}$. One can easily see that the point (23) belongs to the capacity region given by (9). In fact, by this choice the following constraint holds:

$$\sum_{\Gamma \in \Omega} R_\Gamma \le I(\Omega;Y|\Omega^c,Q)$$

for all $\Omega \subseteq \mathbb{M}$. Note that the point (23) can also be achieved using a successive decoding strategy. To see this, considering the encoding graph, the receiver successively decodes the codewords conveying the messages $M_1,\dots,M_{\|\mathbb{M}\|}$. ■

As we previously demonstrated, to achieve the sum-rate capacity for the two-user MACCM a transmission scheme is that both transmitters cooperatively send only the common message and withdraw the other messages. This observation extends to any arbitrary MACCM. In the following, we present a method to determine those messages which should be essentially transmitted to achieve the sum-rate capacity in a $K_1$-transmitter MACCM with any arbitrary distribution of messages. The goal is to determine a set $\mathbb{M}^* \subseteq \mathbb{M}$ such that:

$$\mathcal{C}_{\mathbb{M}-sum}^{MACCM} = \max_{\mathcal{P}_{\mathbb{M}^*-sum}^{MACCM}} I(\mathbb{M}^*; Y|Q) \tag{24}$$

where $\mathcal{P}_{\mathbb{M}^*-sum}^{MACCM}$ is given by (10) except that $\mathbb{M}$ should be replaced by $\mathbb{M}^*$. The key (inspired by the two-user channel) is that if $M_\Delta \in \mathbb{M}^*$, then $M_\Gamma \notin \mathbb{M}^*$, where $\Gamma \subseteq \Delta$. Therefore, we follow the message graph from column $K_1$ toward column 1.

In column $i$, $i \in [1:K_1]$:

1. The message $M_\Delta$ belongs to $\mathbb{M}^*$ if it has not been withdrawn in the previous rounds;
2. All messages $M_\Gamma$ which are connected to $M_\Delta$ by a directed path, i.e., $\Gamma \subseteq \Delta$, are withdrawn from the transmission.

In Fig. 10, we have determined $\mathbb{M}^*$ for the scenario of Example 1 using the above algorithm. A message with the mark “ ” belongs to $\mathbb{M}^*$ while the one with the mark “ ” belongs to $\mathbb{M} - \mathbb{M}^*$. Therefore, to achieve the sum-rate capacity in Example 1, it is only required to consider the transmission of the messages $M_{\{1,2,3\}}$ and $M_{\{2,4\}}$. The sum-rate capacity is given by:

$$\mathcal{C}_{sum}^{EX.1} = \max_{\mathcal{P}^{EX\,1}} I\left(M_{\{1,2,3\}}, M_{\{2,4\}}; Y\middle|Q\right) \tag{25}$$

One can easily show that the expression (25) is equivalent to the following:

$$\mathcal{C}_{sum}^{EX.1} = \max_{P_Q P_{X_1X_3|Q} P_{X_4|Q} P_{X_2|X_1X_3X_4Q}} I(X_1, X_2, X_3, X_4; Y|Q) \tag{26}$$

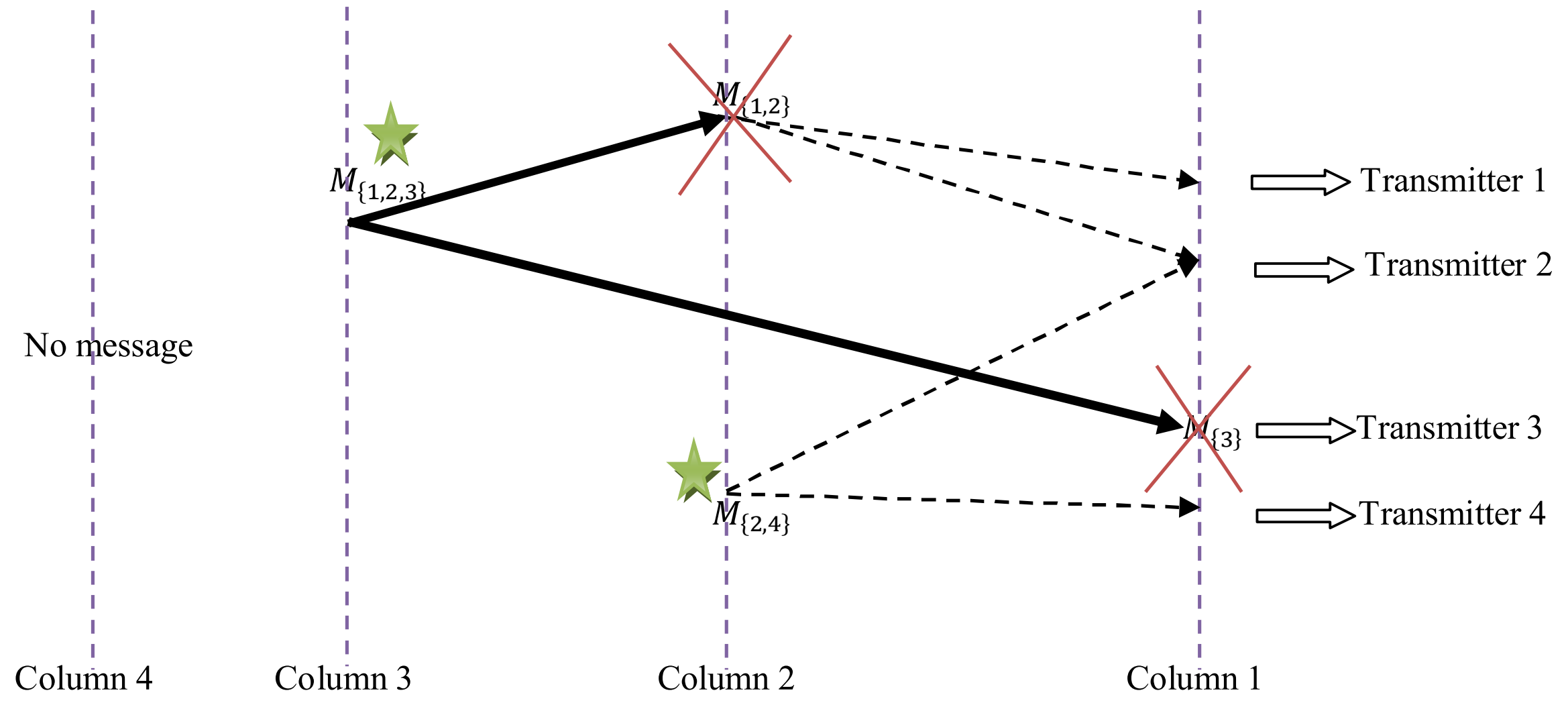

Figure 10. This figure showes the messages $\mathbb{M}^*$ which are required to be considered for transmission scheme for the MACCM of Example 1 to achieve the sum-rate capacity. A message with the mark “ ” belongs to $\mathbb{M}^*$, while the one with the mark “ ” is withdrawn from the transmission.

Finally, we remark that the sum-rate capacity for the MACCM with a given arbitrary set of messages $\mathbb{M}$ can be always written as follows:

$$\mathcal{C}_{\mathbb{M}-sum}^{MACCM} = \max_{\mathcal{P}_{\mathbb{M}^*-sum}^{MACCM}} I\left(X_1, X_2, \ldots, X_{K_1}; Y\middle|Q\right) \tag{27}$$

This formulation does not contain any auxiliary random variable; however, the probability distribution which is induced on the input signals $X_1, X_2, \ldots, X_{K_1}$ should be determined based on the messages $\mathbb{M}^*$ determined in (24). For the scenario of Example 1, this distribution is given by $P_{QX_1X_2X_3X_4} = P_Q P_{X_1X_3|Q} P_{X_4|Q} P_{X_2|X_1X_3X_4Q}$.

## II.B) The BC with Common Messages (BCCM)

Next, we investigate the BC with common messages. This is a communication system with a transmitter and several receivers where (in the general case) the transmitter sends an individual message for each subset of receivers. In the following, we present some new insights regarding the nature of information flow for this system. The results in this subsection are indeed very insightful for our future derivations in the rest of the paper. First, consider the two-user BC with a common message as shown in Fig. 11.

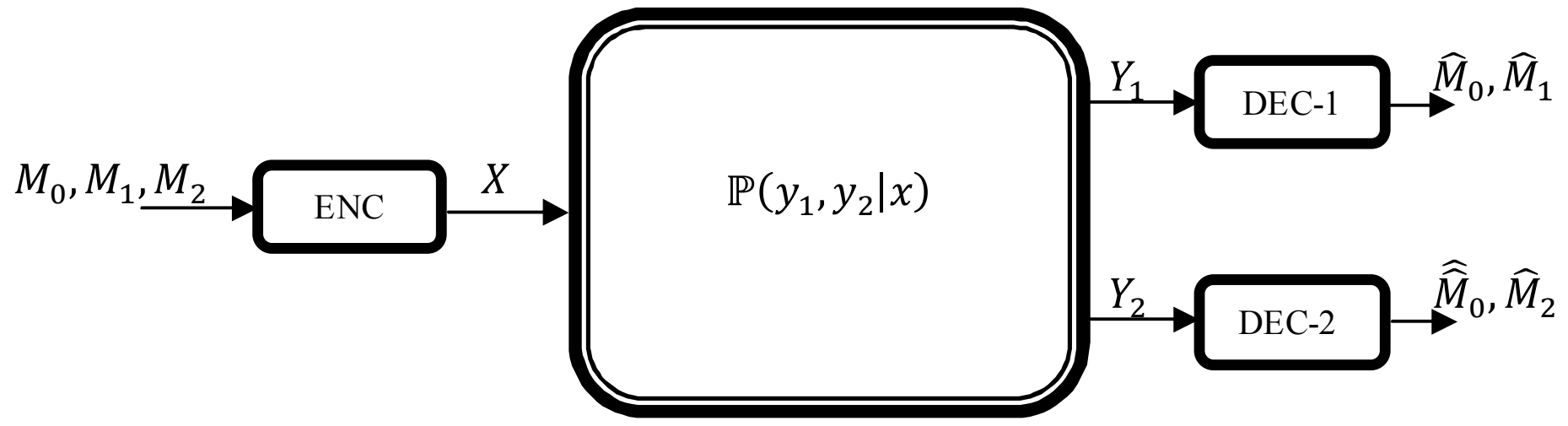

Figure 11. The two-user Broadcast Channel (BC) with common message.

This communication scenario is obtained from the general interference network defined in Section II.B of Part I [1] (see also Fig. 14 in Subsection III) by setting $K_1 = 1, K_2 = 2$, $\mathbb{M} = \{M_0, M_1, M_2\}$, $\mathbb{M}_X = \{M_0, M_1, M_2\}$, $\mathbb{M}_{Y_1} = \{M_0, M_1\}$, and $\mathbb{M}_{Y_2} = \{M_0, M_2\}$, (note that since for the BC there is only one transmitter, we omit the subscripts from $\mathcal{X}_1$, $X_1$, $x_1$).

Unfortunately, unlike the MAC, the capacity region of the BC is unknown except for some special cases. As discussed in Part I [1], the best achievable rate region for the 2-receiver channel is due to Marton [12] which is given below.

$$\mathfrak{R}_i^{Marton} \triangleq \bigcup_{P_{WUVX}(w,u,v,x)} \left\{ \begin{array}{l} (R_0, R_1, R_2) \in \mathbb{R}_+^3 : \\ R_0 + R_1 \leq I(W, U; Y_1) \\ R_0 + R_2 \leq I(W, V; Y_2) \\ R_0 + R_1 + R_2 \leq I(W, U; Y_1) + I(V; Y_2|W) \\ \qquad\qquad -I(U; V|W) \\ R_0 + R_1 + R_2 \leq I(U; Y_1|W) + I(W, V; Y_2) \\ \qquad\qquad -I(U; V|W) \\ 2R_0 + R_1 + R_2 \leq I(W, U; Y_1) + I(W, V; Y_2) \\ \qquad\qquad -I(U; V|W) \end{array} \right\} \tag{28}$$

Here, we briefly discuss the Marton's coding scheme. The key idea in the encoding step is the use of mutual covering lemma [11]. Roughly speaking, the common message $M_0$ is encoded by a codeword $W^n$ generated based on $P_W$. For each of the private messages, a bin of codewords is randomly generated superimposing on the common message codeword $W^n$: The bin corresponding to $M_1$ contains the codewords $U^n$ generated based on $P_{U|W}$ and the one for $M_2$ contains the codeword $V^n$ generated based on $P_{V|W}$. These two bins are explored against each other to find a jointly typical pair of codewords. By using the mutual covering lemma [11], the sizes of the bins are selected sufficiently large such that the existence of such typical pair of codewords is guaranteed. Next, the codeword $X^n$ is superimposed on the designated jointly typical codewords $W^n, U^n, V^n$ according to $P_{X|WUV}$ and is sent over the channel. Each receiver decodes its respective codewords (the first one decodes $W^n, U^n$, and the second one decodes $W^n, V^n$) using a jointly typical decoder. The resultant achievable rate region is further enlarged (leading to (28)) by the fact that if the rate triple $(R_0, R_1, R_2) \in \mathbb{R}_+^3$ is achievable for the BC, then $(R_0 - \tau_1 - \tau_2, R_1 + \tau_1, R_2 + \tau_2) \in \mathbb{R}_+^3$ is also achievable. The graphical representation of the Marton's coding has been shown in Fig. 12.

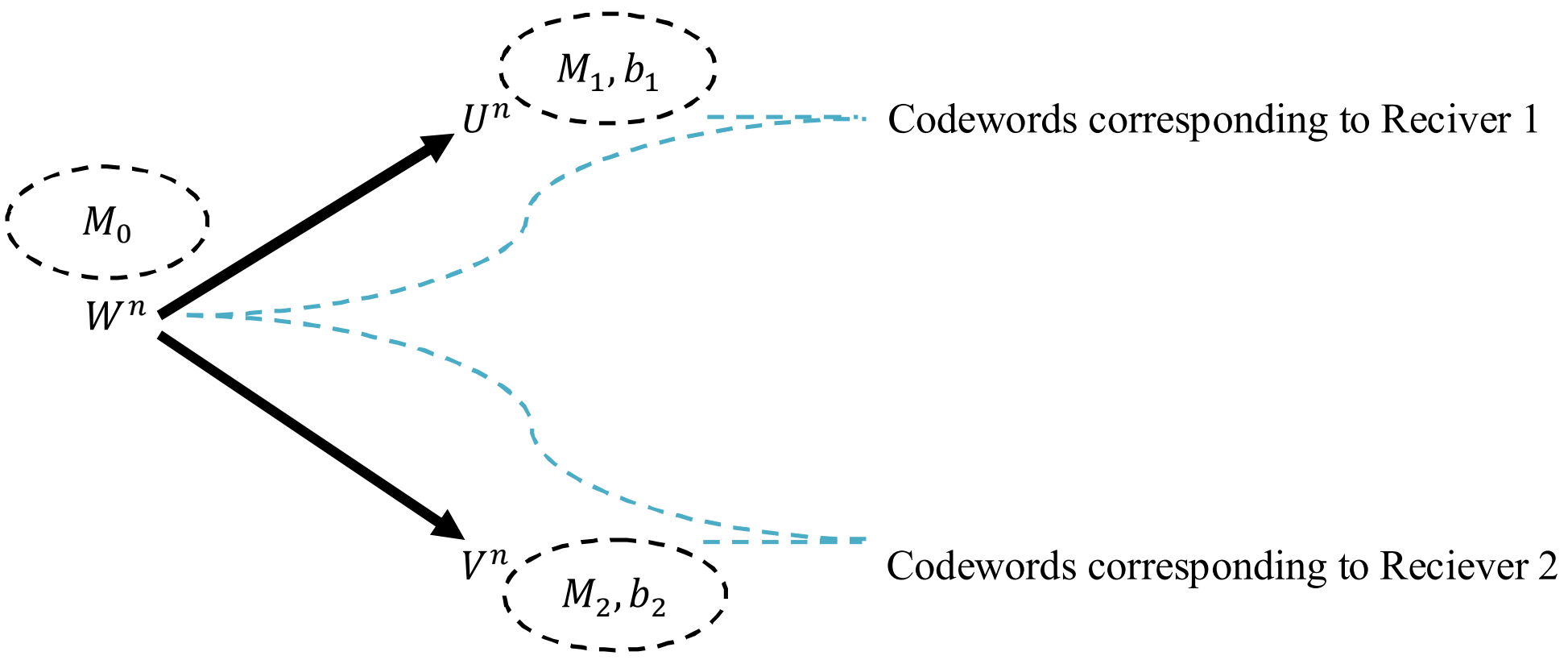

Figure 12. The graphical illustration of the Marton's coding scheme for the two-user BC.

It is interesting to note that the superposition structures among the generated codewords in the Marton's coding scheme for the two-user BCCM are exactly the same as those for the two-user MACCM shown in Fig. 1. The only difference in the encoding scheme is that, unlike the MAC, for the BC we can apply the binning technique since all the messages are available at one transmitter. Using the binning scheme, we can construct the transmitted codewords jointly typical with the PDF $P_{WUVX}$, which yields a larger achievable rate region than the case where the messages are encoded using only the superposition coding according to the PDF $P_W P_{U|W} P_{V|W} P_{X|WUV}$.

As remarked in our previous paper [7] presented in [8], one can consider some new coding schemes for the two-user BC other than the Marton's one. For example, it is possible to encode all the messages (both common and private messages) using only the binning technique, i.e., without superposition coding. In this scheme, a bin of codewords is generated (the bins are generated independently) with respect to each message and then these three bins are explored against each other to find a jointly typical triple. By using the multivariate covering lemma [11], the sizes of the bins are selected large enough to guarantee that there exists such a triple of codewords. The transmitter then generates its codeword superimposing on this jointly typical triple and sends it over the channel. Each receiver decodes its respective messages using a jointly typical decoder. Other coding strategies are also available. We have examined such coding schemes and found that all the resulting achievable rate regions are equivalent to the Marton's region if not its subset. In this paper, we omit the details of the corresponding algebraic computations. Therefore, we can conclude that for broadcasting both common and private messages, it is more beneficial to encode the private messages superimposing on top of the common messages. Using this general insight, we will propose an achievability scheme for transmission of general message sets over the multi-receiver BCCM such that the superposition structures among the generated codewords are exactly similar to the multi-transmitter MACCM (Part V of our paper [5]). To derive these superposition structures, it is sufficient to look at the receivers of the BCCM as the transmitters of a MACCM from the viewpoint of the corresponding messages. The details can be found in [7, 8]. The main scope of the rest of this subsection is to determine the sum-rate capacity for the degraded BCCMs.

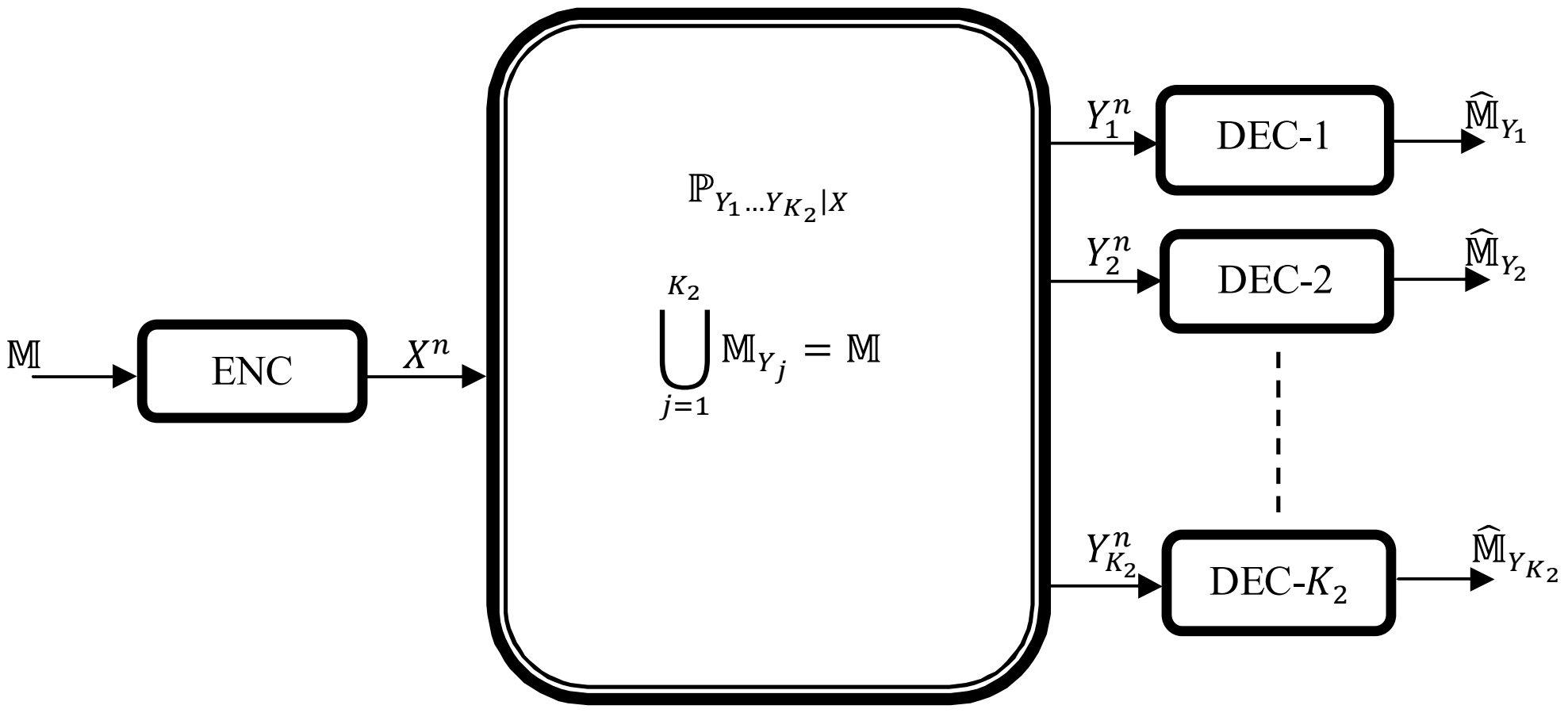

Figure 13. The general multi-user BC with Common Messages (BCCM).

Let us now consider a BCCM with $K_2$ receivers as shown in Fig. 13. This scenario is derived from the general interference network defined in Subsection II.B of Part I [1] (see also Fig. 14 in Section III) by setting $K_1 = 1$ and $\mathbb{M} = \mathbb{M}_X$. In the general case, the transmitter sends an individual message to each nonempty subset of the receivers; therefore $2^{K_2} - 1$ messages are designated for transmission. We refer to this setup as the BCCM with complete set of messages. Similar to the MACCM, we can label (in superscript) each message by a nonempty subset of $\{1, \dots, K_2\}$ such that it shows to which subset of receivers the message is transmitted. For example, $M^{\{1,2,3\}}$ is a message which is sent to the Receivers 1, 2 and 3. Such representation of the messages is very useful for our future derivations.

As mentioned before, the capacity region of the BC is still unknown even for two-receiver channel. An important class of BCs for which the capacity region is known for all distributions of messages, is the degraded BCs.

***Definition 4:*** *The $K_2$-receiver BC, as depicted in Fig. 13, is said to be physically degraded if:*

$$\mathbb{P}_{Y_1 \dots Y_{K_2}|X}(y_1, \dots, y_{K_2}|x) = \mathbb{P}_{Y_1|X}(y_1|x)\mathbb{P}_{Y_2|Y_1}(y_2|y_1) \dots \mathbb{P}_{Y_{K_2}|Y_{K_2-1}}(y_{K_2}|y_{K_2-1}) \tag{29}$$

*In other words, $X \to Y_1 \to \cdots \to Y_{K_2-1} \to Y_{K_2}$ forms a Markov chain.*

A larger class of BCs is the stochastically degraded BCs, which behave in essence similar to physically degraded channels.

***Definition 5:*** *The $K_2$-Receiver BC, as depicted in Fig. 13, is said to be stochastically degraded if there exist some transition probability functions $\tilde{\mathbb{P}}_{Y_2|Y_1}(y_2|y_1), \dots, \tilde{\mathbb{P}}_{Y_{K_2}|Y_{K_2-1}}(y_{K_2}|y_{K_2-1})$ such that:*

$$\mathbb{P}_{Y_j|X}(y_j|x) = \sum_{y_1,\dots,y_{j-1}} \mathbb{P}_{Y_1|X}(y_1|x)\tilde{\mathbb{P}}_{Y_2|Y_1}(y_2|y_1) \times \dots \times \tilde{\mathbb{P}}_{Y_j|Y_{j-1}}(y_j|y_{j-1}), \qquad j = 2, \dots, K_2 \tag{30}$$

Note that the stochastically degraded BCs strictly contain the class of physically degraded channels as a subset [13]. As discussed in Subsection II.B of Part I [1], the capacity region of all interference networks depend only on marginal distributions of network transition probability function; especially for the BC, the capacity region depends on $\mathbb{P}_{Y_j|X}(y_j|x), j = 1, \dots, K_2$. Therefore, no distinction is required to be made between stochastically degraded and physically degraded BCs, and hereafter, we refer to both as the degraded BCs.

The capacity region of the degraded BCs was determined in [14-16] for the case where the transmitter sends an independent message for each receiver. This capacity result, which is derived using superposition coding technique, extends to the scenario of transmitting any arbitrary set of messages.

Our discussion in the rest of this subsection is regarded to derive the sum-rate capacity of the degraded BCCM with an arbitrary distribution of messages. Let us first consider the two-user channel. The sum-rate capacity of the two-user degraded BC with common message shown in Fig. 11 is given below:

$$\max_{P_X(x)} I(X; Y_1)$$

This result can be simply proved as follows. Using the cut-set outer bound for multi-terminal channels [10], we have:

$$\mathcal{C}_{sum} \leq \max_{P_X(x)} I(X; Y_1, Y_2) = \max_{P_X(x)} I(X; Y_1) \tag{31}$$

where the equality above holds because $Y_2$ is a degraded version of $Y_1$. On the other hand, note that $\max_{P_X(x)} I(X; Y_1)$ is the maximum achievable rate for $M_1$, that is the private message with respect to the stronger receiver. Accordingly, to achieve this rate, the transmitter sends the private message corresponding to the stronger receiver at its capacity rate, i.e., $\max_{P_X(x)} I(X; Y_1)$; meanwhile, it does not transmit the other messages. Therefore, the sum-rate capacity is achieved by transmitting only one message out of all messages, which should be chosen intelligently. This is a simple yet fundamental property of the degraded BCs which constitutes one of the keys to derive the sum-rate capacity for any arbitrary degraded interference network.

Now, let us consider the $K_2$-Receiver degraded BCCM with any arbitrary set of messages. In the next proposition, we determine the sum-rat capacity of this channel.

***Proposition 3)*** *Consider the $K_2$-receiver degraded BCCM in (29)-(30). Assume that the messages $\mathbb{M} = \{M^{\nabla_1}, \dots, M^{\nabla_{\|\mathbb{M}\|}}\}$ are designated for transmission, where $\mathbb{M} \subseteq \mathbb{M}^{2^{[1:K_2]}} = \{M^{\nabla} : \nabla \subseteq \{1, \dots, K_2\}, \nabla \neq \emptyset\}$. The sum-rate capacity, denoted by $\mathcal{C}_{sum}^{BCCM-deg}$, is given by:*

$$\mathcal{C}_{sum}^{BCCM-deg} = \max_{P_X(x)} I(X; Y_\Theta) \tag{32}$$

*where,*

$$\Theta \triangleq \min\{\max \nabla : \nabla \subseteq \{1, \dots, K_2\}, M^{\nabla} \in \mathbb{M}\} = \min\{\max \nabla_1, \dots, \max \nabla_{\|\mathbb{M}\|}\} \tag{33}$$

*also, $\max \nabla$ is the maximum elements which belongs to the set $\nabla$.*

*Proof of Proposition 3)* Let us first consider the scenario in which only one message, e.g., $M^{\nabla} \in \mathbb{M}$, is sent over the channel. In this case, the message $M^{\nabla}$ is commonly sent for all receivers $Y_j$ with $j \in \nabla$. Therefore, the capacity is given by:

$$\max_{P_X(x)} \min_{j \in \nabla} \{I(X; Y_j)\} = \max_{P_X(x)} I(X; Y_{\max \nabla}) \tag{34}$$

where the equality in (34) is due to the degradedness property of the channel in (29)-(30). Now, consider the scenario where the messages $\mathbb{M} = \{M^{\nabla_1}, \dots, M^{\nabla_{\|\mathbb{M}\|}}\}$ are transmitted. Let $\vartheta \in \{1, \dots, \|\mathbb{M}\|\}$ be such that $\max \nabla_\vartheta = \Theta$, where $\Theta$ is given by (33). To achieve the rate (32), it is sufficient to transmit the message $M^{\nabla_\vartheta}$ at its capacity rate, i.e., $\max_{P_X(x)} I(X; Y_{\max \nabla_\vartheta})$, and withdraw the other messages. For the converse part, first note that all the messages which belongs to $\{M^{\nabla_1}, \dots, M^{\nabla_{\|\mathbb{M}\|}}\}$ should be decoded at least at one of the receivers $Y_\Theta, Y_{\Theta+1} \dots, Y_{K_2}$. Therefore, using Fano's inequality one can easily derive:

$$\mathcal{C}_{sum}^{BCCM-deg} \leq \max_{P_X(x)} I(X; Y_\Theta, Y_{\Theta+1}, \dots, Y_{K_2}) = \max_{P_X(x)} I(X; Y_\Theta) \tag{35}$$

where the equality in (35) is due to the degradedness property of the channel. This completes the proof. ■

The remarkable consequence of Proposition 3 is that to achieve the sum-rate capacity in the degraded BCCM with an arbitrary distribution of messages, it is sufficient to transmit only one message: the message $M^{\nabla_\vartheta}$ with $\max \nabla_\vartheta = \Theta$ where $\Theta$ is given in (33). This characteristic of the degraded BCCMs is very insightful for establishing the sum-rate capacity of arbitrary degraded interference networks as will be treated in the next section.

# III. Degraded Interference Networks: Explicit Sum-Rate Capacity

Now by establishing the basic tools, we are ready to derive a fundamental result regarding the nature of information flow in arbitrary degraded interference networks. Specifically, we aim at establishing the sum-rate capacity for any degraded interference network with arbitrary number of transmitters, arbitrary number of receivers, and arbitrary distribution of messages among the transmitters and the receivers for both discrete and Gaussian networks. In other words, the only constraint which we impose on the underlying network is the degradedness as defined subsequently.

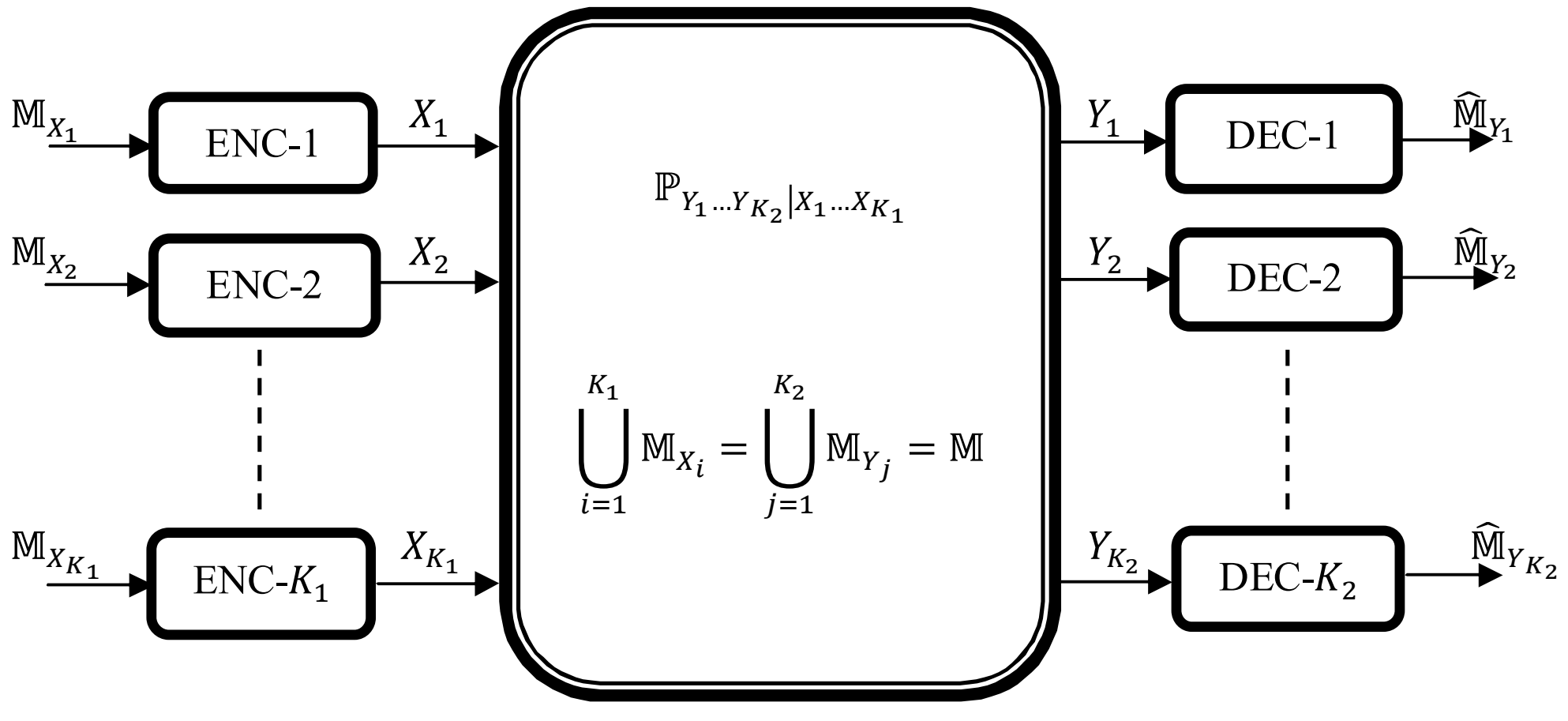

Figure 14. The General Interference Netwrok (GIN).

Let us briefly review the definitive components of the general interference network as shown in Fig. 14. In this network, $K_1$ transmitters send independent messages $\mathbb{M}\triangleq\{M_1,\dots,M_K\}$ to $K_2$ receivers: the transmitter $X_i$ sends the messages $\mathbb{M}_{X_i}$ over the channel, $i=1,\dots,K_1$, and the receiver $Y_j$ decodes the messages $\mathbb{M}_{Y_j}$for $j=1,\dots,K_2$. Therefore, we have:

$$\bigcup_{i=1}^{K_1}\mathbb{M}_{X_i}=\bigcup_{j=1}^{K_2}\mathbb{M}_{Y_j}=\mathbb{M}$$

The network transition probability function $\mathbb{P}_{Y_1\dots Y_{K_2}|X_1\dots X_{K_1}}(y_1,\dots,y_{K_2}|x_1,\dots,x_{K_1})$ describes the relation between the inputs and the outputs. The coding procedure for the network was discussed in details in Subsection II.B of Part I [1]. Also, the general Gaussian interference network with real-valued input and output signals is given below:

$$\begin{bmatrix}Y_1\\ \vdots\\ Y_{K_2}\end{bmatrix}=\begin{bmatrix}a_{11} & \cdots & a_{1K_1}\\ \vdots & \ddots & \vdots\\ a_{K_21} & \cdots & a_{K_2K_1}\end{bmatrix}\begin{bmatrix}X_1\\ \vdots\\ X_{K_1}\end{bmatrix}+\begin{bmatrix}Z_1\\ \vdots\\ Z_{K_2}\end{bmatrix}\tag{36}$$

where the parameters $\{a_{ji}\}_{\substack{j=1,\dots,K_2\\ i=1,\dots,K_1}}$ are (fixed) real-valued numbers, the RVs $\{X_i\}_{i=1}^{K_1}$ are the input symbols and the noise terms $\{Z_j\}_{j=1}^{K_2}$ are zero-mean unit-variance Gaussian RVs. The $i^{th}$ encoder is subject to an average power constraint as: $\mathbb{E}[X_i^2]\leq P_i$, where $P_i\in\mathbb{R}_+$, $i=1,\dots,K_1$.

***Definition 6: Degraded Interference Network***

*The $K_1$-transmitter/$K_2$-receiver interference network shown in Fig. 14 is said to be physically degraded if there exists a permutation $\lambda(.)$ of the element of the set $[1:K_2]$ with:*

$$\mathbb{P}_{Y_1\dots Y_{K_2}|X_1\dots X_{K_1}}(y_1,\dots,y_{K_2}|x_1,\dots,x_{K_1})=\mathbb{P}_{Y_{\lambda(1)}|X_1\dots X_{K_1}}(y_{\lambda(1)}|x_1,\dots,x_{K_1})\mathbb{P}_{Y_{\lambda(2)}|Y_{\lambda(1)}}(y_{\lambda(2)}|y_{\lambda(1)})\dots\mathbb{P}_{Y_{\lambda(K_2)}|Y_{\lambda(K_2-1)}}(y_{\lambda(K_2)}|y_{\lambda(K_2-1)})\tag{37}$$

*Equivalently, $X_1,\dots,X_{K_1}\to Y_{\lambda(1)}\to\cdots\to Y_{\lambda(K_2-1)}\to Y_{\lambda(K_2)}$ forms a Markov chain. In the following analysis, without loss of generality, we consider the case of $\lambda(j)\equiv j$, for $j\in[1:K_2]$.*

A strictly larger class of interference networks, which behave essentially the same as the physically degraded networks, is defined naturally as the *stochastically degraded* ones. The definition of such networks can be easily perceived by comparing (29), (37), and the definition given in (30). As discussed earlier, the capacity region of all interference networks depends only on marginal distributions of network transition probability function and no distinction is required to be made between stochastically and physically degradedness. Thus, we refer to both as the degraded interference networks.

Also, consider the general Gaussian interference network given by (36). In the following lemma, we provide the degradedness condition for this network.

***Lemma 1)*** *The general Gaussian interference network in (36) is stochastically degraded provided that the network gain matrix* $[a_{ji}]_{K_2\times K_1}$ *has rank one.*

*Proof of Lemma 1)* Since the matrix $[a_{ji}]_{K_2\times K_1}$ has rank one, without loss of generality we can assume that,

$$\frac{a_{j1}}{a_{(j+1)1}}=\frac{a_{j2}}{a_{(j+1)2}}=\cdots=\frac{a_{jK_1}}{a_{(j+1)K_1}}=b_{j+1},\qquad b_{j+1}\geq 1,\qquad j=2,\ldots,K_2 \tag{38}$$

Now consider a network with the following outputs:

$$\begin{cases}\tilde{Y}_1\equiv Y_1\\ \tilde{Y}_{j+1}\equiv\dfrac{1}{b_{j+1}}\tilde{Y}_j+\sqrt{1-\dfrac{1}{b_{j+1}^2}}\tilde{Z}_{j+1}, & j=2,\ldots,K_2\end{cases} \tag{39}$$

where $\tilde{Z}_2,\ldots,\tilde{Z}_{K_2}$ are independent Gaussian RVs (also independent of $Z_1$) with zero mean and unit variance. One can see that the marginal distributions of transition probability function for the network (36) is given similar to (39). Therefore, they are equivalent. On the one hand, from (39) it is clear that the following Markov chain holds:

$$X_1,\ldots,X_{K_1}\to\tilde{Y}_1\to\cdots\to\tilde{Y}_{K_2-1}\to\tilde{Y}_{K_2}$$

Thereby, the network is degraded. ■

In the following, we establish the sum-rate capacity for the general degraded interference network with arbitrary message set $\mathbb{M}\triangleq\{M_1,\ldots,M_K\}$. We derive a closed form expression for $\mathcal{C}_{\Sigma_{\mathbb{M}}}^{GIN-deg}$ as defined in Definition II.8 of Part I [1]. It should be noted that since we impose no restriction on the sets $\mathbb{M},\mathbb{M}_{X_i},i=1,\ldots,K_1$, and $\mathbb{M}_{Y_j},j=1,\ldots,K_2$, indeed we derive not only the sum-rate capacity but also *all partial sum-rate capacities* with respect to the network, i.e., $\mathcal{C}_{\Sigma_\Omega}^{GIN-deg}$ where $\Omega\subseteq\mathbb{M}$. The key is the use of Lemma II.5 of Part I [1] based on which $\mathcal{C}_{\Sigma_\Omega}^{GIN-deg}$ is equal to the sum-rate capacity of a network in which only the messages in $\Omega$ are transmitted. Therefore, we derive an explicit characterization for all $2^K-1$ corner points of the capacity region. It is also remarkable to note that in many interference networks, the sum-rate capacity is achieved by transmitting only a certain subset of messages, as given in (24) and (32) for the MACCM and the degraded BCCM, respectively. In the sequel, in addition to providing a complete characterization of the sum-rate capacity for any arbitrary degraded interference network, we will explicitly determine which messages are needed for the transmission scheme to achieve it.

To prove the main result, we first derive an outer bound on the sum-rate capacity of the network. In fact, inspired by the approach we followed for the MACCM to derive the capacity characterization (4), here, we establish an outer bound on the sum-rate capacity of the degraded interference networks which is characterized by the following parameters:

1. The RVs representing the receivers signals, i.e., $Y_j,j=1,\ldots,K_2$.
2. The RVs representing the messages $\mathbb{M}\triangleq\{M_1,\ldots,M_K\}$.
3. The parameter $Q$ which is the time-sharing RV.

We then adapt this characterization to the MACCM graph concept and propose a coding scheme to achieve the established outer bound. Our outer bound on the sum-rate capacity for the degraded interference networks (37) is given in the next lemma.

***Lemma 2)*** *Consider the* $K_1$*-Transmitter/*$K_2$*-Receiver degraded interference network in (37) with the message sets* $\mathbb{M},\mathbb{M}_{X_i},i=1,\ldots,K_1$, *and* $\mathbb{M}_{Y_j},j=1,\ldots,K_2$. *The sum-rate capacity is outer-bounded as follows:*

$$\mathcal{C}_{sum}^{GIN-deg} \leq max_{\mathcal{P}_{sum}^{GIN-deg}}\left(I\left(\mathbb{M}_{Y_1};Y_1\middle|\mathbb{M}_{Y_2},\dots,\mathbb{M}_{Y_{K_2-1}},\mathbb{M}_{Y_{K_2}},Q\right)+\dots+I\left(\mathbb{M}_{Y_{K_2-1}};Y_{K_2-1}\middle|\mathbb{M}_{Y_{K_2}},Q\right)+I\left(\mathbb{M}_{Y_{K_2}};Y_{K_2}\middle|Q\right)\right)$$

(40)

*where* $\mathcal{P}_{sum}^{GIN-deg}$ *denotes the set of all joint PDFs* $P_{QM_1\dots M_KX_1\dots X_{K_1}}(q,m_1,\dots,m_2,x_1,\dots,x_{K_1})$ *satisfying:*

$$P_{QM_1\dots M_KX_1\dots X_{K_1}} = P_QP_{M_1}\dots P_{M_K}P_{X_1|\mathbb{M}_{X_1},Q}\dots P_{X_{K_1}|\mathbb{M}_{X_{K_1}},Q}$$

(41)

*Also, the PDFs* $P_{M_l}, l = 1,\dots,K,$ *are uniformly distributed, and* $P_{X_i|\mathbb{M}_{X_i}Q} \in \{0,1\}$ *for* $i = 1,\dots,K_1$, *i.e.,* $X_i$ *is a deterministic function of* $(\mathbb{M}_{X_i},Q)$.

*Proof of Lemma 2)* Refer to Appendix. ■

Now, we proceed toward a coding scheme to achieve the bound (40). It is interesting to know that the structure of this bound by itself gives insight to the achievability scheme. In fact, the decoding step in the achievability scheme is determined by inspection of the outer bound (40): it is a *successive decoding strategy*. Roughly speaking, in this scheme, the receiver $Y_{K_2}$ decodes its respective messages $\mathbb{M}_{Y_{K_2}}$ using jointly decoding technique. At the receiver $Y_{K_2-1}$, first the messages $\mathbb{M}_{Y_{K_2}}$ are jointly decoded; this is performed without introducing any new rate cost because $Y_{K_2}$ is a degraded version of $Y_{K_2-1}$. Then, it decodes its receptive messages, i.e., $\mathbb{M}_{Y_{K_2-1}}$ (more precisely, the messages $\mathbb{M}_{Y_{K_2-1}} - \mathbb{M}_{Y_{K_2}}$) using a joint decoder. This successive decoding strategy is repeated at other receivers step by step from the weaker receivers towards the stronger ones. To complete the achievability scheme, still one essential issue has remained, that is the encoding scheme. In other words, one how should encode the messages to achieve the rate (40). In the sequel, we will present an exact response to this question.

***MACCM Plan of Messages:*** Now, consider an arbitrary interference network with the message sets $\mathbb{M}, \mathbb{M}_{X_i}, i = 1,\dots,K_1$, and $\mathbb{M}_{Y_j}, j = 1,\dots,K_2$ as shown in Fig. 14. Each subset of transmitters sends at most one message to each subset of receivers. There exist $K_1$ transmitters and $K_2$ receivers. Therefore, we can label each message by a nonempty subset of $\{1,\dots,K_1\}$ to denote which transmitters transmit the message, as well as a nonempty subset of $\{1,\dots,K_2\}$ to determine to which subset of receivers the message is sent. We represent each message of $\mathbb{M}$ as $M_\Delta^\nabla$, where $\Delta \subseteq \{1,\dots,K_1\}$ and $\nabla \subseteq \{1,\dots,K_2\}$. For example, $M_{\{1,2,3\}}^{\{2,4\}}$ indicates a message which is sent by Transmitters 1, 2 and 3 to Receivers 2 and 4.

Now, for each $\Delta \subseteq \{1,\dots,K_1\}$ define:

$$\mathbb{M}_\Delta \triangleq \{M_\Delta^L \in \mathbb{M} : L \subseteq \{1,\dots,K_2\}\}$$

(42)

Using this representation, we can arrange the messages into a MACCM graph-like illustration as shown in Fig. 15. This graph-like illustration is configured exactly similar to the MACCM graph except that instead of each message $M_\Delta$ in the MACCM graph, the set $\mathbb{M}_\Delta$ is situated. Such representation is very useful to both describe and design achievability schemes for the interference networks (see Parts III and V of our multi-part papers [3], [5]). We name such arrangement of messages as the *MACCM Plan of Messages*. In this paper, we make use of the MACCM plan of messages to describe the achievability scheme for the sum-rate capacity of an arbitrary degraded interference network.

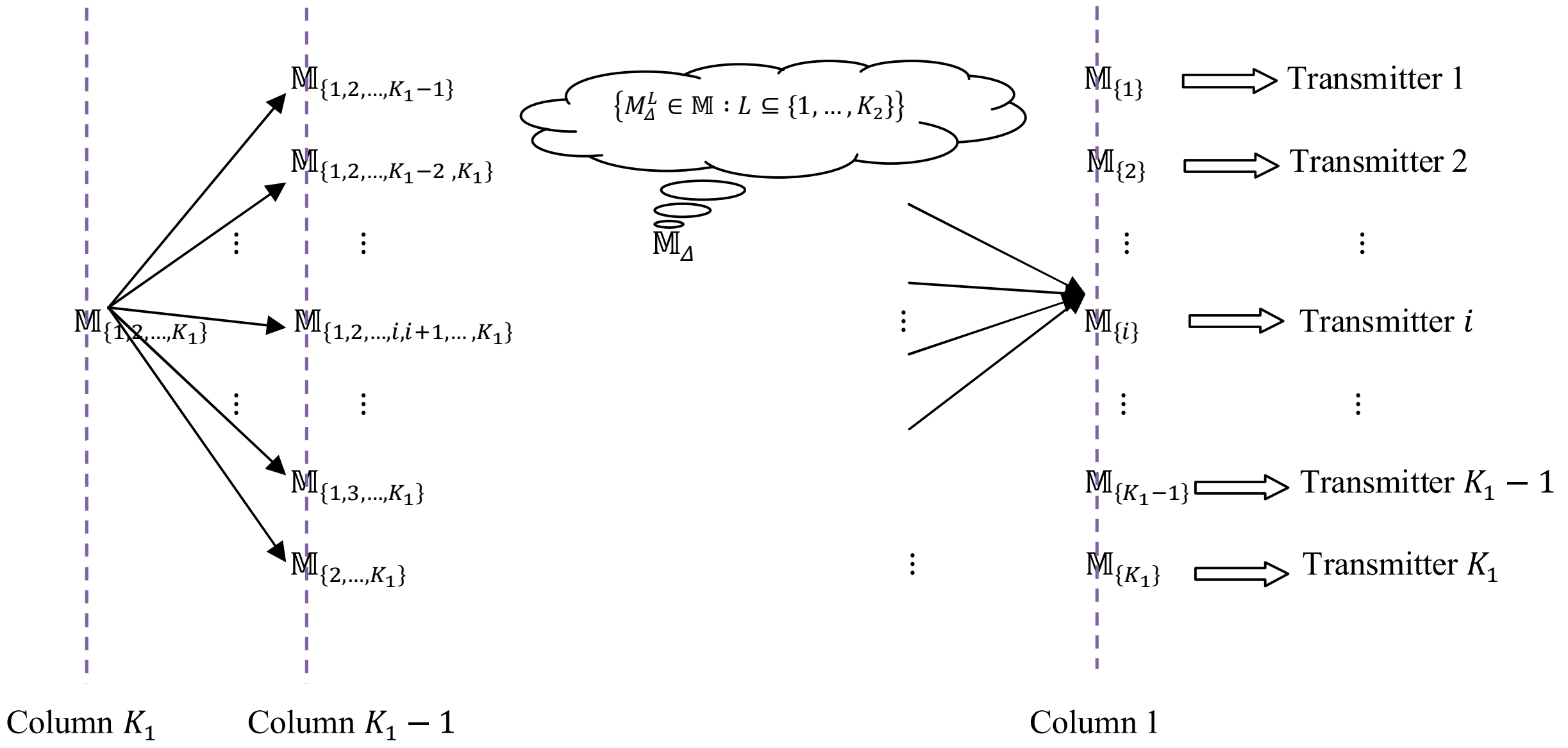

Figure 15. The MACCM Plan of Messages for an arbitrary interference Network.

Let $\Delta$ be an arbitrary nonempty subset of $\{1,\ldots,K_1\}$. According to the above representation, the messages $\mathbb{M}_\Delta$ given in (42) are broadcasted by the transmitters $X_i, i \in \Delta$; meanwhile, no transmitter other than those in $\{X_i, i \in \Delta\}$ has access to these messages. Here, we recall our previous result regarding the degraded BCs with any arbitrary distribution of messages based on which it is sufficient to transmit only a carefully picked message and ignore the others to achieve the sum-rate capacity in these channels. Inspired by this result, in the following, we prove that to achieve the maximum sum-rate in any degraded interference network, it is optimal to transmit only one of the messages in $\mathbb{M}_\Delta$ and ignore the others.

Let $\mathbb{M}_\Delta = \left\{M_\Delta^{\nabla_1},\ldots,M_\Delta^{\nabla_{\|\mathbb{M}_\Delta\|}}\right\}$, where $\nabla_l \subseteq \{1,\ldots,K_2\}$, $l=1,\ldots,K_2$. Assume that $\|\mathbb{M}_\Delta\| \geq 1$; indeed for the case of $\|\mathbb{M}_\Delta\| = 0$, the problem is trivial. Define:

$$\Theta_\Delta \triangleq \min\left\{\max\nabla :\ \nabla \subseteq \{1,\ldots,K_2\}, M_\Delta^\nabla \in \mathbb{M}_\Delta\right\} = \min\left\{\max\nabla_1,\ldots,\max\nabla_{\|\mathbb{M}_\Delta\|}\right\} \tag{43}$$

Let also $\vartheta_\Delta \in \{1,\ldots,\|\mathbb{M}_\Delta\|\}$ be such that $\max\nabla_{\vartheta_\Delta} = \Theta_\Delta$. If there exist multiple choices for $\vartheta_\Delta$, one is selected arbitrarily. Also, define:

$$\mathbb{M}_{Y_j}^\Delta \triangleq \mathbb{M}_{Y_j} \cap \mathbb{M}_\Delta, \qquad \widetilde{\mathbb{M}}_{Y_j}^\Delta \triangleq \mathbb{M}_{Y_j} - \mathbb{M}_\Delta, \qquad j=1,\ldots,K_2 \tag{44}$$

Thus, the messages in $\mathbb{M}_{Y_j}^\Delta$ are those of $\mathbb{M}_{Y_j}$ which belong to $\mathbb{M}_\Delta$, while $\widetilde{\mathbb{M}}_{Y_j}^\Delta$ are those which do not belong to $\mathbb{M}_\Delta$. Now consider the right hand side of (40). Define:

$$I_j \triangleq I\left(\mathbb{M}_{Y_j}; Y_j \middle| \mathbb{M}_{Y_{j+1}},\ldots,\mathbb{M}_{Y_{K_2-1}},\mathbb{M}_{Y_{K_2}},Q\right), \qquad j=1,\ldots,K_2 \tag{45}$$

Then we have:

$$\begin{aligned}
I_{Y_{K_2-1}} + I_{Y_{K_2}} &= I\left(\mathbb{M}_{Y_{K_2-1}}; Y_{K_2-1} \middle| \mathbb{M}_{Y_{K_2}}, Q\right) + I\left(\mathbb{M}_{Y_{K_2}}; Y_{K_2} \middle| Q\right) \\
&= I\left(\widetilde{\mathbb{M}}_{Y_{K_2-1}}^\Delta, \mathbb{M}_{Y_{K_2-1}}^\Delta; Y_{K_2-1} \middle| \widetilde{\mathbb{M}}_{Y_{K_2}}^\Delta, \mathbb{M}_{Y_{K_2}}^\Delta, Q\right) + I\left(\widetilde{\mathbb{M}}_{Y_{K_2}}^\Delta, \mathbb{M}_{Y_{K_2}}^\Delta; Y_{K_2} \middle| Q\right) \\
&= I\left(\widetilde{\mathbb{M}}_{Y_{K_2-1}}^\Delta, \mathbb{M}_{Y_{K_2-1}}^\Delta; Y_{K_2-1} \middle| \widetilde{\mathbb{M}}_{Y_{K_2}}^\Delta, \mathbb{M}_{Y_{K_2}}^\Delta, Q\right) + I\left(\mathbb{M}_{Y_{K_2}}^\Delta; Y_{K_2} \middle| \widetilde{\mathbb{M}}_{Y_{K_2}}^\Delta, Q\right) + I\left(\widetilde{\mathbb{M}}_{Y_{K_2}}^\Delta; Y_{K_2} \middle| Q\right)
\end{aligned}$$

$$\overset{(a)}{\leq} I\left(\widetilde{\mathbb{M}}_{Y_{K_2-1}}^{\Delta},\mathbb{M}_{Y_{K_2-1}}^{\Delta};Y_{K_2-1}\middle|\widetilde{\mathbb{M}}_{Y_{K_2}}^{\Delta},\mathbb{M}_{Y_{K_2}}^{\Delta},Q\right)+I\left(\mathbb{M}_{Y_{K_2}}^{\Delta};Y_{K_2-1}\middle|\widetilde{\mathbb{M}}_{Y_{K_2}}^{\Delta},Q\right)+I\left(\widetilde{\mathbb{M}}_{Y_{K_2}}^{\Delta};Y_{K_2}\middle|Q\right)$$

$$=I\left(\widetilde{\mathbb{M}}_{Y_{K_2-1}}^{\Delta},\mathbb{M}_{Y_{K_2-1}}^{\Delta},\mathbb{M}_{Y_{K_2}}^{\Delta};Y_{K_2-1}\middle|\widetilde{\mathbb{M}}_{Y_{K_2}}^{\Delta},Q\right)+I\left(\widetilde{\mathbb{M}}_{Y_{K_2}}^{\Delta};Y_{K_2}\middle|Q\right)$$

(46)

where inequality (a) holds because $Y_{K_2}$ is a degraded version of $Y_{K_2-1}$. Next by using (46), we proceed as follows:

$$I_{Y_{K_2-2}}+I_{Y_{K_2-1}}+I_{Y_{K_2}}\leq I\left(\widetilde{\mathbb{M}}_{Y_{K_2-2}}^{\Delta},\mathbb{M}_{Y_{K_2-2}}^{\Delta};Y_{K_2-2}\middle|\widetilde{\mathbb{M}}_{Y_{K_2-1}}^{\Delta},\mathbb{M}_{Y_{K_2-1}}^{\Delta},\widetilde{\mathbb{M}}_{Y_{K_2}}^{\Delta},\mathbb{M}_{Y_{K_2}}^{\Delta},Q\right)+I\left(\widetilde{\mathbb{M}}_{Y_{K_2-1}}^{\Delta},\mathbb{M}_{Y_{K_2-1}}^{\Delta},\mathbb{M}_{Y_{K_2}}^{\Delta};Y_{K_2-1}\middle|\widetilde{\mathbb{M}}_{Y_{K_2}}^{\Delta},Q\right)$$
$$+I\left(\widetilde{\mathbb{M}}_{Y_{K_2}}^{\Delta};Y_{K_2}\middle|Q\right)$$

$$=I\left(\widetilde{\mathbb{M}}_{Y_{K_2-2}}^{\Delta},\mathbb{M}_{Y_{K_2-2}}^{\Delta};Y_{K_2-2}\middle|\widetilde{\mathbb{M}}_{Y_{K_2-1}}^{\Delta},\mathbb{M}_{Y_{K_2-1}}^{\Delta},\widetilde{\mathbb{M}}_{Y_{K_2}}^{\Delta},\mathbb{M}_{Y_{K_2}}^{\Delta},Q\right)+I\left(\mathbb{M}_{Y_{K_2-1}}^{\Delta},\mathbb{M}_{Y_{K_2}}^{\Delta};Y_{K_2-1}\middle|\widetilde{\mathbb{M}}_{Y_{K_2-1}}^{\Delta},\widetilde{\mathbb{M}}_{Y_{K_2}}^{\Delta},Q\right)$$
$$+I\left(\widetilde{\mathbb{M}}_{Y_{K_2-1}}^{\Delta};Y_{K_2-1}\middle|\widetilde{\mathbb{M}}_{Y_{K_2}}^{\Delta},Q\right)+I\left(\widetilde{\mathbb{M}}_{Y_{K_2}}^{\Delta};Y_{K_2}\middle|Q\right)$$

$$\overset{(a)}{\leq} I\left(\widetilde{\mathbb{M}}_{Y_{K_2-2}}^{\Delta},\mathbb{M}_{Y_{K_2-2}}^{\Delta};Y_{K_2-2}\middle|\widetilde{\mathbb{M}}_{Y_{K_2-1}}^{\Delta},\mathbb{M}_{Y_{K_2-1}}^{\Delta},\widetilde{\mathbb{M}}_{Y_{K_2}}^{\Delta},\mathbb{M}_{Y_{K_2}}^{\Delta},Q\right)+I\left(\mathbb{M}_{Y_{K_2-1}}^{\Delta},\mathbb{M}_{Y_{K_2}}^{\Delta};Y_{K_2-2}\middle|\widetilde{\mathbb{M}}_{Y_{K_2-1}}^{\Delta},\widetilde{\mathbb{M}}_{Y_{K_2}}^{\Delta},Q\right)$$
$$+I\left(\widetilde{\mathbb{M}}_{Y_{K_2-1}}^{\Delta};Y_{K_2-1}\middle|\widetilde{\mathbb{M}}_{Y_{K_2}}^{\Delta},Q\right)+I\left(\widetilde{\mathbb{M}}_{Y_{K_2}}^{\Delta};Y_{K_2}\middle|Q\right)$$

$$=I\left(\widetilde{\mathbb{M}}_{Y_{K_2-2}}^{\Delta},\mathbb{M}_{Y_{K_2-2}}^{\Delta},\mathbb{M}_{Y_{K_2-1}}^{\Delta},\mathbb{M}_{Y_{K_2}}^{\Delta};Y_{K_2-2}\middle|\widetilde{\mathbb{M}}_{Y_{K_2-1}}^{\Delta},\widetilde{\mathbb{M}}_{Y_{K_2}}^{\Delta},Q\right)+I\left(\widetilde{\mathbb{M}}_{Y_{K_2-1}}^{\Delta};Y_{K_2-1}\middle|\widetilde{\mathbb{M}}_{Y_{K_2}}^{\Delta},Q\right)+I\left(\widetilde{\mathbb{M}}_{Y_{K_2}}^{\Delta};Y_{K_2}\middle|Q\right)$$

(47)

where (a) holds because $Y_{K_2-1}$ is a degraded version of $Y_{K_2-2}$. By repeating this procedure $K_2-\Theta_\Delta$ times ($\Theta_\Delta$ is given by (43)), one can accumulate the sets $\mathbb{M}_{Y_{\Theta_\Delta}}^{\Delta},\mathbb{M}_{Y_{\Theta_\Delta-1}}^{\Delta},\ldots,\mathbb{M}_{Y_{K_2}}^{\Delta}$ in the term $I_{Y_{\Theta_\Delta}}$ and derive:

$$I_{Y_{\Theta_\Delta}}+\cdots+I_{Y_{K_2-1}}+I_{Y_{K_2}}$$
$$\leq I\left(\widetilde{\mathbb{M}}_{Y_{\Theta_\Delta}}^{\Delta},\mathbb{M}_{Y_{\Theta_\Delta}}^{\Delta},\mathbb{M}_{Y_{\Theta_\Delta+1}}^{\Delta},\ldots,\mathbb{M}_{Y_{K_2}}^{\Delta};Y_{\Theta_\Delta}\middle|\widetilde{\mathbb{M}}_{Y_{\Theta_\Delta+1}}^{\Delta},\ldots,\widetilde{\mathbb{M}}_{Y_{K_2}}^{\Delta},Q\right)+I\left(\widetilde{\mathbb{M}}_{Y_{\Theta_\Delta+1}}^{\Delta};Y_{\Theta_\Delta+1}\middle|\widetilde{\mathbb{M}}_{Y_{\Theta_\Delta+2}}^{\Delta},\ldots,\widetilde{\mathbb{M}}_{Y_{K_2}}^{\Delta},Q\right)+\cdots+I\left(\widetilde{\mathbb{M}}_{Y_{K_2}}^{\Delta};Y_{K_2}\middle|Q\right)$$

(48)

It should be noted that according to the definition of $\Theta_\Delta$ in (43), each message which belongs to $\mathbb{M}_\Delta$ should be decoded at least by one of the receivers $Y_{\Theta_\Delta},Y_{\Theta_\Delta+1},\ldots,Y_{K_2}$. Therefore, we have:

$$\mathbb{M}_{Y_{\Theta_\Delta}}^{\Delta}\cup\mathbb{M}_{Y_{\Theta_\Delta+1}}^{\Delta}\cup\ldots\cup\mathbb{M}_{Y_{K_2}}^{\Delta}=\mathbb{M}_\Delta$$

(49)

Then, from (48) and (49), we obtain:

$$I_{Y_{\Theta_\Delta}}+\cdots+I_{Y_{K_2-1}}+I_{Y_{K_2}}$$
$$\leq I\left(\widetilde{\mathbb{M}}_{Y_{\Theta_\Delta}}^{\Delta},\mathbb{M}_\Delta;Y_{\Theta_\Delta}\middle|\widetilde{\mathbb{M}}_{Y_{\Theta_\Delta+1}}^{\Delta},\ldots,\widetilde{\mathbb{M}}_{Y_{K_2}}^{\Delta},Q\right)+I\left(\widetilde{\mathbb{M}}_{Y_{\Theta_\Delta+1}}^{\Delta};Y_{\Theta_\Delta+1}\middle|\widetilde{\mathbb{M}}_{Y_{\Theta_\Delta+2}}^{\Delta},\ldots,\widetilde{\mathbb{M}}_{Y_{K_2}}^{\Delta},Q\right)+\cdots+I\left(\widetilde{\mathbb{M}}_{Y_{K_2}}^{\Delta};Y_{K_2}\middle|Q\right)$$

(50)

Now consider the terms $I_{Y_1},\ldots,I_{Y_{\Theta_\Delta-1}}$. We have ($j=1,\ldots,\Theta_\Delta-1$):

$$I_{Y_j}=I\left(\mathbb{M}_{Y_j};Y_j\middle|\mathbb{M}_{Y_{j+1}},\ldots,\mathbb{M}_{Y_{K_2-1}},\mathbb{M}_{Y_{K_2}},Q\right)$$
$$=I\left(\widetilde{\mathbb{M}}_{Y_j}^{\Delta},\mathbb{M}_{Y_j}^{\Delta};Y_j\middle|\widetilde{\mathbb{M}}_{Y_{j+1}}^{\Delta},\mathbb{M}_{Y_{j+1}}^{\Delta},\ldots,\widetilde{\mathbb{M}}_{Y_{K_2-1}}^{\Delta},\mathbb{M}_{Y_{K_2-1}}^{\Delta},\widetilde{\mathbb{M}}_{Y_{K_2}}^{\Delta},\mathbb{M}_{Y_{K_2}}^{\Delta},Q\right)$$

$$= I\left(\widetilde{\mathbb{M}}_{Y_j}^{\Delta}, \mathbb{M}_{Y_j}^{\Delta}; Y_j \middle| \widetilde{\mathbb{M}}_{Y_{j+1}}^{\Delta}, \dots, \widetilde{\mathbb{M}}_{Y_{K_2-1}}^{\Delta}, \widetilde{\mathbb{M}}_{Y_{K_2}}^{\Delta}, \underbrace{\mathbb{M}_{Y_{j+1}}^{\Delta}, \dots, \mathbb{M}_{Y_{\Theta_\Delta}}^{\Delta}, \mathbb{M}_{Y_{\Theta_\Delta+1}}^{\Delta}, \dots, \mathbb{M}_{Y_{K_2-1}}^{\Delta}, \mathbb{M}_{Y_{K_2}}^{\Delta}}_{\mathbb{M}_\Delta}, Q\right)$$

$$= I\left(\widetilde{\mathbb{M}}_{Y_j}^{\Delta}, \mathbb{M}_{Y_j}^{\Delta}; Y_j \middle| \widetilde{\mathbb{M}}_{Y_{j+1}}^{\Delta}, \dots, \widetilde{\mathbb{M}}_{Y_{K_2-1}}^{\Delta}, \widetilde{\mathbb{M}}_{Y_{K_2}}^{\Delta}, \mathbb{M}_\Delta, Q\right)$$

$$\overset{(a)}{=} I\left(\widetilde{\mathbb{M}}_{Y_j}^{\Delta}; Y_j \middle| \widetilde{\mathbb{M}}_{Y_{j+1}}^{\Delta}, \dots, \widetilde{\mathbb{M}}_{Y_{K_2-1}}^{\Delta}, \widetilde{\mathbb{M}}_{Y_{K_2}}^{\Delta}, \mathbb{M}_\Delta, Q\right) \tag{51}$$

where equality (a) holds because $\mathbb{M}_{Y_j}^{\Delta}$ is a subset of $\mathbb{M}_\Delta$. Therefore, we have:

$$\mathcal{C}_{sum}^{GIN-deg} \leq \max_{\mathcal{P}_{sum}^{GIN-deg}} \left(I_{Y_1} + \dots + I_{Y_{\Theta_\Delta-1}} + I_{Y_{\Theta_\Delta}} + \dots + I_{Y_{K_2}}\right)$$

$$\leq \max_{\mathcal{P}_{sum}^{GIN-deg}} \left(\begin{array}{c} I\left(\widetilde{\mathbb{M}}_{Y_1}^{\Delta}; Y_1 \middle| \widetilde{\mathbb{M}}_{Y_2}^{\Delta}, \dots, \widetilde{\mathbb{M}}_{Y_{K_2}}^{\Delta}, \mathbb{M}_\Delta, Q\right) + \dots + I\left(\widetilde{\mathbb{M}}_{Y_{\Theta_\Delta-1}}^{\Delta}; Y_{\Theta_\Delta-1} \middle| \widetilde{\mathbb{M}}_{Y_{\Theta_\Delta}}^{\Delta}, \dots, \widetilde{\mathbb{M}}_{Y_{K_2}}^{\Delta}, \mathbb{M}_\Delta, Q\right) + \\ I\left(\widetilde{\mathbb{M}}_{Y_{\Theta_\Delta}}^{\Delta}, \mathbb{M}_\Delta; Y_{\Theta_\Delta} \middle| \widetilde{\mathbb{M}}_{Y_{\Theta_\Delta+1}}^{\Delta}, \dots, \widetilde{\mathbb{M}}_{Y_{K_2}}^{\Delta}, Q\right) + \dots + I\left(\widetilde{\mathbb{M}}_{Y_{K_2}}^{\Delta}; Y_{K_2} \middle| Q\right) \end{array}\right) \tag{52}$$

Note that according to (41) and (42) if $i \in \Delta$, then $\mathbb{M}_\Delta \subseteq \mathbb{M}_{X_i}$; thus,

$$P_{X_i|\mathbb{M}_{X_i}Q} = P_{X_i|\mathbb{M}_\Delta, \mathbb{M}_{X_i}-\mathbb{M}_\Delta, Q} \tag{53}$$

Otherwise, if $i \notin \Delta$, then $\mathbb{M}_\Delta \cap \mathbb{M}_{X_i} = \emptyset$. As we see in the maximization (52), all the messages in $\mathbb{M}_\Delta$ appear next to each other for both the mutual information function terms and the PDFs $P_{X_i|\mathbb{M}_{X_i}Q}, i = 1, \dots, K_1$. Therefore, as a solution of the maximization (53), we can collectively represent the messages $\mathbb{M}_\Delta$ (which are auxiliary RVs in (40)) using a new RV. Note that one can replace the random variables $\mathbb{M}_\Delta$ (collectively) by any arbitrary RV which is independent of $\mathbb{M} - \mathbb{M}_\Delta$ and $Q$. We have:

$$\mathcal{C}_{sum}^{GIN-deg}$$

$$\leq \max_{\mathcal{P}_{sum}^{GIN_{deg}}} \left(\begin{array}{c} I\left(\widetilde{\mathbb{M}}_{Y_1}^{\Delta}; Y_1 \middle| \widetilde{\mathbb{M}}_{Y_2}^{\Delta}, \dots, \widetilde{\mathbb{M}}_{Y_{K_2}}^{\Delta}, \underbrace{\mathbb{M}_\Delta}_{V}, Q\right) + \dots + I\left(\widetilde{\mathbb{M}}_{Y_{\Theta_\Delta-1}}^{\Delta}; Y_{\Theta_\Delta-1} \middle| \widetilde{\mathbb{M}}_{Y_{\Theta_\Delta}}^{\Delta}, \dots, \widetilde{\mathbb{M}}_{Y_{K_2}}^{\Delta}, \underbrace{\mathbb{M}_\Delta}_{V}, Q\right) + \\ I\left(\widetilde{\mathbb{M}}_{Y_{\Theta_\Delta}}^{\Delta}, \underbrace{\mathbb{M}_\Delta}_{V}; Y_{\Theta_\Delta} \middle| \widetilde{\mathbb{M}}_{Y_{\Theta_\Delta+1}}^{\Delta}, \dots, \widetilde{\mathbb{M}}_{Y_{K_2}}^{\Delta}, Q\right) + \dots + I\left(\widetilde{\mathbb{M}}_{Y_{K_2}}^{\Delta}; Y_{K_2} \middle| Q\right) \end{array}\right)$$

$$= \max_{\mathcal{P}_{sum:\mathbb{M}_\Delta \leftrightarrow V}^{GIN_{deg}}} \left(\begin{array}{c} I\left(\widetilde{\mathbb{M}}_{Y_1}^{\Delta}; Y_1 \middle| \widetilde{\mathbb{M}}_{Y_2}^{\Delta}, \dots, \widetilde{\mathbb{M}}_{Y_{K_2}}^{\Delta}, V, Q\right) + \dots + I\left(\widetilde{\mathbb{M}}_{Y_{\Theta_\Delta-1}}^{\Delta}; Y_{\Theta_\Delta-1} \middle| \widetilde{\mathbb{M}}_{Y_{\Theta_\Delta}}^{\Delta}, \dots, \widetilde{\mathbb{M}}_{Y_{K_2}}^{\Delta}, V, Q\right) + \\ I\left(\widetilde{\mathbb{M}}_{Y_{\Theta_\Delta}}^{\Delta}, V; Y_{\Theta_\Delta} \middle| \widetilde{\mathbb{M}}_{Y_{\Theta_\Delta+1}}^{\Delta}, \dots, \widetilde{\mathbb{M}}_{Y_{K_2}}^{\Delta}, Q\right) + \dots + I\left(\widetilde{\mathbb{M}}_{Y_{K_2}}^{\Delta}; Y_{K_2} \middle| Q\right) \end{array}\right)$$

$$= \max_{\mathcal{P}_{sum,\Delta}^{GIN-deg}} \left(\begin{array}{c} I\left(\widetilde{\mathbb{M}}_{Y_1}^{\Delta}; Y_1 \middle| \widetilde{\mathbb{M}}_{Y_2}^{\Delta}, \dots, \widetilde{\mathbb{M}}_{Y_{K_2-1}}^{\Delta}, \widetilde{\mathbb{M}}_{Y_{K_2}}^{\Delta}, M_\Delta^{\nabla_{\vartheta_\Delta}}, Q\right) + \dots + I\left(\widetilde{\mathbb{M}}_{Y_{\Theta_\Delta-1}}^{\Delta}; Y_{\Theta_\Delta-1} \middle| \widetilde{\mathbb{M}}_{Y_{\Theta_\Delta}}^{\Delta}, \dots, \widetilde{\mathbb{M}}_{Y_{K_2-1}}^{\Delta}, \widetilde{\mathbb{M}}_{Y_{K_2}}^{\Delta}, M_\Delta^{\nabla_{\vartheta_\Delta}}, Q\right) + \\ I\left(\widetilde{\mathbb{M}}_{Y_{\Theta_\Delta}}^{\Delta}, M_\Delta^{\nabla_{\vartheta_\Delta}}; Y_{\Theta_\Delta} \middle| \widetilde{\mathbb{M}}_{Y_{\Theta_\Delta+1}}^{\Delta}, \dots, \widetilde{\mathbb{M}}_{Y_{K_2}}^{\Delta}, Q\right) + I\left(\widetilde{\mathbb{M}}_{Y_{\Theta_\Delta+1}}^{\Delta}; Y_{\Theta_\Delta+1} \middle| \widetilde{\mathbb{M}}_{Y_{\Theta_\Delta+2}}^{\Delta}, \dots, \widetilde{\mathbb{M}}_{Y_{K_2}}^{\Delta}, Q\right) + \dots + I\left(\widetilde{\mathbb{M}}_{Y_{K_2}}^{\Delta}; Y_{K_2} \middle| Q\right) \end{array}\right) \tag{54}$$

where $\mathcal{P}_{sum:\mathbb{M}_\Delta \leftrightarrow V}^{GIN_{deg}}$ denotes the set of all joint PDFs similar to (41) except that $\mathbb{M}_\Delta$ is replaced everywhere by the random variable $V$. Also, $\mathcal{P}_{sum,\Delta}^{GIN-deg}$ is the set of all joint PDFs similar to (41) except that $\mathbb{M}_\Delta$ is replaced everywhere by $M_\Delta^{\nabla_{\vartheta_\Delta}}$. In fact, in the second equality of (54), the random variable $V$ has been re-named by $M_\Delta^{\nabla_{\vartheta_\Delta}}$. Note that this substitution is well-defined because $\vartheta_\Delta \in \{1, \dots, \|\mathbb{M}_\Delta\|\}$ is such that $\max \nabla_{\vartheta_\Delta} = \Theta_\Delta$, where $\Theta_\Delta$ is given by (43); therefore, the message $M_\Delta^{\nabla_{\vartheta_\Delta}}$ should be decoded at the receiver $Y_{\Theta_\Delta}$, while it is not decoded at any receiver $Y_j$ with $j > \Theta_\Delta$.

In fact, this is an advantage of our methodology where we can treat $M_{\Delta}^{\nabla}$ either as a message or an auxiliary RV in the outer bound (40). The relation (54) is derived by switching between these two views. In order to derive the first equality in (54), we look at $\mathbb{M}_{\Delta}$ as a set of auxiliary RVs; while for deriving the second equality, we again replace the auxiliary random variable $V$ with a suitable message (consequently, the configuration of the maximization is preserved as the bound (40)). Note that choosing $M_{\Delta}^{\nabla_{\vartheta_\Delta}}$ to re-express the random variable $V$ is beneficial because, as we see subsequently, it leads to achievability scheme for the resulting outer bound. From another point of view, one can deduce that a solution to the maximization (40) is to nullify the random variables $\mathbb{M}_{\Delta}-\left\{M_{\Delta}^{\nabla_{\vartheta_\Delta}}\right\}$.

Let us revisit the outer bound (54). This is indeed the outer bound (40) for an interference network where the messages $\mathbb{M}-\left(\mathbb{M}_{\Delta}-\left\{M_{\Delta}^{\nabla_{\vartheta_\Delta}}\right\}\right)$ are transmitted. Therefore, by performing the above procedure for all $\Delta\subseteq\{1,\ldots,K_1\}$, we can derive the following result. Define the sets $\widetilde{\mathbb{M}}$, $\widetilde{\mathbb{M}}_{X_i}, i=1,\ldots,K_1$, and $\widetilde{\mathbb{M}}_{Y_j}, j=1,\ldots,K_2$, as follows:

$$\begin{cases}\widetilde{\mathbb{M}}\triangleq\mathbb{M}-\bigcup\limits_{\Delta\subseteq\{1,\ldots,K_1\}}\left(\mathbb{M}_{\Delta}-\left\{M_{\Delta}^{\nabla_{\vartheta_\Delta}}\right\}\right)\\ \widetilde{\mathbb{M}}_{X_i}\triangleq\mathbb{M}_{X_i}-\bigcup\limits_{\Delta\subseteq\{1,\ldots,K_1\}}\left(\mathbb{M}_{\Delta}-\left\{M_{\Delta}^{\nabla_{\vartheta_\Delta}}\right\}\right)\\ \widetilde{\mathbb{M}}_{Y_j}\triangleq\mathbb{M}_{Y_j}-\bigcup\limits_{\Delta\subseteq\{1,\ldots,K_1\}}\left(\mathbb{M}_{\Delta}-\left\{M_{\Delta}^{\nabla_{\vartheta_\Delta}}\right\}\right)\end{cases}\tag{55}$$

***Lemma 3)***

*Consider the $K_1$-transmitter/$K_2$-receiver degraded interference network in (37) with the message sets $\mathbb{M},\mathbb{M}_{X_i}, i=1,\ldots,K_1$, and $\mathbb{M}_{Y_j}, j=1,\ldots,K_2$. The sum-rate capacity is outer-bounded by:*

$$\mathcal{C}_{sum}^{GIN-deg}\leq max_{\tilde{\mathcal{P}}_{sum}^{GIN-deg}}\left(I\left(\widetilde{\mathbb{M}}_{Y_1};Y_1\middle|\widetilde{\mathbb{M}}_{Y_2},\ldots,\widetilde{\mathbb{M}}_{Y_{K_2-1}},\widetilde{\mathbb{M}}_{Y_{K_2}},Q\right)+\cdots+I\left(\widetilde{\mathbb{M}}_{Y_{K_2-1}};Y_{K_2-1}\middle|\widetilde{\mathbb{M}}_{Y_{K_2}},Q\right)+I\left(\widetilde{\mathbb{M}}_{Y_{K_2}};Y_{K_2}\middle|Q\right)\right)\tag{56}$$

*where $\tilde{\mathcal{P}}_{sum}^{GIN-deg}$ denotes the set of all joint PDFs as follows:*

$$P_Q\times\prod_{M_{\Delta}^{\nabla}\in\widetilde{\mathbb{M}}}P_{M_{\Delta}^{\nabla}}\times\prod_{i\in[1:K_1]}P_{X_i|\widetilde{\mathbb{M}}_{X_i}Q}\tag{57}$$

*Also, the PDFs $P_{M_{\Delta}^{\nabla}}$, $M_{\Delta}^{\nabla}\in\widetilde{\mathbb{M}}$, are uniformly distributed, and $P_{X_i|\widetilde{\mathbb{M}}_{X_i}Q}\in\{0,1\}$ for $i=1,\ldots,K_1$, i.e., $X_i$ is a deterministic function of $\left(\widetilde{\mathbb{M}}_{X_i},Q\right)$.*

***Example 2;***

Consider a 4-transmitter/3-receiver interference network with the following message set:

$$\mathbb{M}=\left\{M_{\{1,2,4\}}^{\{3\}},M_{\{1,2,4\}}^{\{1,3\}},M_{\{1,2,4\}}^{\{2,3\}},M_{\{1,2\}}^{\{2,3\}},M_{\{1,2\}}^{\{1,3\}},M_{\{3,4\}}^{\{1,2\}},M_{\{3,4\}}^{\{1,3\}},M_{\{3,4\}}^{\{2,3\}},M_{\{1\}}^{\{1,3\}},M_{\{1\}}^{\{2,3\}},M_{\{2\}}^{\{3\}},M_{\{2\}}^{\{1,3\}},M_{\{3\}}^{\{2\}},M_{\{3\}}^{\{2,3\}},M_{\{4\}}^{\{1\}},M_{\{4\}}^{\{2\}},M_{\{4\}}^{\{1,2\}}\right\}\tag{58}$$

The network has been depicted in Fig. 16.

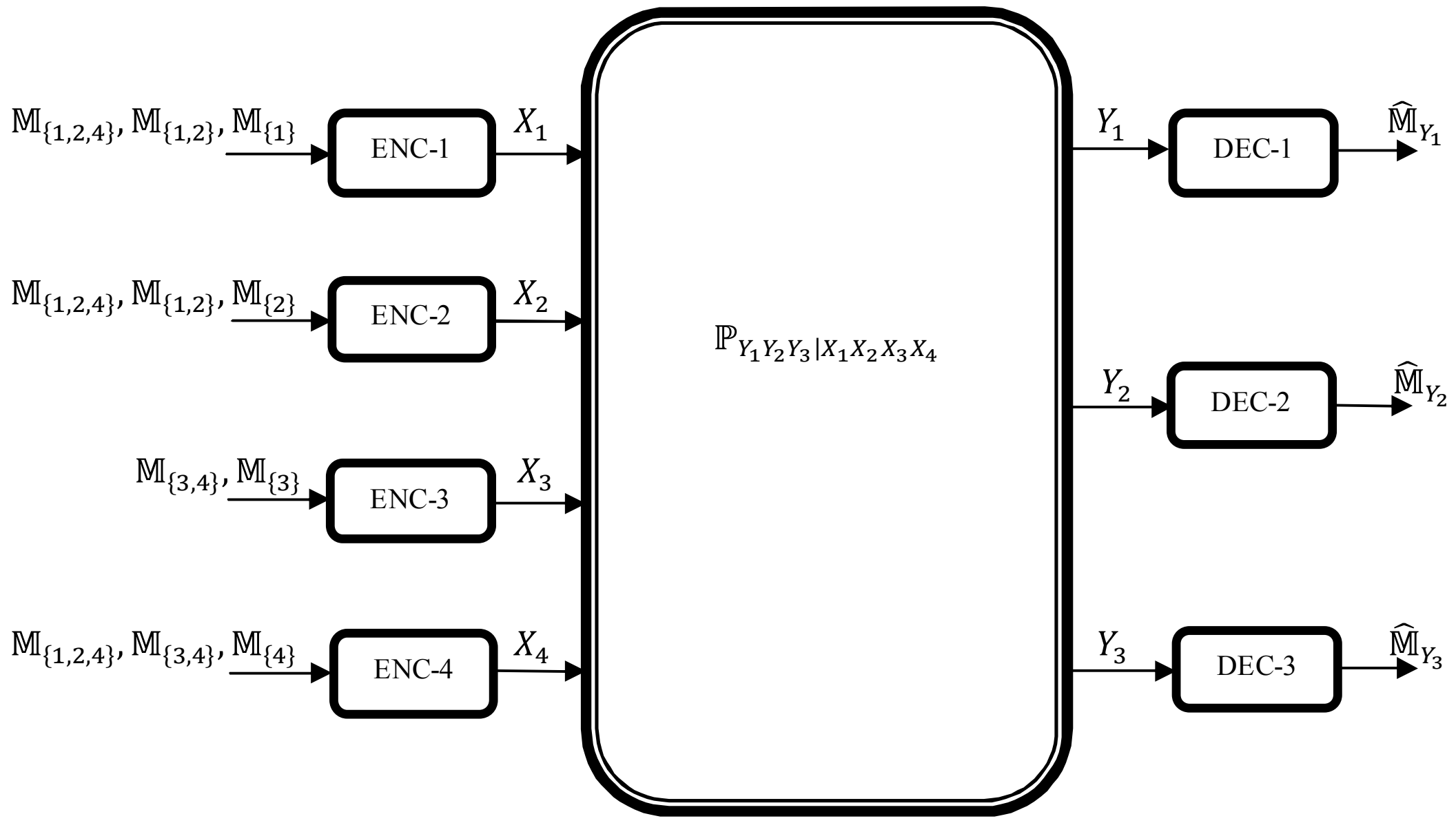

Figure 16. A 4-transmitter/3-receiver interference network with messages in (58).

Note that according to the definition (42) the message set $\mathbb{M}$ in (58) is partitioned into the following subsets:

$$\begin{cases} \mathbb{M}_{\{1,2,4\}} = \left\{M_{\{1,2,4\}}^{\{3\}}, M_{\{1,2,4\}}^{\{1,3\}}, M_{\{1,2,4\}}^{\{2,3\}}\right\} \\ \mathbb{M}_{\{1,2\}} = \left\{M_{\{1,2\}}^{\{2,3\}}, M_{\{1,2\}}^{\{1,3\}}\right\}, \qquad \mathbb{M}_{\{3,4\}} = \left\{M_{\{3,4\}}^{\{1,2\}}, M_{\{3,4\}}^{\{1,3\}}, M_{\{3,4\}}^{\{2,3\}}\right\} \\ \mathbb{M}_{\{1\}} = \left\{M_{\{1\}}^{\{1,3\}}, M_{\{1\}}^{\{2,3\}}\right\}, \qquad \mathbb{M}_{\{2\}} = \left\{M_{\{2\}}^{\{3\}}, M_{\{2\}}^{\{1,3\}}\right\}, \qquad \mathbb{M}_{\{3\}} = \left\{M_{\{3\}}^{\{3\}}, M_{\{3\}}^{\{2,3\}}\right\}, \qquad \mathbb{M}_{\{4\}} = \left\{M_{\{4\}}^{\{1\}}, M_{\{4\}}^{\{2\}}, M_{\{4\}}^{\{1,2\}}\right\} \end{cases}$$

(59)

As we see these subsets all have more than one element. Nevertheless, in order to achieve the sum-rate capacity it is optimal to transmit only one message from each set. Using the definition (43), we choose the desired messages as follows:

$$\widetilde{\mathbb{M}} = \left\{M_{\{1,2,4\}}^{\{3\}}, M_{\{1,2\}}^{\{2,3\}}, M_{\{3,4\}}^{\{1,2\}}, M_{\{1\}}^{\{1,3\}}, M_{\{2\}}^{\{3\}}, M_{\{3\}}^{\{3\}}, M_{\{4\}}^{\{1\}}\right\}$$

(60)

Therefore, according to (55), we have:

$$\begin{cases} \widetilde{\mathbb{M}}_{Y_1} = \left\{M_{\{3,4\}}^{\{1,2\}}, M_{\{1\}}^{\{1,3\}}, M_{\{4\}}^{\{1\}}\right\} \\ \widetilde{\mathbb{M}}_{Y_2} = \left\{M_{\{1,2\}}^{\{2,3\}}, M_{\{3,4\}}^{\{1,2\}}\right\} \\ \widetilde{\mathbb{M}}_{Y_3} = \left\{M_{\{1,2,4\}}^{\{3\}}, M_{\{1,2\}}^{\{2,3\}}, M_{\{1\}}^{\{1,3\}}, M_{\{2\}}^{\{3\}}, M_{\{3\}}^{\{3\}}\right\} \end{cases}$$

(61)

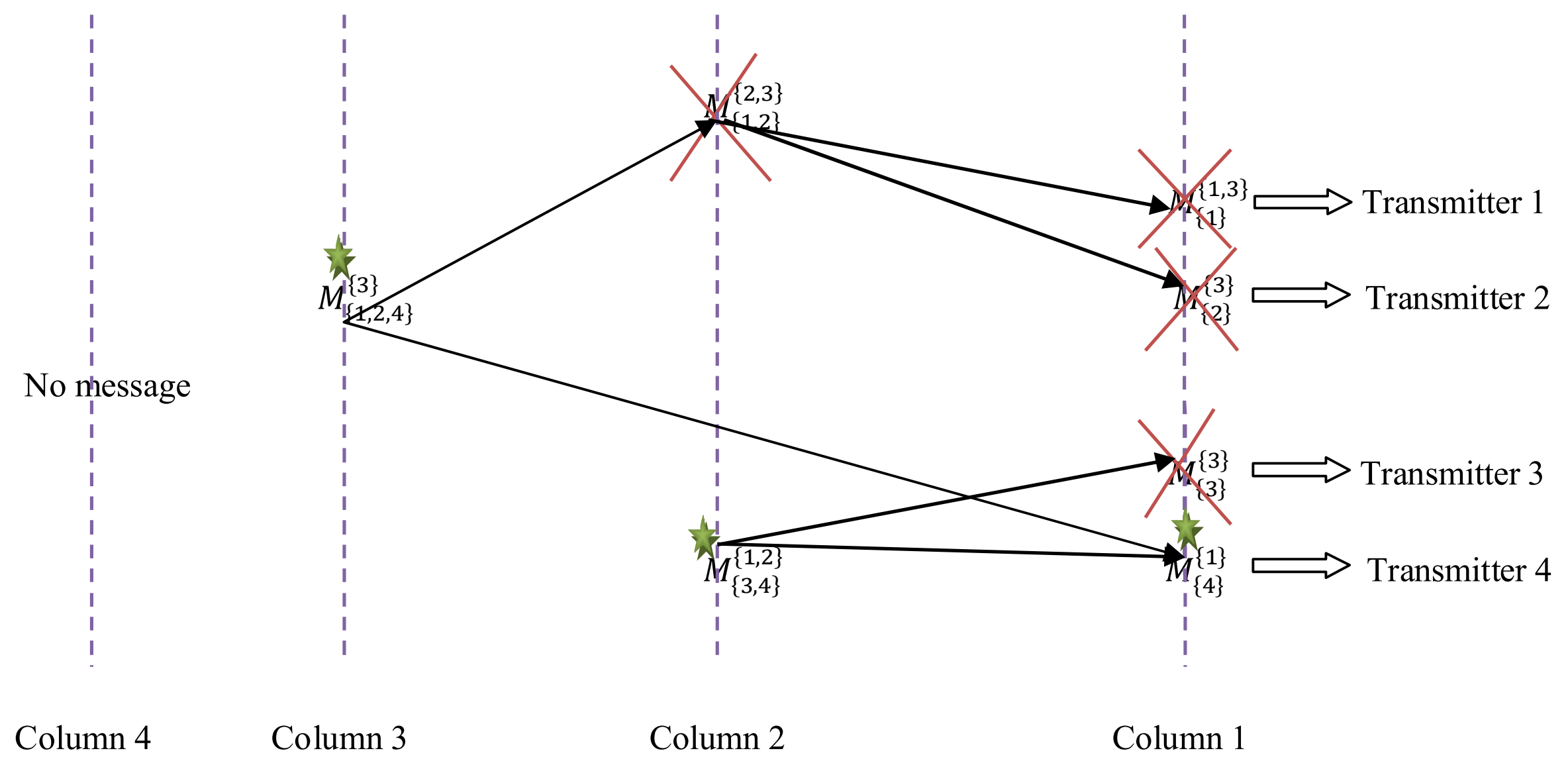

Figure 17. The MACCM message graph with respect to the messages (60) for Example 2. To achieve the sum-rate capacity, it is only required to transmit the messages with the mark "★" and ignore the others.

To constitute the set $\widetilde{\mathbb{M}}$ in Lemma 3, from each subset $\mathbb{M}_\Delta$ in (42), a message $M_\Delta^{\nabla_{\vartheta_\Delta}}$ with $\max \nabla_{\vartheta_\Delta} = \Theta_\Delta$ is selected, where $\Theta_\Delta$ is given in (43). For some subsets, for example $\mathbb{M}_{\{1,2,4\}}$ in (59), there exist multiple choices for $M_\Delta^{\nabla_{\vartheta_\Delta}}$ where one is selected arbitrarily. Now consider the message set $\widetilde{\mathbb{M}}$. In this set, there exist no messages $M_{\Delta_1}^{\nabla_1}$ and $M_{\Delta_2}^{\nabla_2}$ with $\Delta_1 = \Delta_2$. Therefore, we can treat $\widetilde{\mathbb{M}}$ as a message set designated for transmission over a MACCM. We expand these messages over the MACCM graph. Then, we configure the encoding graph generated by the message graph. Based on the encoding graph, the encoding scheme is now determined. One may expect that the rate (56) is achievable by the successive decoding strategy described earlier. However, it is still an important issue. Note that based on the successive decoding strategy, as described earlier, the receiver $K_2$ decodes the codewords with respect to the messages $\widetilde{\mathbb{M}}_{Y_{K_2}}$ using a jointly typical decoder. Remember the encoding graph based on which we have encoded the messages. In some situations, there exist messages $M_{\Delta_1}^{\nabla_1}, M_{\Delta_2}^{\nabla_2} \in \widetilde{\mathbb{M}}$ with $\Delta_2 \subseteq \Delta_1$ where $M_{\Delta_2}^{\nabla_2} \in \widetilde{\mathbb{M}}_{Y_{K_2}}$ but $M_{\Delta_1}^{\nabla_1} \notin \widetilde{\mathbb{M}}_{Y_{K_2}}$. From the view point of the encoding graph, the codeword with respect to $M_{\Delta_2}^{\nabla_2}$ is a satellite codeword and the codeword with respect to $M_{\Delta_1}^{\nabla_1}$ is its cloud center. We know that if two codewords construct a superposition structure, it is required that the cloud center is initially decoded before decoding the satellite. Therefore, we should also decode $M_{\Delta_1}^{\nabla_1}$ at the receiver $K_2$ while it does not belong to $\widetilde{\mathbb{M}}_{Y_{K_2}}$. Such inconsistency may occur also at the other receivers. For more explanation, again consider the network in Example 2. We have expanded these messages into the MACCM graph shown in Fig. 17. Note that the messages are expanded on the message graph based on their subscript sets. Let us consider the messages. We know that the message $M_{\{3\}}^{\{3\}}$ belong to $\widetilde{\mathbb{M}}_{Y_3}$ and the message $M_{\{3,4\}}^{\{1,2\}}$ is a cloud center for it. Therefore, at the receiver $Y_3$, decoding of the messages $M_{\{3\}}^{\{3\}}$ is not possible unless the message $M_{\{3,4\}}^{\{1,2\}}$ is decoded first. But $M_{\{3,4\}}^{\{1,2\}}$ does not belong to $\widetilde{\mathbb{M}}_{Y_3}$. Thereby, in order to achieve the bound (56), the successive decoding strategy fails.

How does one resolve this inconsistency? Recall that for the MACCM with an arbitrary distribution of messages, we demonstrated that it is required to transmit only a certain subset of messages (see (24)) to achieve the sum-rate capacity. Let the messages intended for transmission over a MACCM be $\mathbb{M}$; the desired subset of $\mathbb{M}$, denoted by $\mathbb{M}^*$ in (24), is composed of all messages $M_\Delta \in \mathbb{M}$ for which there is no $M_\Gamma \in \mathbb{M}$ with $\Delta \subsetneq \Gamma$. In other words, no message in $\mathbb{M}^*$ is a satellite for any other message in $\mathbb{M}$. Next, we derive a similar result for the degraded interference networks. Consider again the outer bound (56). Define:

$$\widetilde{\mathbb{M}}_{\overset{\mapsto}{Y_j}} \triangleq \widetilde{\mathbb{M}}_{Y_j} - \left(\widetilde{\mathbb{M}}_{Y_{j+1}} \cup \ldots \cup \widetilde{\mathbb{M}}_{Y_{K_2}}\right), \qquad j = 1, \ldots, K_2 \tag{62}$$

Note that $\widetilde{\mathbb{M}}_{\overset{\mapsto}{Y_j}}, j = 1, \ldots, K_2$, constitute a partition for the set $\widetilde{\mathbb{M}}$, i.e.,

$$\widetilde{\mathbb{M}} = \bigcup_{j\in[1:K_2]} \widetilde{\mathbb{M}}_{\overset{\mapsto}{Y_j}}, \qquad \widetilde{\mathbb{M}}_{\overset{\mapsto}{Y_{j_1}}} \bigcap \widetilde{\mathbb{M}}_{\overset{\mapsto}{Y_{j_2}}} = \emptyset, \qquad j_1, j_2 \in [1:K_2], \qquad j_1 \neq j_2 \tag{63}$$

Also, using (63), we have:

$$I\left(\widetilde{\mathbb{M}}_{Y_j}; Y_j \middle| \widetilde{\mathbb{M}}_{Y_{j+1}}, \dots, \widetilde{\mathbb{M}}_{Y_{K_2-1}}, \widetilde{\mathbb{M}}_{Y_{K_2}}, Q\right) = I\left(\widetilde{\mathbb{M}}_{\overset{\mapsto}{Y_j}}; Y_j \middle| \widetilde{\mathbb{M}}_{\overset{\mapsto}{Y_{j+1}}}, \dots, \widetilde{\mathbb{M}}_{\overset{\mapsto}{Y_{K_2-1}}}, \widetilde{\mathbb{M}}_{\overset{\mapsto}{Y_{K_2}}}, Q\right), \qquad j = 1, \dots, K_2 \tag{64}$$

Now by substituting (64) in (56), we derive:

$$\mathcal{C}_{sum}^{GIN-deg} \leq \max_{\tilde{\mathcal{P}}_{sum}^{GIN-deg}} \left( I\left(\widetilde{\mathbb{M}}_{\overset{\mapsto}{Y_1}}; Y_1 \middle| \widetilde{\mathbb{M}}_{\overset{\mapsto}{Y_2}}, \dots, \widetilde{\mathbb{M}}_{\overset{\mapsto}{Y_{K_2-1}}}, \widetilde{\mathbb{M}}_{\overset{\mapsto}{Y_{K_2}}}, Q\right) + \dots + I\left(\widetilde{\mathbb{M}}_{\overset{\mapsto}{Y_{K_2-1}}}; Y_{K_2-1} \middle| \widetilde{\mathbb{M}}_{\overset{\mapsto}{Y_{K_2}}}, Q\right) + I\left(\widetilde{\mathbb{M}}_{\overset{\mapsto}{Y_{K_2}}}; Y_{K_2} \middle| Q\right) \right) \tag{65}$$

where $\tilde{\mathcal{P}}_{sum}^{GIN-deg}$ is the set of all joint PDFs (57). Then, consider (65). Define:

$$\mathbb{M}_{Y_j}^{*} \triangleq \left\{ M_{\Delta}^{\nabla} \in \widetilde{\mathbb{M}}_{\overset{\mapsto}{Y_j}} : \text{There is no } M_{\Gamma}^{L} \in \widetilde{\mathbb{M}} - \left(\widetilde{\mathbb{M}}_{\overset{\mapsto}{Y_{j+1}}} \cup \dots \cup \widetilde{\mathbb{M}}_{\overset{\mapsto}{Y_{K_2}}}\right) \text{ with } \Delta \subsetneq \Gamma \right\}, \qquad \mathbb{M}_{Y_j}^{\times} \triangleq \widetilde{\mathbb{M}}_{\overset{\mapsto}{Y_j}} - \mathbb{M}_{Y_j}^{*}, \qquad j = 1, \dots, K_2 \tag{66}$$

According to the definition (66), the messages $\mathbb{M}_{Y_j}^{*}$ are those of $\widetilde{\mathbb{M}}_{\overset{\mapsto}{Y_j}}$ which are not satellite for any message in $\widetilde{\mathbb{M}} - \left(\widetilde{\mathbb{M}}_{\overset{\mapsto}{Y_{j+1}}} \cup \dots \cup \widetilde{\mathbb{M}}_{\overset{\mapsto}{Y_{K_2}}}\right)$, while $\mathbb{M}_{Y_j}^{\times}$ are those of $\widetilde{\mathbb{M}}_{\overset{\mapsto}{Y_j}}$ which are a satellite for at least one message in $\widetilde{\mathbb{M}} - \left(\widetilde{\mathbb{M}}_{\overset{\mapsto}{Y_{j+1}}} \cup \dots \cup \widetilde{\mathbb{M}}_{\overset{\mapsto}{Y_{K_2}}}\right)$.

In fact, given a certain message of $\widetilde{\mathbb{M}}_{\overset{\mapsto}{Y_j}}$ to determine whether it belongs to $\mathbb{M}_{Y_j}^{*}$ or to $\mathbb{M}_{Y_j}^{\times}$, it is required to explore among all the messages in $\widetilde{\mathbb{M}} - \left(\widetilde{\mathbb{M}}_{\overset{\mapsto}{Y_{j+1}}} \cup \dots \cup \widetilde{\mathbb{M}}_{\overset{\mapsto}{Y_{K_2}}}\right)$. If there is no cloud center for the message (equivalently, the message is not a satellite for any other message in $\widetilde{\mathbb{M}} - \left(\widetilde{\mathbb{M}}_{\overset{\mapsto}{Y_{j+1}}} \cup \dots \cup \widetilde{\mathbb{M}}_{\overset{\mapsto}{Y_{K_2}}}\right)$), it belongs to $\mathbb{M}_{Y_j}^{*}$; otherwise, it belongs to $\mathbb{M}_{Y_j}^{\times}$. By using the MACCM graph, the sets $\mathbb{M}_{Y_j}^{*}$ and $\mathbb{M}_{Y_j}^{\times}$ can be readily determined by inspection. For example, consider the message graph in Fig. 17. Based on (61) and (62), we have:

$$\begin{cases} \widetilde{\mathbb{M}}_{\overset{\mapsto}{Y_1}} = \left\{ M_{\{4\}}^{\{1\}} \right\} \\ \widetilde{\mathbb{M}}_{\overset{\mapsto}{Y_2}} = \left\{ M_{\{3,4\}}^{\{1,2\}} \right\} \\ \widetilde{\mathbb{M}}_{\overset{\mapsto}{Y_3}} = \left\{ M_{\{1,2,4\}}^{\{3\}}, M_{\{1,2\}}^{\{2,3\}}, M_{\{1\}}^{\{1,3\}}, M_{\{2\}}^{\{3\}}, M_{\{3\}}^{\{3\}} \right\} \end{cases} \tag{67}$$

Now consider the messages $\widetilde{\mathbb{M}}_{\overset{\mapsto}{Y_3}}, \widetilde{\mathbb{M}}_{\overset{\mapsto}{Y_2}}, \widetilde{\mathbb{M}}_{\overset{\mapsto}{Y_1}}$. As we see from Fig. 17,

1. The message $M_{\{1,2,4\}}^{\{3\}}$ is a cloud center for the messages $M_{\{1,2\}}^{\{2,3\}}, M_{\{1\}}^{\{1,3\}}, M_{\{2\}}^{\{3\}}$. Also, the message $M_{\{3,4\}}^{\{1,2\}}$ is a cloud center for $M_{\{3\}}^{\{3\}}$. Therefore, from the set $\widetilde{\mathbb{M}}_{\overset{\mapsto}{Y_3}}$ only the message $M_{\{1,2,4\}}^{\{3\}}$ belongs to $\widetilde{\mathbb{M}}_{Y_3}^{*}$. Thus, $\widetilde{\mathbb{M}}_{Y_3}^{*} = \left\{ M_{\{1,2,4\}}^{\{3\}} \right\}$.
2. There is no cloud center for the message $M_{\{3,4\}}^{\{1,2\}}$ in $\widetilde{\mathbb{M}} - \left(\widetilde{\mathbb{M}}_{\overset{\mapsto}{Y_3}}\right) = \left\{ M_{\{3,4\}}^{\{1,2\}}, M_{\{4\}}^{\{1\}} \right\}$. Thus, $\widetilde{\mathbb{M}}_{Y_2}^{*} = \left\{ M_{\{3,4\}}^{\{1,2\}} \right\}$.
3. The message $M_{\{4\}}^{\{1\}}$ is a satellite for both $M_{\{1,2,4\}}^{\{3\}}$ and $M_{\{3,4\}}^{\{1,2\}}$, but not for a message in $\widetilde{\mathbb{M}} - \left(\widetilde{\mathbb{M}}_{\overset{\mapsto}{Y_2}} \cup \widetilde{\mathbb{M}}_{\overset{\mapsto}{Y_3}}\right) = \left\{ M_{\{4\}}^{\{1\}} \right\}$. Thus, $\widetilde{\mathbb{M}}_{Y_1}^{*} = \left\{ M_{\{4\}}^{\{1\}} \right\}$.

In Fig. 17, we have marked the messages belonging to $\mathbb{M}^{*} = \widetilde{\mathbb{M}}_{Y_3}^{*} \cup \widetilde{\mathbb{M}}_{Y_2}^{*} \cup \widetilde{\mathbb{M}}_{Y_1}^{*}$ by "★" and those in $\mathbb{M}^{\times} = \mathbb{M}_{Y_1}^{\times} \cup \mathbb{M}_{Y_2}^{\times} \cup \mathbb{M}_{Y_3}^{\times}$ by "✗".

In the following, we prove that in order to achieve the maximum sum-rate in the degraded interference networks, it is optimal to transmit only the messages $\mathbb{M}_{Y_j}^{*}$ from the set $\widetilde{\mathbb{M}}_{\overset{\mapsto}{Y_j}}$ and ignore the others, i.e., $\mathbb{M}_{Y_j}^{\times}$, for all $j \in [1:K_2]$. In other words, if a message of $\widetilde{\mathbb{M}}_{\overset{\mapsto}{Y_j}}$ has a cloud center which should be decoded at the receiver $Y_j$ or a stronger receiver, to achieve the maximum sum-rate capacity, that message is not transmitted.

Define:

$$\tilde{I}_j \triangleq I\left(\widetilde{\mathbb{M}}_{\overset{\mapsto}{Y_j}};Y_j\middle|\widetilde{\mathbb{M}}_{\overset{\mapsto}{Y_{j+1}}},\ldots,\widetilde{\mathbb{M}}_{\overset{\mapsto}{Y_{K_2-1}}},\widetilde{\mathbb{M}}_{\overset{\mapsto}{Y_{K_2}}},Q\right), \qquad j=1,\ldots,K_2 \tag{68}$$

First, consider the set $\widetilde{\mathbb{M}}_{\overset{\mapsto}{Y_{K_2}}}$. Let $\mathbb{M}_{Y_{K_2}}^{\times} = \left\{M_{\Delta_1}^{\nabla_1},\ldots,M_{\Delta_\delta}^{\nabla_\delta}\right\}$, where $\delta = \left\|\mathbb{M}_{Y_{K_2}}^{\times}\right\|$. Note that according to the definition, for each message of $\mathbb{M}_{Y_{K_2}}^{\times}$, there exists at least one cloud center in $\widetilde{\mathbb{M}} - \mathbb{M}_{Y_{K_2}}^{\times}$, i.e.,

$$\forall \;\; l \in [1:\delta], \qquad \exists \;\; M_{\Gamma_l}^{L_l} \in \widetilde{\mathbb{M}} - \mathbb{M}_{Y_{K_2}}^{\times} \;:\; \Delta_l \subsetneq \Gamma_l \tag{69}$$

Consider the argument of the maximization (65). Let $M_{\Gamma_1}^{L_1} \in \widetilde{\mathbb{M}} - \mathbb{M}_{Y_{K_2}}^{\times}$ be a cloud center for the message $M_{\Delta_1}^{\nabla_1}$ in the sense of $\Delta_1 \subsetneq \Gamma_1$. Using (68), we can write:

$$\begin{aligned}
&\tilde{I}_{K_2-1} + \tilde{I}_{K_2}\\
&\quad= I\left(\widetilde{\mathbb{M}}_{\overset{\mapsto}{Y_{K_2-1}}};Y_{K_2-1}\middle|\mathbb{M}_{Y_{K_2}}^{*},\left\{M_{\Delta_1}^{\nabla_1},\ldots,M_{\Delta_\delta}^{\nabla_\delta}\right\},Q\right) + I\left(\mathbb{M}_{Y_{K_2}}^{*},\left\{M_{\Delta_1}^{\nabla_1},\ldots,M_{\Delta_\delta}^{\nabla_\delta}\right\};Y_{K_2}\middle|Q\right)\\
&\quad= I\left(\widetilde{\mathbb{M}}_{\overset{\mapsto}{Y_{K_2-1}}};Y_{K_2-1}\middle|\mathbb{M}_{Y_{K_2}}^{*},\left\{M_{\Delta_1}^{\nabla_1},\ldots,M_{\Delta_\delta}^{\nabla_\delta}\right\},Q\right) + I\left(M_{\Delta_1}^{\nabla_1};Y_{K_2}\middle|\mathbb{M}_{Y_{K_2}}^{*},\left\{M_{\Delta_2}^{\nabla_2},\ldots,M_{\Delta_\delta}^{\nabla_\delta}\right\},Q\right) + I\left(\mathbb{M}_{Y_{K_2}}^{*},\left\{M_{\Delta_2}^{\nabla_2},\ldots,M_{\Delta_\delta}^{\nabla_\delta}\right\};Y_{K_2}\middle|Q\right)\\
&\quad\overset{(a)}{\leq} I\left(\widetilde{\mathbb{M}}_{\overset{\mapsto}{Y_{K_2-1}}};Y_{K_2-1}\middle|\mathbb{M}_{Y_{K_2}}^{*},\left\{M_{\Delta_1}^{\nabla_1},\ldots,M_{\Delta_\delta}^{\nabla_\delta}\right\},Q\right) + I\left(M_{\Delta_1}^{\nabla_1};Y_{K_2-1}\middle|\mathbb{M}_{Y_{K_2}}^{*},\left\{M_{\Delta_2}^{\nabla_2},\ldots,M_{\Delta_\delta}^{\nabla_\delta}\right\},Q\right) + I\left(\mathbb{M}_{Y_{K_2}}^{*},\left\{M_{\Delta_2}^{\nabla_2},\ldots,M_{\Delta_\delta}^{\nabla_\delta}\right\};Y_{K_2}\middle|Q\right)\\
&\quad= I\left(\widetilde{\mathbb{M}}_{\overset{\mapsto}{Y_{K_2-1}}},M_{\Delta_1}^{\nabla_1};Y_{K_2-1}\middle|\mathbb{M}_{Y_{K_2}}^{*},\left\{M_{\Delta_2}^{\nabla_2},\ldots,M_{\Delta_\delta}^{\nabla_\delta}\right\},Q\right) + I\left(\mathbb{M}_{Y_{K_2}}^{*},\left\{M_{\Delta_2}^{\nabla_2},\ldots,M_{\Delta_\delta}^{\nabla_\delta}\right\};Y_{K_2}\middle|Q\right)
\end{aligned} \tag{70}$$

where (a) holds because $Y_{K_2}$ is a degraded version of $Y_{K_2-1}$. By repeating this step, one can derive:

$$\begin{aligned}
&\tilde{I}_{\max L_1} + \cdots + \tilde{I}_{K_2-1} + \tilde{I}_{K_2}\\
&\quad\leq I\left(\widetilde{\mathbb{M}}_{\overset{\mapsto}{Y_{\max L_1}}},M_{\Delta_1}^{\nabla_1};Y_{\max L_1}\middle|\widetilde{\mathbb{M}}_{\overset{\mapsto}{Y_{\max L_1+1}}},\ldots,\widetilde{\mathbb{M}}_{\overset{\mapsto}{Y_{K_2-1}}},\mathbb{M}_{Y_{K_2}}^{*},\left\{M_{\Delta_2}^{\nabla_2},\ldots,M_{\Delta_\delta}^{\nabla_\delta}\right\},Q\right) + \cdots + I\left(\mathbb{M}_{Y_{K_2}}^{*},\left\{M_{\Delta_2}^{\nabla_2},\ldots,M_{\Delta_\delta}^{\nabla_\delta}\right\};Y_{K_2}\middle|Q\right)
\end{aligned} \tag{71}$$

Also, let $\mathcal{P}_{sum \nrightarrow M_{\Delta_1}^{\nabla_1}}^{GIN-deg}$ be the set of all joint PDFs $P_{Q\widetilde{\mathbb{M}}X_1\ldots X_{K_1}}$ which are factorized as follows:

$$P_Q \times \prod_{M_\Delta^\nabla \in \widetilde{\mathbb{M}}} P_{M_\Delta^\nabla} \times \prod_{\substack{i\in[1:K_1],\\ i \notin \Gamma_1-\Delta_1}} P_{X_i|\widetilde{\mathbb{M}}_{X_i},Q} \times \prod_{\substack{i\in[1:K_1],\\ i \in \Gamma_1-\Delta_1}} P_{X_i|\widetilde{\mathbb{M}}_{X_i},M_{\Delta_1}^{\nabla_1},Q} \tag{72}$$

By comparing (57) and (72), it is clear that $\tilde{\mathcal{P}}_{sum}^{GIN-deg} \subseteq \mathcal{P}_{sum \nrightarrow M_{\Delta_1}^{\nabla_1}}^{GIN-deg}$. Considering this inclusion and by substituting (71) in (56), we derive:

$$\begin{aligned}
&\mathcal{C}_{sum}^{GIN-deg}\\
&\quad\leq \max\nolimits_{\tilde{\mathcal{P}}_{sum}^{GIN-deg}}\left(\tilde{I}_1 + \cdots + \tilde{I}_{K_2-1} + \tilde{I}_{K_2}\right)
\end{aligned}$$

$$\leq \max_{\mathcal{P}^{GIN-deg}_{sum \nrightarrow M^{\nabla_1}_{\Delta_1}}} \left(\begin{array}{c} I\left(\widetilde{\mathbb{M}}_{\overset{\mapsto}{Y_1}};Y_1\middle|\widetilde{\mathbb{M}}_{\overset{\mapsto}{Y_2}},\ldots,\widetilde{\mathbb{M}}_{\overset{\mapsto}{Y_{K_2-1}}},\widetilde{\mathbb{M}}_{\overset{\mapsto}{Y_{K_2}}},Q\right)+\cdots+I\left(\widetilde{\mathbb{M}}_{\overset{\mapsto}{Y_{\max L_1-1}}};Y_{\max L_1-1}\middle|\widetilde{\mathbb{M}}_{\overset{\mapsto}{Y_{\max L_1}}},\ldots,\widetilde{\mathbb{M}}_{\overset{\mapsto}{Y_{K_2-1}}},\widetilde{\mathbb{M}}_{\overset{\mapsto}{Y_{K_2}}},Q\right) \\ I\left(\widetilde{\mathbb{M}}_{\overset{\mapsto}{Y_{\max L_1}}},M^{\nabla_1}_{\Delta_1};Y_{\max L_1}\middle|\widetilde{\mathbb{M}}_{\overset{\mapsto}{Y_{\max L_1+1}}},\ldots,\widetilde{\mathbb{M}}_{\overset{\mapsto}{Y_{K_2-1}}},\mathbb{M}^{*}_{Y_{K_2}},\left\{M^{\nabla_2}_{\Delta_2},\ldots,M^{\nabla_\delta}_{\Delta_\delta}\right\},Q\right) \\ I\left(\widetilde{\mathbb{M}}_{\overset{\mapsto}{Y_{\max L_1+1}}};Y_{\max L_1+1}\middle|\widetilde{\mathbb{M}}_{\overset{\mapsto}{Y_{\max L_1+2}}},\ldots,\widetilde{\mathbb{M}}_{\overset{\mapsto}{Y_{K_2-1}}},\mathbb{M}^{*}_{Y_{K_2}},\left\{M^{\nabla_2}_{\Delta_2},\ldots,M^{\nabla_\delta}_{\Delta_\delta}\right\},Q\right)+\cdots+I\left(\mathbb{M}^{*}_{Y_{K_2}},\left\{M^{\nabla_2}_{\Delta_2},\ldots,M^{\nabla_\delta}_{\Delta_\delta}\right\};Y_{K_2}\middle|Q\right) \end{array}\right)$$

(73)

Now, consider the maximization (73). Note that $M^{L_1}_{\Gamma_1}$ belongs to $\widetilde{\mathbb{M}}_{\overset{\mapsto}{Y_{\max L_1}}}$ and it does not belong to $\widetilde{\mathbb{M}}_{\overset{\mapsto}{Y_j}}$ for $j \neq \max L_1$. Therefore, in all mutual information functions in the first row of the argument (73), both messages $M^{L_1}_{\Gamma_1}, M^{\nabla_1}_{\Delta_1}$ appear next to each other and after the conditioning operator. In the second row, these messages appear next to each other before the conditioning operator, and in the third row none of these messages appears. Moreover, in the distributions $\mathcal{P}^{GIN-deg}_{sum \nrightarrow M^{\nabla_1}_{\Delta_1}}$ given by (72), if $i \in \Gamma_1$, then $X_i$ depends on both $M^{L_1}_{\Gamma_1}, M^{\nabla_1}_{\Delta_1}$; otherwise, if $i \notin \Gamma_1$, then $X_i$ does not depend on any of them. Therefore, as a solution of the maximization (73), one can collectively represent both messages $\left(M^{L_1}_{\Gamma_1}, M^{\nabla_1}_{\Delta_1}\right)$ (which are auxiliary RVs in the characterization of (56)) using a new RV. It should be noted that one can replace the pair $\left(M^{L_1}_{\Gamma_1}, M^{\nabla_1}_{\Delta_1}\right)$ by any arbitrary RV which is independent of $\widetilde{\mathbb{M}} - \left\{M^{L_1}_{\Gamma_1}, M^{\nabla_1}_{\Delta_1}\right\}$ and $Q$. We represent these RVs by $M^{L_1}_{\Gamma_1}$. In fact, a solution to the maximization (73) is to nullify the random variable $M^{\nabla_1}_{\Delta_1}$. The benefit of this choice is that it leads to achievability scheme for the resultant outer bound as we shall see subsequently.

By repeating the procedure discussed in the above for all messages in $\mathbb{M}^{\times}_{Y_{K_2}} = \left\{M^{\nabla_1}_{\Delta_1},\ldots,M^{\nabla_\delta}_{\Delta_\delta}\right\}$, they can be removed them from the bound (56) to yield:

$$\mathcal{C}^{GIN-deg}_{sum} \leq \max_{\tilde{\mathcal{P}}^{GIN-deg}_{sum\backslash\mathbb{M}^{\times}_{Y_{K_2}}}} \left(I\left(\widetilde{\mathbb{M}}_{\overset{\mapsto}{Y_1}};Y_1\middle|\widetilde{\mathbb{M}}_{\overset{\mapsto}{Y_2}},\ldots,\widetilde{\mathbb{M}}_{\overset{\mapsto}{Y_{K_2-1}}},\mathbb{M}^{*}_{Y_{K_2}},Q\right)+\cdots+I\left(\widetilde{\mathbb{M}}_{\overset{\mapsto}{Y_{K_2-1}}};Y_{K_2-1}\middle|\mathbb{M}^{*}_{Y_{K_2}},Q\right)+I\left(\mathbb{M}^{*}_{Y_{K_2}};Y_{K_2}\middle|Q\right)\right)$$

(74)

where $\tilde{\mathcal{P}}^{GIN-deg}_{sum\backslash\mathbb{M}^{\times}_{Y_{K_2}}}$ denotes the set of all joint PDFs as in (57) except that the messages $\mathbb{M}^{\times}_{Y_{K_2}}$ are nullified everywhere. In fact by following the above approach, we can eliminate also the message sets $\mathbb{M}^{\times}_{Y_{K_2-1}},\ldots,\mathbb{M}^{\times}_{Y_1}$ from the bound (56) step by step, i.e., after $\mathbb{M}^{\times}_{Y_j}$, the set $\mathbb{M}^{\times}_{Y_{j-1}}$ is removed. Thereby, we can derive the following theorem.

Define:

$$\begin{cases} \mathbb{M}^{*} \triangleq \bigcup_{j\in[1:K_2]} \mathbb{M}^{*}_{Y_j} \\ \mathbb{M}^{*}_{X_i} \triangleq \widetilde{\mathbb{M}}_{X_i} - \left(\mathbb{M}^{\times}_{Y_1} \cup \ldots \cup \mathbb{M}^{\times}_{Y_{K_2}}\right), \qquad i = 1,\ldots,K_1 \end{cases}$$

(75)

where $\mathbb{M}^{*}_{Y_j}, \mathbb{M}^{\times}_{Y_j}, j = 1,\ldots,K_2$, are given by (66), and $\widetilde{\mathbb{M}}_{X_i}, i = 1,\ldots,K_1$ is given by (55).

***Theorem 1) Information Flow in Degraded Interference Networks;***

Consider the $K_1$-Transmitter/$K_2$-Receiver degraded interference network given in (37) with the message sets $\mathbb{M}, \mathbb{M}_{X_i}, i = 1, \dots, K_1$, and $\mathbb{M}_{Y_j}, j = 1, \dots, K_2$. The sum-rate capacity is given by:

$$\mathcal{C}_{sum}^{GIN-deg} = \max_{\overset{*}{\mathcal{P}}_{sum}^{GIN-deg}} \left( I\left(\mathbb{M}_{Y_1}^{*}; Y_1 \middle| \mathbb{M}_{Y_2}^{*}, \dots, \mathbb{M}_{Y_{K_2-1}}^{*}, \mathbb{M}_{Y_{K_2}}^{*}, Q\right) + \cdots + I\left(\mathbb{M}_{Y_{K_2-1}}^{*}; Y_{K_2-1} \middle| \mathbb{M}_{Y_{K_2}}^{*}, Q\right) + I\left(\mathbb{M}_{Y_{K_2}}^{*}; Y_{K_2} \middle| Q\right) \right)$$

(76)

where $\overset{*}{\mathcal{P}}_{sum}^{GIN-deg}$ denotes the set of all joint PDFs as follows:

$$P_Q \times \prod_{M_\Delta^\nabla \in \mathbb{M}^*} P_{M_\Delta^\nabla} \times \prod_{i \in [1:K_1]} P_{X_i | \mathbb{M}_{X_i}^* Q}$$

(77)

also, the PDFs $P_{M_\Delta^\nabla}$, $M_\Delta^\nabla \in \mathbb{M}^*$ are uniformly distributed, and $P_{X_i|\mathbb{M}_{X_i}^* Q} \in \{0,1\}$ for $i = 1, \dots, K_1$, i.e., $X_i$ is a deterministic function of $\left(\mathbb{M}_{X_i}^*, Q\right)$. The messages $\mathbb{M}^*$ are given in (75).

*Proof of Theorem 1)*

The converse part is derived by removing the message sets $\mathbb{M}_{Y_{K_2}}^{\times}, \dots, \mathbb{M}_{Y_1}^{\times}$ from the bound (56) step by step as described before. To prove the achievability, we expand the messages $\mathbb{M}^*$ over the MACCM graph. Then, we configure the encoding graph generated by the message graph. The encoding scheme is perfectly determined by the encoding graph. At the receivers, the messages are decoded using successive decoding strategy described below:

The receiver $Y_{K_2}$ decodes its respective messages $\mathbb{M}_{Y_{K_2}}^*$ using joint decoding technique. Note that since the messages $\mathbb{M}_{Y_{K_2}}^*$ have no cloud center in $\mathbb{M}^*$, this decoding does work successfully. The partial sum-rate due to this step is:

$$I\left(\mathbb{M}_{Y_{K_2}}^*; Y_{K_2} \middle| Q\right)$$

At the receiver $Y_{K_2-1}$, the messages $\mathbb{M}_{Y_{K_2}}^*$ are jointly decoded first; this is performed without introducing any new rate cost because $Y_{K_2}$ is a degraded version of $Y_{K_2-1}$. Then, the receiver jointly decodes its respective messages $\mathbb{M}_{Y_{K_2-1}}^*$ using the received sequence and the previously decoded codewords. We know that there is no cloud center in $\mathbb{M}^* - \mathbb{M}_{Y_{K_2}}^*$ for the messages $\mathbb{M}_{Y_{K_2-1}}^*$; also, their cloud centers which possibly exist in $\mathbb{M}_{Y_{K_2}}^*$ have been already decoded. Thereby, successful decoding with the following partial sum-rate is attainable:

$$I\left(\mathbb{M}_{Y_{K_2-1}}^*; Y_{K_2-1} \middle| \mathbb{M}_{Y_{K_2}}^*, Q\right)$$

This successive decoding strategy is repeated at other receivers step by step from the weaker receivers towards the stronger ones. Consequently, the sum-rate (76) is achievable.

It should be noted that, in the achievability scheme described above, each receiver jointly decodes its respective messages (after decoding the messages corresponding to weaker receivers); nevertheless, it is possible to use successive decoding also at each receiver locally. In this case, the decoding scheme at each receiver is designed similar to the one mentioned in Proposition 2 for the MACCM to achieve the point (23). Accordingly, we have a two-layer successive decoding strategy: Each receiver first decodes the messages corresponding to the weaker receivers in the same order as to themselves; then it decodes its respective messages successively. ■

***Remarks 1:***

1. Theorem 1 establishes the sum-rate capacity for all degraded interference networks with arbitrary number of transmitters, arbitrary number of receivers, and arbitrary distribution of messages among transmitters and receivers. In addition, it explicitly determines which subset of messages is required to be sent in order to achieve the sum-capacity.

2. It should be mentioned that the sum-rate capacity given in (76) can be readily re-expressed using the encoding graph. This is treated exactly similar to the one described in Subsection II.A for the MACCM capacity region.
3. The sum-rate capacity result (76) holds also for the Gaussian degraded interference networks. For these channels, one can easily solve the corresponding optimization. To this end, first the sum-rate capacity (76) is re-described based on the encoding graph. Then, in the resulting characterization, one can prove that (for example, using the EPI but other ways are also possible) the Gaussian input distributions are always optimal. Subsequently, we illustrate this point by some examples. Nevertheless, providing a closed-form expression for the general case is undesirable due to its high formulation and computational complexities.

Also, it should be remarked that for those networks with no cost-constraint on their input signals, the time-sharing parameter $Q$ can be dropped from the characterization (76). This is due to the following argument. Given any joint PDF of the form (77), we can write:

$$\begin{aligned}
&I\left(\mathbb{M}_{Y_1}^*;Y_1\middle|\mathbb{M}_{Y_2}^*,\dots,\mathbb{M}_{Y_{K_2-1}}^*,\mathbb{M}_{Y_{K_2}}^*,Q\right)+\cdots+I\left(\mathbb{M}_{Y_{K_2-1}}^*;Y_{K_2-1}\middle|\mathbb{M}_{Y_{K_2}}^*,Q\right)+I\left(\mathbb{M}_{Y_{K_2}}^*;Y_{K_2}\middle|Q\right)\\
&\quad=\sum_q P_Q(q)\left(I\left(\mathbb{M}_{Y_1}^*;Y_1\middle|\mathbb{M}_{Y_2}^*,\dots,\mathbb{M}_{Y_{K_2-1}}^*,\mathbb{M}_{Y_{K_2}}^*,q\right)+\cdots+I\left(\mathbb{M}_{Y_{K_2-1}}^*;Y_{K_2-1}\middle|\mathbb{M}_{Y_{K_2}}^*,q\right)+I\left(\mathbb{M}_{Y_{K_2}}^*;Y_{K_2}\middle|q\right)\right)\\
&\quad\le\max_q\left(I\left(\mathbb{M}_{Y_1}^*;Y_1\middle|\mathbb{M}_{Y_2}^*,\dots,\mathbb{M}_{Y_{K_2-1}}^*,\mathbb{M}_{Y_{K_2}}^*,q\right)+\cdots+I\left(\mathbb{M}_{Y_{K_2-1}}^*;Y_{K_2-1}\middle|\mathbb{M}_{Y_{K_2}}^*,q\right)+I\left(\mathbb{M}_{Y_{K_2}}^*;Y_{K_2}\middle|q\right)\right)\\
&\quad\le\max_{\overset{*}{\mathcal{P}}_{sum\backslash Q}^{GIN-deg}}\left(I\left(\mathbb{M}_{Y_1}^*;Y_1\middle|\mathbb{M}_{Y_2}^*,\dots,\mathbb{M}_{Y_{K_2-1}}^*,\mathbb{M}_{Y_{K_2}}^*\right)+\cdots+I\left(\mathbb{M}_{Y_{K_2-1}}^*;Y_{K_2-1}\middle|\mathbb{M}_{Y_{K_2}}^*\right)+I\left(\mathbb{M}_{Y_{K_2}}^*;Y_{K_2}\right)\right)
\end{aligned}\tag{78}$$

and $\overset{*}{\mathcal{P}}_{sum\backslash Q}^{GIN-deg}$ the set of all joint PDFs given below:

$$\prod_{M_\Delta^\nabla\in\mathbb{M}^*}P_{M_\Delta^\nabla}\times\prod_{i\in[1:K_1]}P_{X_i|\mathbb{M}_{X_i}^*}\tag{79}$$

Therefore, for $\mathcal{C}_{sum}^{GIN-deg}$ in (76) we have:

$$\mathcal{C}_{sum}^{GIN-deg}\le\max_{\overset{*}{\mathcal{P}}_{sum\backslash Q}^{GIN-deg}}\left(I\left(\mathbb{M}_{Y_1}^*;Y_1\middle|\mathbb{M}_{Y_2}^*,\dots,\mathbb{M}_{Y_{K_2-1}}^*,\mathbb{M}_{Y_{K_2}}^*\right)+\cdots+I\left(\mathbb{M}_{Y_{K_2-1}}^*;Y_{K_2-1}\middle|\mathbb{M}_{Y_{K_2}}^*\right)+I\left(\mathbb{M}_{Y_{K_2}}^*;Y_{K_2}\right)\right)\tag{80}$$

The inequality (80) actually holds with equality because the expression on its right side is derived by setting $Q\equiv\emptyset$ in (76).

Note that the above argument is no longer valid for the networks with cost-constraint on their input signals, e.g., for the Gaussian networks with input power constraint.

For Example 2, we have depicted the message and the encoding graphs with respect to the message set $\mathbb{M}^*=\left\{M_{\{1,2,4\}}^{\{3\}},M_{\{3,4\}}^{\{1,2\}},M_{\{4\}}^{\{1\}}\right\}$, conjugated with each other in Fig. 18. Note that for simplicity we have used the letters $U$ and $V$ for codewords of the messages $M_{\{3,4\}}^{\{1,2\}}$ and $M_{\{1,2,4\}}^{\{3\}}$, respectively.

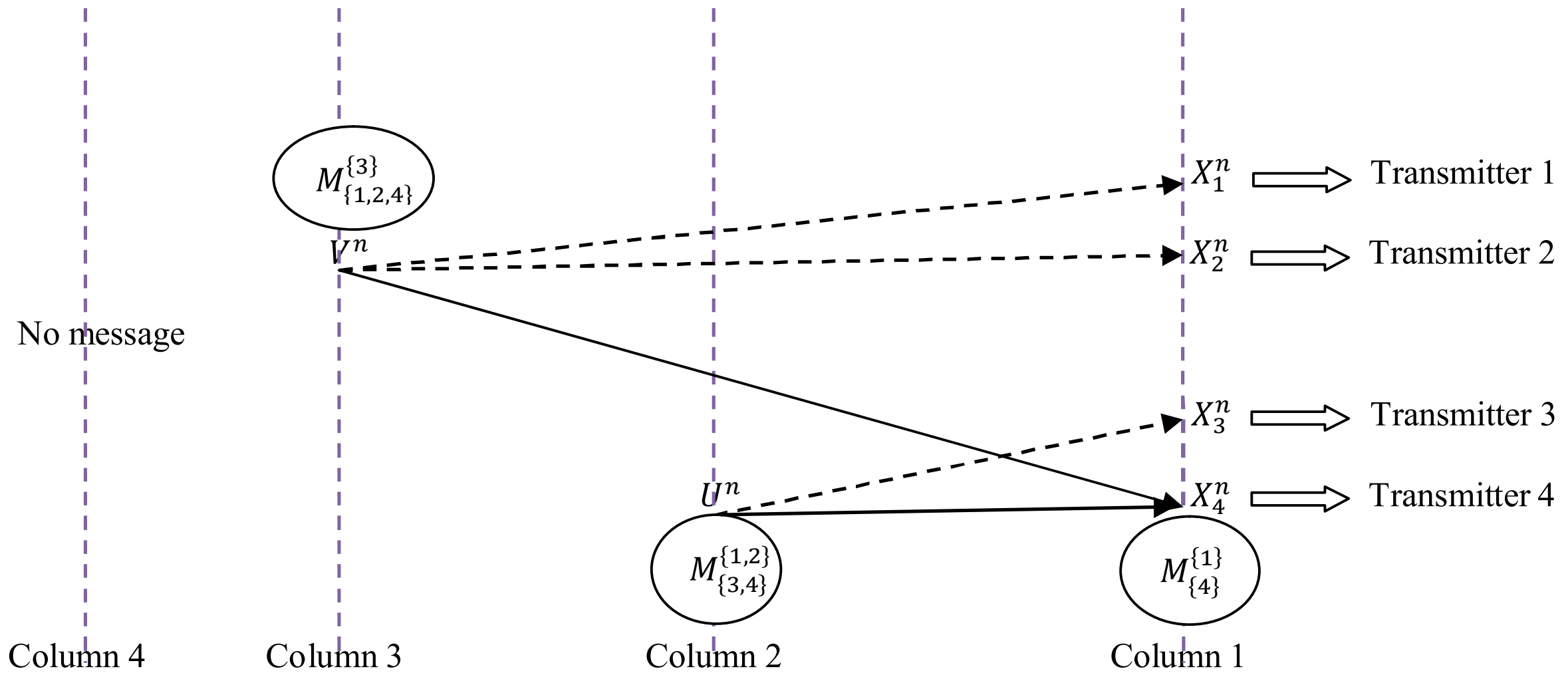

Figure 18. The message and the encoding graphs with respect to the message set $\mathbb{M}^*$ for the interference network in Fig. 16, where $\mathbb{M}^*$ is the set of those messages which are essentially transmitted to achieve the sum-rate capcity for the degraded network.

By considering Fig. 18, we now can express the sum-rate capacity as follows:

$$\mathcal{C}_{sum}^{EX.2} = \max_{\overset{*}{\mathcal{P}}_{sum}^{EX.2}} \left( I\left(M_{\{4\}}^{\{1\}}; Y_1 \middle| M_{\{3,4\}}^{\{1,2\}}, M_{\{1,2,4\}}^{\{3\}}, Q\right) + I\left(M_{\{3,4\}}^{\{1,2\}}; Y_2 \middle| M_{\{1,2,4\}}^{\{3\}}, Q\right) + I\left(M_{\{1,2,4\}}^{\{3\}}; Y_3 \middle| Q\right)\right) \tag{81}$$

where $\overset{*}{\mathcal{P}}_{sum}^{EX.2}$ denotes the set of all joint PDFs as:

$$P_Q \times P_{M_{\{1,2,4\}}^{\{3\}}} \times P_{M_{\{3,4\}}^{\{1,2\}}} \times P_{M_{\{4\}}^{\{1\}}} \times P_{X_1|M_{\{1,2,4\}}^{\{3\}},Q} \times P_{X_2|M_{\{1,2,4\}}^{\{3\}},Q} \times P_{X_3|M_{\{3,4\}}^{\{1,2\}},Q} \times P_{X_4|M_{\{4\}}^{\{1\}} M_{\{1,2,4\}}^{\{3\}}, M_{\{3,4\}}^{\{1,2\}} Q} \tag{82}$$

Using MG/EG translation, we can re-express (simplify) the sum-rate capacity as follows:

$$\mathcal{C}_{sum}^{EX.2} = \max_{\mathcal{P}_{sum}^{EX.2}} (I(X_4; Y_1|U, V, Q) + I(U; Y_2|V, Q) + I(V; Y_3|Q)) \tag{83}$$

where $\mathcal{P}_{sum}^{EX.2}$ denotes the set of all joint PDFs given below:

$$P_Q P_{U|Q} P_{V|Q} P_{X_1|VQ} P_{X_2|VQ} P_{X_3|UQ} P_{X_4|UVQ}, \qquad P_{X_1|VQ}, P_{X_2|VQ}, P_{X_3|UQ} \in \{0,1\} \tag{84}$$

The expression (83) can be further simplified as follows:

$$\begin{aligned}
\mathcal{C}_{sum}^{EX.2} &= \max\nolimits_{\mathcal{P}_{sum}^{EX.2}} (I(X_4; Y_1|U, V, Q) + I(U; Y_2|V, Q) + I(V; Y_3|Q)) \\
&\overset{(a)}{=} \max\nolimits_{\mathcal{P}_{sum}^{EX.2}} (I(X_4; Y_1|X_1, X_2, X_3, U, V, Q) + I(X_3, U; Y_2|X_1, X_2, V, Q) + I(X_1, X_2, V; Y_3|Q)) \\
&= \max\nolimits_{\mathcal{P}_{sum}^{EX.2}} (I(X_4; Y_1|X_1, X_2, X_3, U, V, Q) + I(X_3, U; Y_2|X_1, X_2, V, Q) + I(V; Y_3|X_1, X_2, Q) + I(X_1, X_2; Y_3|Q)) \\
&\overset{(b)}{\leq} \max\nolimits_{\mathcal{P}_{sum}^{EX.2}} (I(X_4; Y_1|X_1, X_2, X_3, U, V, Q) + I(X_3, U; Y_2|X_1, X_2, V, Q) + I(V; Y_2|X_1, X_2, Q) + I(X_1, X_2; Y_3|Q)) \\
&= \max\nolimits_{\mathcal{P}_{sum}^{EX.2}} (I(X_4; Y_1|X_1, X_2, X_3, U, V, Q) + I(X_3, U, V; Y_2|X_1, X_2, Q) + I(X_1, X_2; Y_3|Q)) \\
&= \max\nolimits_{\mathcal{P}_{sum}^{EX.2}} (I(X_4; Y_1|X_1, X_2, X_3, U, V, Q) + I(U, V; Y_2|X_1, X_2, X_3, Q) + I(X_3; Y_2|X_1, X_2, Q) + I(X_1, X_2; Y_3|Q)) \\
&\overset{(c)}{\leq} \max\nolimits_{\mathcal{P}_{sum}^{EX.2}} (I(X_4; Y_1|X_1, X_2, X_3, U, V, Q) + I(U, V; Y_1|X_1, X_2, X_3, Q) + I(X_3; Y_2|X_1, X_2, Q) + I(X_1, X_2; Y_3|Q)) \\
&= \max\nolimits_{\mathcal{P}_{sum}^{EX.2}} (I(X_4, U, V; Y_1|X_1, X_2, X_3, Q) + I(X_3; Y_2|X_1, X_2, Q) + I(X_1, X_2; Y_3|Q)) \\
&= \max\nolimits_{\mathcal{P}_{sum}^{EX.2}} (I(X_4; Y_1|X_1, X_2, X_3, Q) + I(X_3; Y_2|X_1, X_2, Q) + I(X_1, X_2; Y_3|Q))
\end{aligned} \tag{85}$$

where equality (a) holds because $X_1$ and $X_2$ are given by deterministic functions of $(V,Q)$, and $X_3$ is given by a deterministic function of $(U,Q)$; also the inequalities (b) and (c) are due to the degradedness of the network. Note that according to (84), conditioning on $Q$, the pair $(X_1,X_2)$ is independent of $X_3$. On the other hand, it is clear that by setting $U \equiv X_3$ and $V \equiv (X_1,X_2)$ in (83), the expression at the right side of the last equality in (85) is derived. Therefore, the sum-rate capacity is given by:

$$\mathcal{C}_{sum}^{EX.2} = \max_{P_Q P_{X_1X_2|Q} P_{X_3|Q} P_{X_4|X_1X_2X_3Q}} \left( I(X_4;Y_1|X_1,X_2,X_3,Q) + I(X_3;Y_2|X_1,X_2,Q) + I(X_1,X_2;Y_3|Q) \right) \tag{86}$$

Now let us consider the Gaussian version of Example 2: the degraded network is formulated as follows:

$$\begin{cases} Y_1 = a_1X_1 + a_2X_2 + a_3X_3 + a_4X_4 + Z_1 \\ Y_2 = \dfrac{a_1}{b_2}X_1 + \dfrac{a_2}{b_2}X_2 + \dfrac{a_3}{b_2}X_3 + \dfrac{a_4}{b_2}X_4 + Z_2 \qquad , \qquad b_2, b_3 \geq 1 \\ Y_3 = \dfrac{a_1}{b_2b_3}X_1 + \dfrac{a_2}{b_2b_3}X_2 + \dfrac{a_3}{b_2b_3}X_3 + \dfrac{a_4}{b_2b_3}X_4 + Z_3 \end{cases} \tag{87}$$

where $Z_1, Z_2, Z_3$ are Gaussian noises with zero mean and unit variance, and the input signals are subject to power constraints $\mathbb{E}[X_i^2] \leq P_i, i = 1,2,3,4$. As shown in Lemma 1, the channel (87) is equivalent to the following:

$$\begin{cases} \tilde{Y}_1 = Y_1 \\ \tilde{Y}_2 = \dfrac{1}{b_2}\tilde{Y}_1 + \sqrt{1 - \dfrac{1}{b_2^2}}\,\tilde{Z}_2 \\ \tilde{Y}_3 = \dfrac{1}{b_3}\tilde{Y}_2 + \sqrt{1 - \dfrac{1}{b_3^2}}\,\tilde{Z}_3 \end{cases} \tag{88}$$

where $\tilde{Z}_2, \tilde{Z}_3$ are independent Gaussian RVs (also independent of $Z_1$) with zero mean and unit variance. We now present an explicit characterization for the sum-rate capacity of the Gaussian network in (87).

***Proposition 4)*** *Consider the multi-message interference network in Fig. 16. The sum-rate capacity of the Gaussian network in (87) is given in the following:*

$$\mathcal{C}_{sum}^{EX.2\sim G} = \max_{\substack{\alpha,\beta \in [0,1] \\ \alpha^2+\beta^2 \leq 1}} \left( \begin{array}{c} \psi\left(a_4^2(1-(\alpha^2+\beta^2))P_4\right) + \psi\left( \dfrac{\frac{1}{b_2^2}\left(|a_3| + |a_4|\beta\sqrt{\frac{P_4}{P_3}}\right)^2 P_3}{\frac{a_4^2}{b_2^2}(1-(\alpha^2+\beta^2))P_4 + 1} \right) \\ +\psi\left( \dfrac{\frac{1}{b_2^2b_3^2}\left(a_1^2P_1 + a_2^2P_2 + a_4^2\alpha^2P_4 + 2|a_1a_2|\sqrt{P_1P_2} + 2|a_1a_4|\alpha\sqrt{P_1P_4} + 2|a_2a_4|\alpha\sqrt{P_2P_4}\right)}{\frac{1}{b_2^2b_3^2}\left(\left(|a_3| + |a_4|\beta\sqrt{\frac{P_4}{P_3}}\right)^2 P_3 + a_4^2(1-(\alpha^2+\beta^2))P_4\right) + 1} \right) \end{array} \right) \tag{89}$$

*Proof of Proposition 4)* Let $X_1, X_3$ and $Z$ be independent Gaussian RVs with zero means and variances $P_1, P_3$, and 1, respectively. Moreover, define $X_2$ and $X_4$ as follows:

$$\begin{cases} X_2 \triangleq \text{sign}(a_1a_2)\sqrt{\frac{P_2}{P_1}}X_1 \\ X_4 \triangleq \text{sign}(a_1a_4)\alpha\sqrt{\frac{P_4}{P_1}}X_1 + \text{sign}(a_3a_4)\beta\sqrt{\frac{P_4}{P_3}}X_3 + \sqrt{(1-(\alpha^2+\beta^2))P_4}\,Z \end{cases} \tag{90}$$

where $\alpha, \beta \in [0,1]$ are arbitrary real numbers with $\alpha^2 + \beta^2 \leq 1$; also, for a real number $x$, $\text{sign}(x)$ is equal to 1 if $x$ is nonnegative and -1 otherwise. Then, by setting $X_1, X_2, X_3, X_4$ given by (90) and $Q \equiv \emptyset$ in (86), we derive the achievability of (89). The proof of the optimality of Gaussian input distributions is given in the Appendix. ■

***Remark 2:*** The method of our proof presented in Appendix for the optimality of Gaussian distributions to achieve the sum-rate capacity of the Gaussian network of Example 2 can be adapted to other scenarios. Nonetheless, as the network is degraded, by manipulating the mutual information functions in (76), one may present other arguments for the optimality of Gaussian inputs.

Consider once more the general interference network in Fig.14 with the message set $\mathbb{M}$. In order to achieve the sum-rate capacity for the degraded networks, it is optimal to consider the transmission of only a subset of $\mathbb{M}$ denoted by $\mathbb{M}^*$ and ignore the other messages. Let us summarize the algorithm presented above to extract the messages $\mathbb{M}^*$ from the set $\mathbb{M}$. We first expand the messages into the MACCM plan of messages as in Fig. 15. Then, we show that from each message set $\mathbb{M}_\Delta$ in Fig. 15, where $\Delta$ is an arbitrary subset of $\{1, \dots, K_1\}$, only one message belongs to $\mathbb{M}^*$. The procedure of determining this desired message out of $\mathbb{M}_\Delta$ is identical to the one proposed in Subsection II.A to derive the sum-rate capacity of a degraded BCCM with a given arbitrary set of messages (see (32) and (33)). In the second phase of the algorithm, from the messages corresponding to each receiver, the ones which are a satellite for the messages that should be decoded at either that receiver or some stronger receivers are removed. The algorithm in this phase does work essentially similar to that one we proposed in Subsection II.A to obtain the sum-rate capacity of the MACCM with a given arbitrary set of messages (see equation (24) and Fig. 10). Therefore, the algorithm is relevant to both broadcast and multiple access characteristics of the network; both of these characteristics are suitably represented by the MACCM plan of messages. In general, one can deduce the following principle from our algorithm:

***Principle of Information Flow in Degraded Networks:*** *In order to achieve the maximum sum-rate, the transmitters try to send information for the stronger receivers and, if possible, avoid sending the messages with respect to the weaker receivers.*

Lastly, we shall provide some other examples on our result. Note that in the following examples we assume that the channel is degraded in the sense of (37).

***Example 3; K-User Classical Interference Channel***

Consider the $K$-User CIC as shown in Fig. 19. In this network, each transmitter sends a message to its respective receiver and there is no cooperation among transmitters.

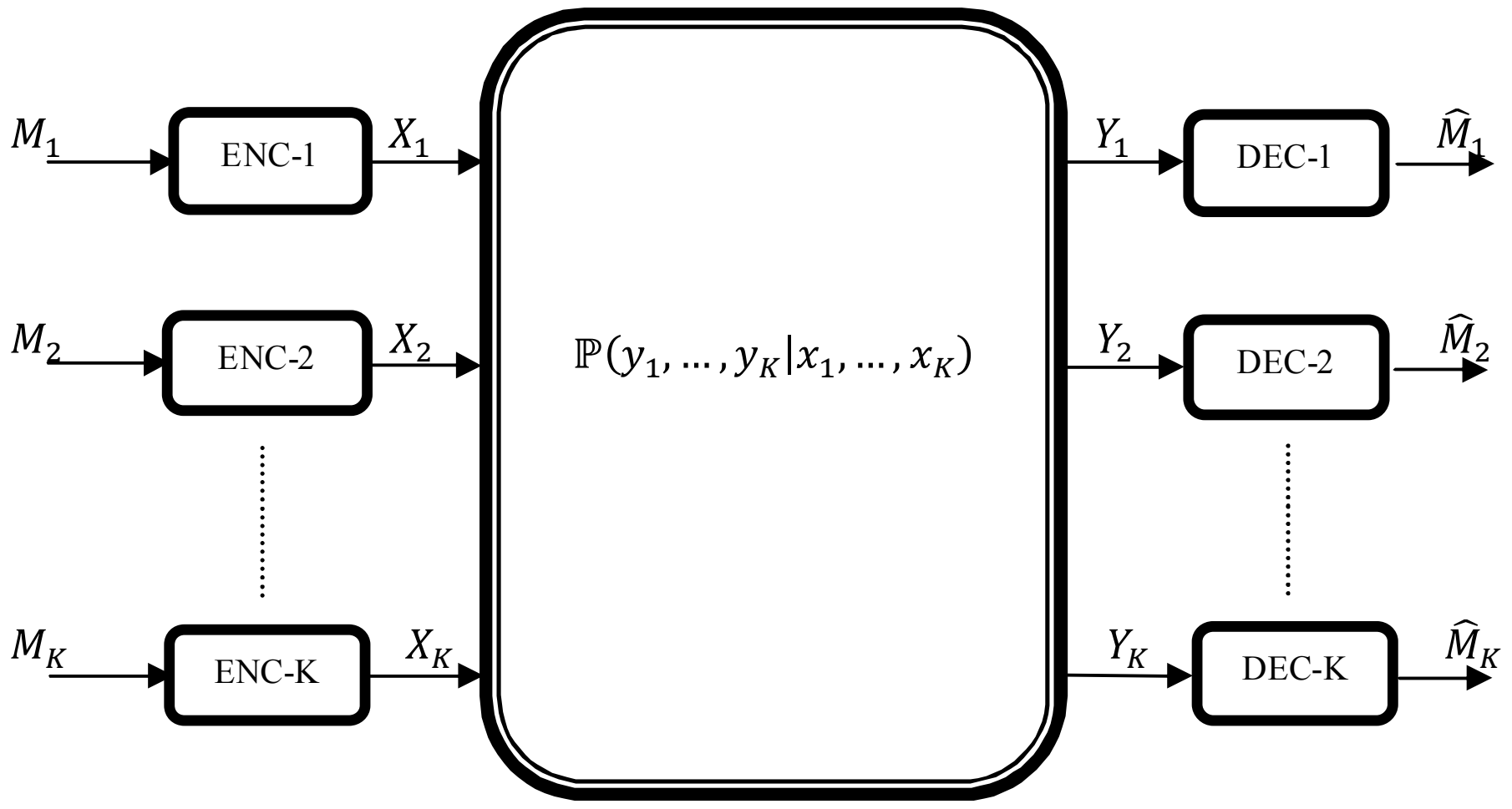

Figure 19. The $K$-user Classical Interference Channel (CIC).

The message and the encoding graphs (conjugated with each other) with respect to this scenario have been depicted in Fig. 20.

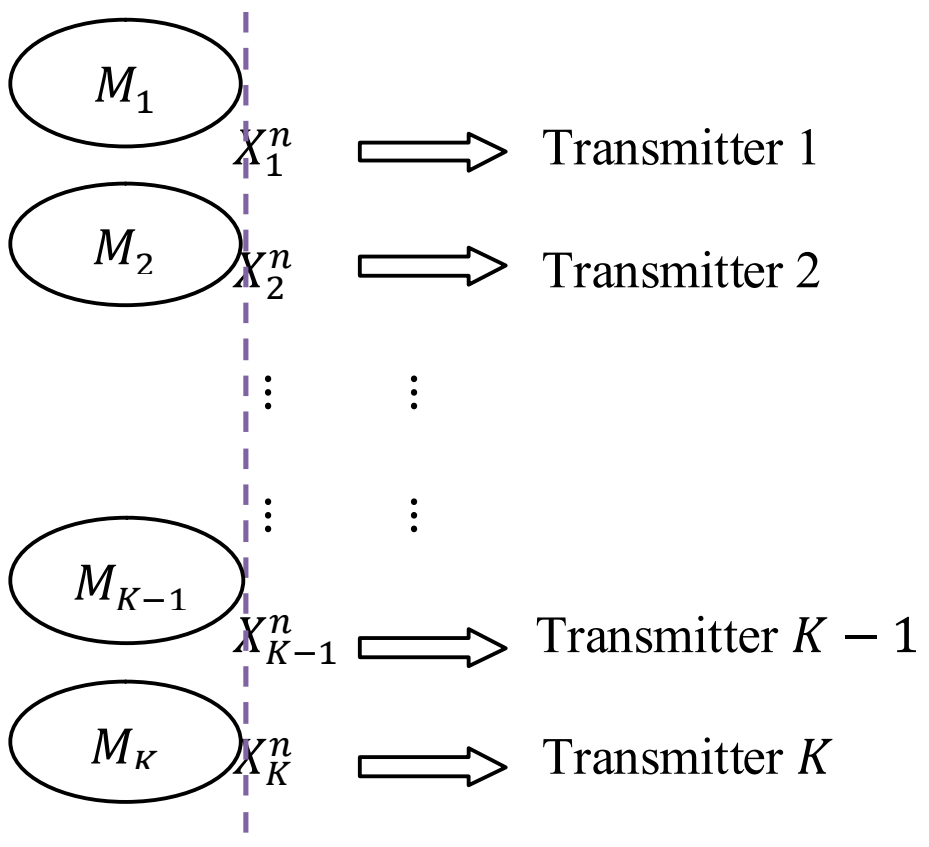

Figure 20. The message and the encoding graphs with respect to the $K$-user CIC.

The sum-rate capacity of the degraded network is directly derived from Theorem 1 which is given as follows:

$$\mathcal{C}_{sum}^{CIC_{deg}:K-user} = \max_{\mathcal{P}_{sum}^{CIC_{deg}:K-user}} \big(I(M_1;Y_1|M_2,\dots,M_{K-1},M_K,Q) + \cdots + I(M_{K-1};Y_{K-1}|M_K,Q) + I(M_K;Y_K|Q)\big) \tag{91}$$

where $\mathcal{P}_{sum}^{CIC_{deg}:K-user}$ is the set of all joint PDFs given by:

$$P_Q \prod_{l=1}^{K} P_{M_l} P_{X_l|M_l,Q} \tag{92}$$

Also, $P_{X_l|M_l,Q} \in \{0,1\}, l = 1,\dots,K$; in other words, $X_l$ is a deterministic function of $(M_l,Q)$. Using the MG/EG translation, one can readily re-write the sum-rate capacity as follows:

$$\mathcal{C}_{sum}^{CIC_{deg}:K-user} = \max_{P_Q \prod_{l=1}^{K} P_{X_l|Q}} \big(I(X_1;Y_1|X_2,\dots,X_{K-1},X_K,Q) + \cdots + I(X_{K-1};Y_{K-1}|X_K,Q) + I(X_K;Y_K|Q)\big) \tag{93}$$

Note that to derive the expression (93), it is sufficient to replace each message by the RV from which the codeword corresponding to that message is constructed in the encoding graph. It is remarkable that this is the first sum-rate capacity result for the $K$-user CIC which is derived for *both discrete and Gaussian networks*. The sum-rate capacity of the degraded Gaussian CIC has been recently derived in [17] using a rather complicated approach (based on genie-aided techniques). It is not difficult to show that for the Gaussian network, the optimal solution to the optimization (93) is attained by Gaussian distributions.

***Example 4;***

As we see from the expressions (86) and (93) for the scenarios in Examples 2 and 3, to describe the sum-rate capacity there is no need to make use of auxiliary random variables. One may think we have the same consequence for other networks. In fact, this is the case for many network scenarios but not for the general case. Here, we provide an example where it is essential to use auxiliary random variables. Specifically, consider the 2-transmitter/2-receiver interference network shown in Fig. 21.

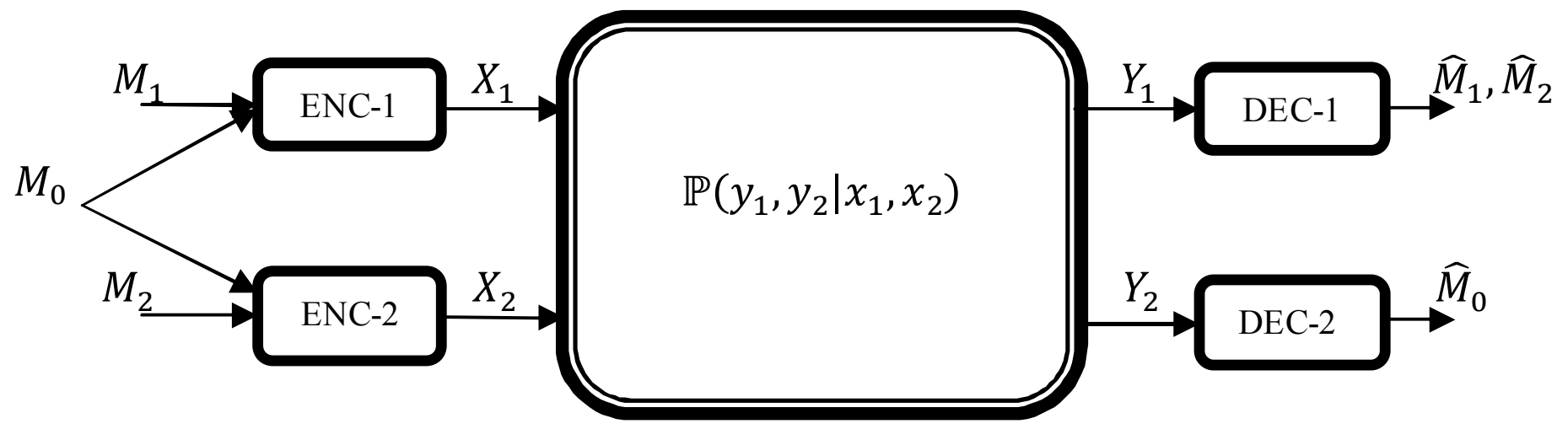

Figure 21. A 2-transmitter/2-receiver interference network.

In this network, two transmitters cooperatively send a message to the second receiver, and also each transmitter sends separately a message to the first receiver. Using (75), one can readily show that for this network $\mathbb{M}^* = \mathbb{M}$. The message and the encoding graphs have been depicted for this scenario in Fig. 22.

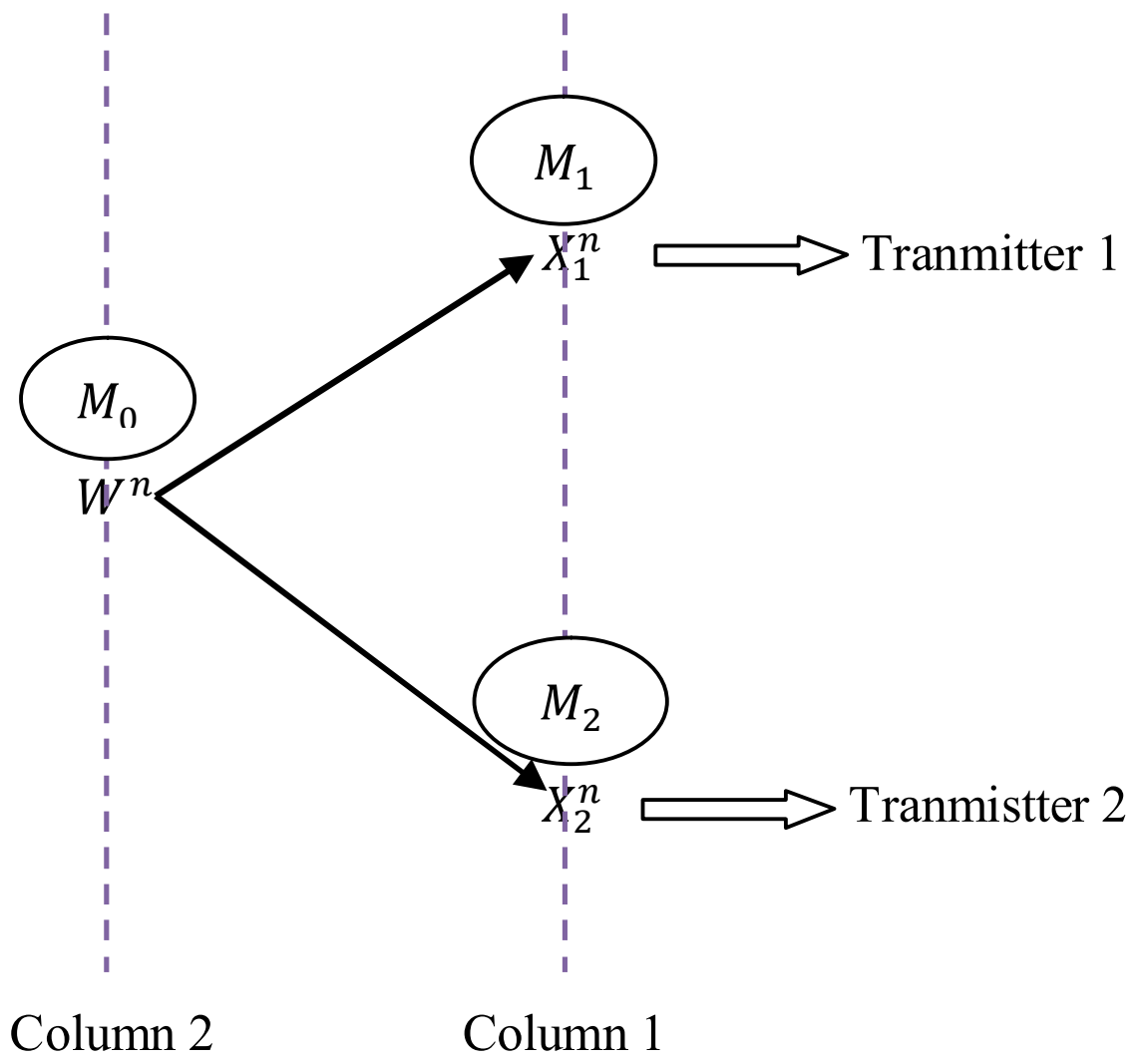

Figure 22. The message and the encoding graph for the interference network of Fig. 21.

Based on Theorem 1, the sum-rate capacity of the degraded channel is given by:

$$\mathcal{C}_{sum}^{EX.4} = \max_{\mathcal{P}_{sum}^{EX.4}} \big(I(M_1, M_2; Y_1|M_0, Q) + I(M_0; Y_2|Q)\big) \tag{94}$$

where $\mathcal{P}_{sum}^{EX.4}$ denotes the set of all joint PDFs as:

$$P_Q P_{M_0} P_{M_1} P_{M_2} P_{X_1|M_0 M_1 Q} P_{X_2|M_0 M_2 Q} \tag{95}$$

Also, by the MG/EG translation, the sum-rate capacity can be re-written as:

$$\mathcal{C}_{sum}^{EX.4} = \max_{P_Q P_{W|Q} P_{X_1|WQ} P_{X_2|WQ}} \big(I(X_1, X_2; Y_1|W, Q) + I(W; Y_2|Q)\big) = \max_{P_W P_{X_1|W} P_{X_2|W}} \big(I(X_1, X_2; Y_1|W) + I(W; Y_2)\big) \tag{96}$$

Note that, in general, none of the choices $W \equiv X_1$, $W \equiv X_2$ or $W \equiv \emptyset$ is optimal for the right side of (96). Therefore, it is unavoidable to use the auxiliary random variable $W$ to describe the sum-rate capacity. Let us now consider the Gaussian channel which is formulated as follows:

$$\begin{cases} Y_1 = X_1 + aX_2 + Z_1 \\ Y_2 = bX_1 + X_2 + Z_2 \end{cases} \tag{97}$$

where $Z_1$ and $Z_2$ are independent Gaussian random variables with zero means and unit variances. The channel is degraded ($Y_2$ is a degraded version of $Y_1$) provided that $ab = 1$ as well as $|a| \geq 1$. In this case, the channel is equivalent to the following:

$$\begin{cases} Y_1 = X_1 + aX_2 + Z_1 \\ \tilde{Y}_2 = bY_1 + \sqrt{1 - b^2}\,\tilde{Z}_2 \end{cases} \tag{98}$$

where $\tilde{Z}_2$ is Gaussian RV with zero mean and unit variance and independent of $Z_1$. In the next proposition, we derive the sum-rate capacity for this network.

***Proposition 5)*** *The sum-rate capacity of the Gaussian interference network (97) in Fig. 21 with* $ab = 1$ *and* $|a| \geq 1$ *is given below:*

$$\mathcal{C}_{sum}^{EX.4\sim G} = \max_{-1\leq\alpha,\beta\leq 1}\left(\psi\big((1-\alpha^2)P_1 + a^2(1-\beta^2)P_2\big) + \psi\left(\frac{b^2\alpha^2P_1 + \beta^2P_2 + 2b\alpha\beta\sqrt{P_1P_2}}{b^2(1-\alpha^2)P_1 + (1-\beta^2)P_2 + 1}\right)\right) \tag{99}$$

*Proof of Proposition 5)* Let $W, \tilde{X}_1, \tilde{X}_2$ be independent Gaussian RVs with zero mean and unit variances. Define:

$$\begin{cases} X_1 \triangleq \alpha\sqrt{P_1}W + \sqrt{(1-\alpha^2)P_1}\tilde{X}_1 \\ X_2 \triangleq \beta\sqrt{P_2}W + \sqrt{(1-\beta^2)P_2}\tilde{X}_2 \end{cases}, \qquad -1 \leq \alpha,\beta \leq 1 \tag{100}$$

By setting $W, X_1$ and $X_2$ in (96), we obtain the achievability. The converse proof is given in Appendix. ■

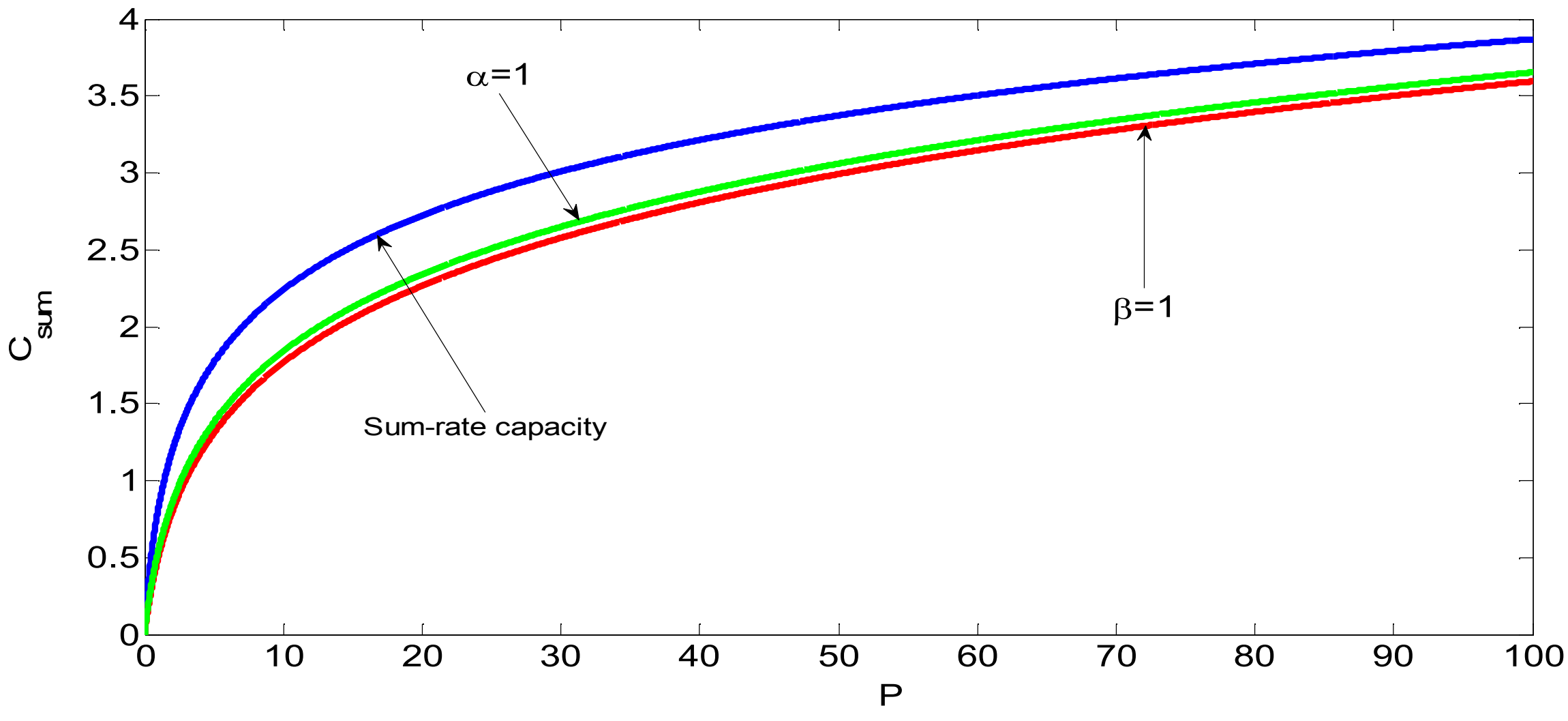

Figure 23. The sum-rate capcaity of the Gaussian degraded interference network (97) of Fig. 21 in terms of $P$, where $P = P_1 = 200P_2$, and $a = {}^1/_b = 15$.

In Fig. 23, we have numerically evaluated the sum-rate capacity for the Gaussian degraded interference network (97) of Fig. 21 in terms of $P$, where $P = P_1 = 200P_2$. Also, we have depicted the expression in the right side of (99) for $\alpha = 1$ as well as for $\beta = 1$. As we see, both the choices are sub-optimal. Therefore, both $W \equiv X_1$ and $W \equiv X_2$ in (96) are suboptimal, verifying that the use of auxiliary random variable is unavoidable to describe the sum-rate capacity.

# IV. GENERAL INTERFERENCE NETWORKS: UNIFIED OUTER BOUNDS

One of the origins of the difficulty in the analysis of the interference networks, specifically networks with more than two receivers, is the lack of non-trivial capacity outer bounds with a satisfactory performance. In this section, we take advantage from our result for degraded networks to establish useful outer bounds for the general non-degraded interference networks. Specifically, in Subsection IV.A a unified outer bound on the sum-rate capacity and in Subsection IV.B a unified outer bound on the entire capacity region of general networks is provided.

## IV.A) Unified Outer Bound on the Sum-Rate Capacity

In this subsection, we directly make use of our general expression in Theorem 1 for the sum-rate capacity of degraded networks to derive a unified outer bound on the sum-rate capacity of the general interference networks in Fig. 14. We then present several classes of non-degraded networks for which the derived outer bound is sum-rate optimal.

***Theorem 2)*** *Consider the $K_1$-Transmitter/$K_2$-Receiver interference network in Fig.14 with the message sets $\mathbb{M}, \mathbb{M}_{X_i}, i = 1, \dots, K_1$, and $\mathbb{M}_{Y_j}, j = 1, \dots, K_2$. The sum-rate capacity is outer-bounded by:*

$$\mathcal{C}_{sum}^{GIN} \le \max_{\overset{*}{\mathcal{P}}_{sum}^{GIN}} \left( I\left(\mathbb{M}_{Y_1}^*; \bar{\bar{Y}}_1 \middle| \mathbb{M}_{Y_2}^*, \dots, \mathbb{M}_{Y_{K_2-1}}^*, \mathbb{M}_{Y_{K_2}}^*, Q\right) + \dots + I\left(\mathbb{M}_{Y_{K_2-1}}^*; \bar{\bar{Y}}_{K_2-1} \middle| \mathbb{M}_{Y_{K_2}}^*, Q\right) + I\left(\mathbb{M}_{Y_{K_2}}^*; \bar{\bar{Y}}_{K_2} \middle| Q\right) \right) \tag{101}$$

*where $\bar{\bar{Y}}_1, \bar{\bar{Y}}_2, \dots, \bar{\bar{Y}}_{K_2}$ are given below:*

$$\bar{\bar{Y}}_j \triangleq \left(Y_j, Y_{j+1}, \dots, Y_{K_2}\right), \qquad j = 1, \dots, K_2 \tag{102}$$

*also, $\overset{*}{\mathcal{P}}_{sum}^{GIN}$ denotes the set of all joint PDFs as:*

$$P_Q \times \prod_{M_\Delta^\nabla \in \mathbb{M}^*} P_{M_\Delta^\nabla} \times \prod_{i \in [1:K_1]} P_{X_i|\mathbb{M}_{X_i}^*, Q} \tag{103}$$

*Moreover, the PDFs $P_{M_\Delta^\nabla}$, $M_\Delta^\nabla \in \mathbb{M}^*$ are uniformly distributed, and $P_{X_i|\mathbb{M}_{X_i}^*,Q} \in \{0,1\}$ for $i = 1, \dots, K_1$, i.e., $X_i$ is a deterministic function of $\left(\mathbb{M}_{X_i}^*, Q\right)$. The messages $\mathbb{M}^*$ are given in (66) and (75); in other words, the messages are determined based on the procedure we developed to achieve the sum-rate capacity of the degraded networks in Section III.*

*Proof of Theorem 2)* The proof is easily understood from the structure of the outer bound as well as the sum-rate capacity of the degraded networks given in Theorem 1. Consider a virtual network with receivers $\bar{\bar{Y}}_1, \bar{\bar{Y}}_2, \dots, \bar{\bar{Y}}_{K_2}$ as given in (102). It is clear that the capacity region of this virtual network contains that of the original network as a subset. Moreover, the virtual network is degraded in the sense of (37). Therefore, according to Theorem 1 its sum-rate capacity is given by (101). The proof is thus complete. ■

***Remark 3:*** Let $\lambda(.)$ be an arbitrary permutation of the elements of the set $\{1, \dots, K_2\}$. By exchanging the indices $1, \dots, K_2$ with $\lambda(1), \dots, \lambda(K_2)$ in (101), respectively, we actually derive other outer bounds on the sum-rate capacity. Although, it is very important to note that the messages $\mathbb{M}^*$ and the corresponding $\mathbb{M}_{X_i}^*, i = 1, \dots, K_1$, and $\mathbb{M}_{Y_j}^*, j = 1, \dots, K_2$, may vary by the permutation $\lambda(.)$; for each permutation $\lambda(.)$, the messages $\mathbb{M}^*$ should be discriminated from the set $\mathbb{M}$ based on the order of the degradedness of the corresponding virtual network.

Let us provide an example. Consider the network of Example 2 shown in Fig. 16 but without the condition of being degraded. Based on Theorem 2, the following relation constitutes an outer bound on the sum-rate capacity:

$$C_{sum}^{IN\to \text{Fig.16}} \leq \max_{P_Q P_{X_1X_2|Q} P_{X_3|Q} P_{X_4|X_1X_2X_3Q}} \left(I(X_4;Y_1,Y_2,Y_3|X_1,X_2,X_3,Q) + I(X_3;Y_2,Y_3|X_1,X_2,Q) + I(X_1,X_2;Y_3|Q)\right) \quad (104)$$

It is clear that the outer bound (104) is tighter than the general cut-set outer bound [10]. In fact, five other bounds such as (104) can be established on the sum-rate capacity.

As indicated by Example 4, to describe the bound (101) for the general interference network, the use of auxiliary random variables is unavoidable; nevertheless, in what follows, we will present a class of networks for which one can derive a representation of the outer bound (101) without any auxiliary random variables.

## IV.A.1) Multiple-Access-Interference Networks (MAIN)

Let us concentrate on network scenarios where some groups of transmitters interested in transmitting information to their respective receiver while causing interference to the other receivers. Clearly, consider an interference network with $K_1 = \sum_{l=1}^{K_2} \mu_l$ transmitters and $K_2$ receivers, where $\mu_1, \dots, \mu_{K_2}$ are arbitrary natural numbers. The transmitters are partitioned into $K_2$ sets labeled $\mathbb{X}_1, \dots, \mathbb{X}_{K_2}$ such that those in $\mathbb{X}_j = \{X_{j,1}, \dots, X_{j,\mu_j}\}$ send the messages $\mathbb{M}_{Y_j}$ to the receiver $Y_j, j = 1, \dots, K_2$, but they have no message for the other receivers; in other words, the message sets $\mathbb{M}_{Y_1}, \dots, \mathbb{M}_{Y_{K_2}}$ are pairwise disjoint. Also, the distribution of messages $\mathbb{M}_{Y_j}$ among the transmitters $\mathbb{X}_j = \{X_{j,1}, \dots, X_{j,\mu_j}\}$ is arbitrary. The network model has been shown in Fig. 24.

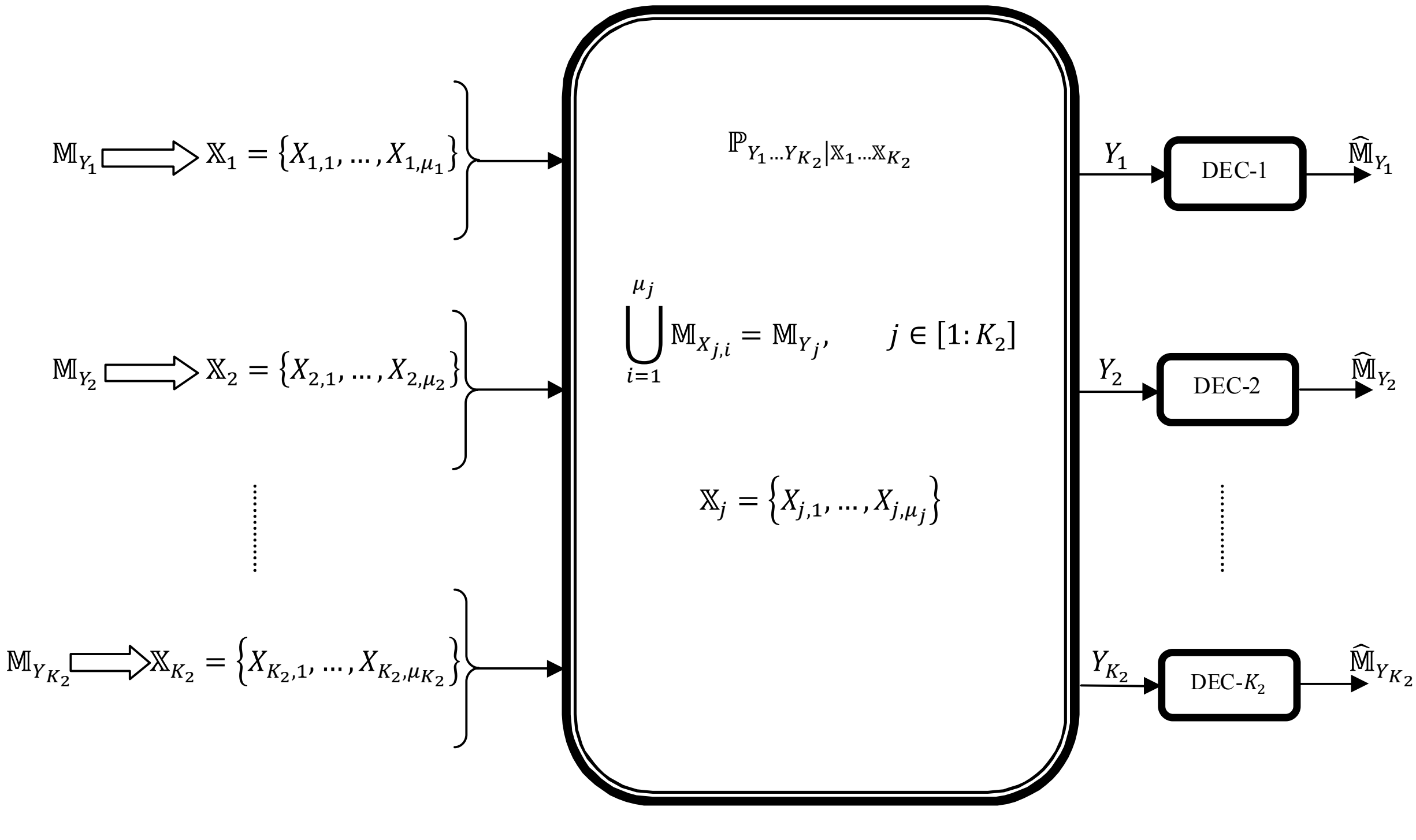

Figure 24. The general Multiple-Access-Interference Networks (MAIN). For $j = 1, \dots, K_2$, $\mathbb{X}_j$ denotes a set of arbitrary transmitters which send (in an arbitrary order) the messages $\mathbb{M}_{Y_j}$ to the receiver $Y_j$.

These scenarios do not contain broadcasting messages to multiple receivers because each transmitter sends information only to a single receiver. In fact, such networks are composed of several interfering MACs; hence, we call them as *Multiple-Access-Interference Networks (MAIN)*. In the following proposition, we prove that for the MAINs, the outer bound (101) can be simply represented by using only the input and output signals.

***Proposition 6)*** *Consider the general MAIN shown in Fig. 24. For this network the sum-rate outer bound in (101) is simplified as follows:*

$$\mathcal{C}_{sum}^{MAIN} \leq \max_{\overset{*}{\mathcal{P}}_{sum}^{MAIN}} \left( I\left(\mathbb{X}_1; \bar{\bar{Y}}_1 \middle| \mathbb{X}_2, \dots, \mathbb{X}_{K_2-1}, \mathbb{X}_{K_2}, Q\right) + \dots + I\left(\mathbb{X}_{K_2-1}; \bar{\bar{Y}}_{K_2-1} \middle| \mathbb{X}_{K_2}, Q\right) + I\left(\mathbb{X}_{K_2}; \bar{\bar{Y}}_{K_2} \middle| Q\right) \right) \tag{105}$$

*where* $\bar{\bar{Y}}_1, \bar{\bar{Y}}_2, \dots, \bar{\bar{Y}}_{K_2}$ *are given in (102), and* $\overset{*}{\mathcal{P}}_{sum}^{MAIN}$ *denotes the set of all joint PDFs which are induced on the input signals* $Q, \mathbb{X}_1, \mathbb{X}_2, \dots, \mathbb{X}_{K_2-1}, \mathbb{X}_{K_2}$, *by the following PDFs:*

$$P_Q \times \prod_{M_\Delta^\nabla \in \mathbb{M}^*} P_{M_\Delta^\nabla} \times \prod_{i \in \left[1:\sum_{l=1}^{K_2} \mu_l\right]} P_{X_i | \mathbb{M}_{X_i}^*, Q} \tag{106}$$

*The messages* $\mathbb{M}^*$ *are a subset of* $\mathbb{M}$ *which are determined using the algorithm we presented in Subsection II.A to derive the sum-rate capacity of MACCM (see (24) and Fig. 10). In fact,* $\mathbb{M}^*$ *consists of those elements of* $\mathbb{M}$ *which are not a satellite for any other message. Moreover, if the network transition probability function implies the following Markov chains:*

$$\mathbb{X}_j \to Y_j, \mathbb{X}_{j+1}, \dots, \mathbb{X}_{K_2} \to Y_{j+1}, Y_{j+2}, \dots, Y_{K_2}, \qquad j = 1, \dots, K_2 - 1 \tag{107}$$

*then, the outer bound is further simplified as:*

$$\mathcal{C}_{sum}^{MAIN} \leq \max_{\overset{*}{\mathcal{P}}_{sum}^{MAIN}} \left( I\left(\mathbb{X}_1; Y_1 \middle| \mathbb{X}_2, \dots, \mathbb{X}_{K_2-1}, \mathbb{X}_{K_2}, Q\right) + \dots + I\left(\mathbb{X}_{K_2-1}; Y_{K_2-1} \middle| \mathbb{X}_{K_2}, Q\right) + I\left(\mathbb{X}_{K_2}; Y_{K_2} \middle| Q\right) \right) \tag{108}$$

*Proof of Proposition 6)* Consider the outer bound (101). For the MAIN scenario in Fig. 24, since the message sets $\mathbb{M}_{Y_1}, \dots, \mathbb{M}_{Y_{K_2}}$ are pairwise disjoint and also each of these sets are sent only to a single receiver, the messages $\mathbb{M}^*$ are those elements of $\mathbb{M}$ which are not a satellite for any other message. Moreover, we have the following joint PDF for the inputs and outputs:

$$P_Q \prod_{M_\Delta^\nabla \in \mathbb{M}^*} P_{M_\Delta^\nabla} \prod_{j=1}^{K_2} P_{\mathbb{X}_j | \mathbb{M}_{Y_j}^*, Q} \mathbb{P}_{Y_1 \dots Y_{K_2} | \mathbb{X}_1 \dots \mathbb{X}_{K_2}} \tag{109}$$

where $P_{\mathbb{X}_j | \mathbb{M}_{Y_j}^*, Q} \in \{0,1\}$. One can readily check that the distribution in (109) implies the following Markov relations:

$$\mathbb{M}_{Y_j}^* \to \mathbb{X}_j, Q \to Y_1, \dots, Y_{K_2}, \qquad j = 1, \dots, K_2 \tag{110}$$

Therefore, for the argument of the maximization in (101), we have:

$$\begin{aligned}
& I\left(\mathbb{M}_{Y_1}^*; \bar{\bar{Y}}_1 \middle| \mathbb{M}_{Y_2}^*, \dots, \mathbb{M}_{Y_{K_2-1}}^*, \mathbb{M}_{Y_{K_2}}^*, Q\right) + \dots + I\left(\mathbb{M}_{Y_{K_2-1}}^*; \bar{\bar{Y}}_{K_2-1} \middle| \mathbb{M}_{Y_{K_2}}^*, Q\right) + I\left(\mathbb{M}_{Y_{K_2}}^*; \bar{\bar{Y}}_{K_2} \middle| Q\right) \\
& \overset{(a)}{=} \begin{pmatrix} I\left(\mathbb{X}_1, \mathbb{M}_{Y_1}^*; \bar{\bar{Y}}_1 \middle| \mathbb{M}_{Y_2}^*, \dots, \mathbb{M}_{Y_{K_2}}^*, \mathbb{X}_2, \dots, \mathbb{X}_{K_2}, Q\right) + I\left(\mathbb{X}_2, \mathbb{M}_{Y_2}^*; \bar{\bar{Y}}_2 \middle| \mathbb{M}_{Y_3}^*, \dots, \mathbb{M}_{Y_{K_2}}^*, \mathbb{X}_3, \dots, \mathbb{X}_{K_2}, Q\right) \\ + \dots + I\left(\mathbb{X}_{K_2-1}, \mathbb{M}_{Y_{K_2-1}}^*; \bar{\bar{Y}}_{K_2-1} \middle| \mathbb{M}_{Y_{K_2}}^*, \mathbb{X}_{K_2}, Q\right) + I\left(\mathbb{X}_{K_2}, \mathbb{M}_{Y_{K_2}}^*; \bar{\bar{Y}}_{K_2} \middle| Q\right) \end{pmatrix} \\
& \overset{(b)}{=} I\left(\mathbb{X}_1; \bar{\bar{Y}}_1 \middle| \mathbb{X}_2, \dots, \mathbb{X}_{K_2-1}, \mathbb{X}_{K_2}, Q\right) + \dots + I\left(\mathbb{X}_{K_2-1}; \bar{\bar{Y}}_{K_2-1} \middle| \mathbb{X}_{K_2}, Q\right) + I\left(\mathbb{X}_{K_2}; \bar{\bar{Y}}_{K_2} \middle| Q\right)
\end{aligned} \tag{111}$$

where equality (a) holds because the inputs $\mathbb{X}_j$ are given by deterministic functions of $\left(\mathbb{M}_{Y_j}^*, Q\right)$, and the equality (b) is due to (109)-(110). Also, one can readily check that if the Markov relations (107) hold, then each mutual information function in (111) is reduced to the corresponding one in (108). ■

***Remarks 4:***

1. Note that for the MAIN in Fig. 24, all the auxiliary random variables (the messages) are removed from the outer bound (105). Nevertheless, the joint PDF which is imposed on the input signals should be carefully determined based on arrangement of the messages $\mathbb{M}^*$ among the transmitters. The MACCM plan of messages is very helpful for this propose. Compare (25) and (26) for an example.
2. It is clear that for the MAINs which satisfy the degraded condition (37), the sum-rate capacity is given by (108).

We now intend to provide several classes of networks, other than the degraded ones, for which the outer bound of Theorem 2 is sum-rate optimal. Specifically, we will introduce two new interference networks coined as "*Generalized Z-Interference Networks*" and "*Many-to-One Interference Networks*". We identify noisy interference regimes for these networks. Also, for the first time, interesting networks are introduced for which a combination of successive decoding and treating interference as noise is sum-rate optimal.

## IV.A.2) Generalized Z-Interference Networks

Consider a special case of the MAINs in Fig. 24 for which the transition probability function of the network is factorized as follows:

$$\mathbb{P}_{Y_1\ldots Y_{K_2}|\mathbb{X}_1\ldots\mathbb{X}_{K_2}} = \mathbb{P}_{Y_1|\mathbb{X}_1}\mathbb{P}_{Y_2|\mathbb{X}_1\mathbb{X}_2}\mathbb{P}_{Y_3|\mathbb{X}_1\mathbb{X}_2\mathbb{X}_3}\cdots\mathbb{P}_{Y_{K_2}|\mathbb{X}_1\mathbb{X}_2\mathbb{X}_3\ldots\mathbb{X}_{K_2}} \tag{112}$$

The network model is shown in Fig. 25. In this figure, each receiver has been linked to its connected transmitters (see Definition II.9 of Part I [1]) by a dashed line.

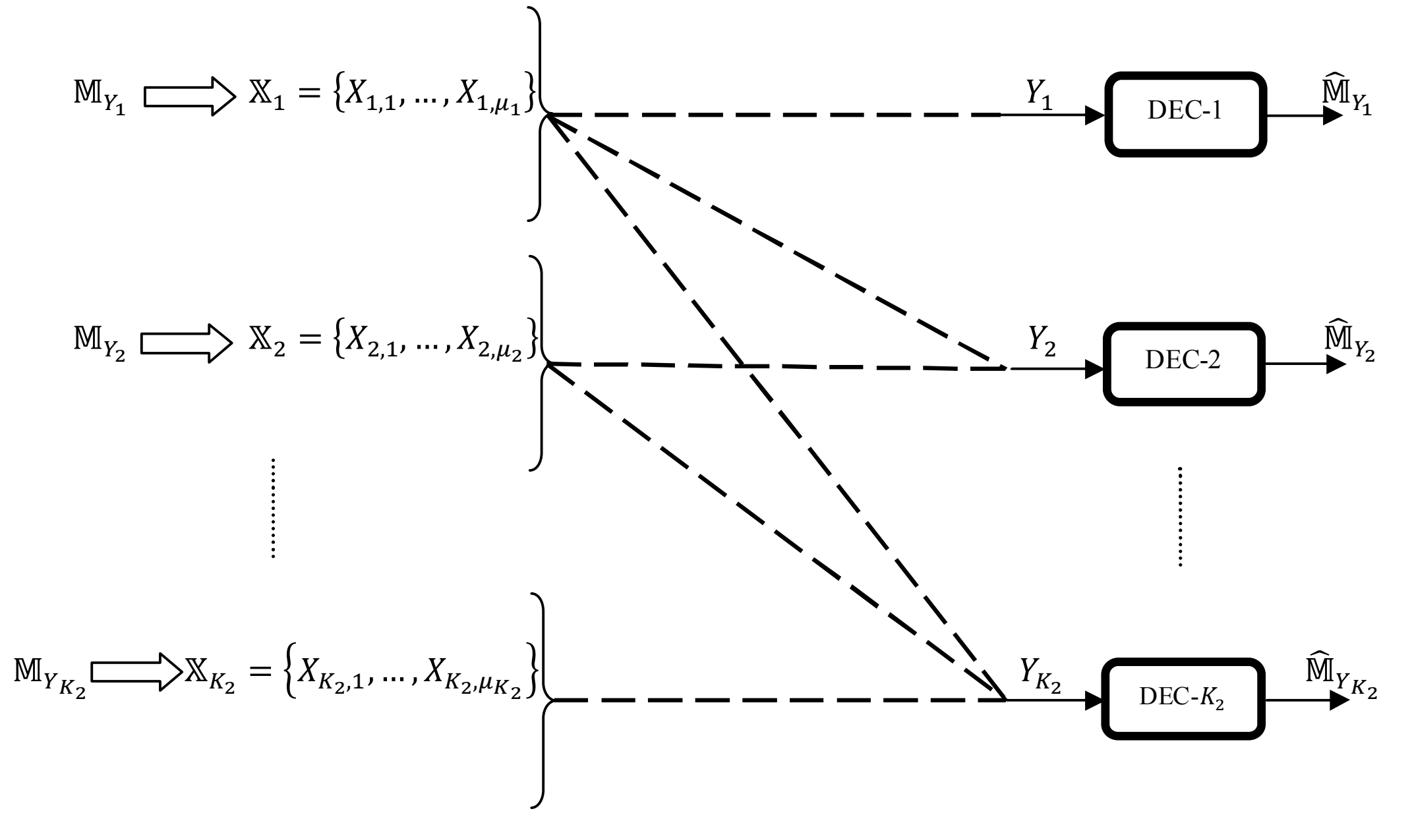

Figure 25. The Generalized Z-Interference Network. For $j = 1, \ldots, K_2$, $\mathbb{X}_j$ denotes a set of arbitrary transmitters which send (in an arbitrary order) the messages $\mathbb{M}_{Y_j}$ to the receiver $Y_j$.

Note that such networks can be considered as a natural generalization of the Z-interference channel in Part I [1, Eq. III~31]; hence, the name of "Generalized Z-Interference Network". Now we derive a sum rate capacity result for these networks. Specifically, let the transition probability function of the network satisfies the following degradedness condition as well:

$$\mathbb{P}_{Y_1\ldots Y_{K_2}|\mathbb{X}_1\ldots\mathbb{X}_{K_2}} = \mathbb{P}_{Y_1|\mathbb{X}_1}\mathbb{P}_{Y_2|Y_1\mathbb{X}_2}\mathbb{P}_{Y_3|Y_2\mathbb{X}_3}\cdots\mathbb{P}_{Y_{K_2}|Y_{K_2-1}\mathbb{X}_{K_2}} \tag{113}$$

Note that the degradedness condition in (113) does not imply (37), and the networks in (112)-(113) in general differ from those in (37). In the next theorem, we will prove that for the generalized Z-interference networks with (113), treating interference as noise achieves the sum-rate capacity.

***Theorem 3)*** *Consider the generalized Z-interference network in (112) with the degraded condition (113). The sum-rate capacity of the network is given by:*

$$\mathcal{C}_{sum}^{Z-IN_{deg}} = \max_{\overset{*}{\mathcal{P}}_{sum}^{MAIN}} \Big( I(\mathbb{X}_1;Y_1|Q) + I(\mathbb{X}_2;Y_2|Q) + \cdots + I\big(\mathbb{X}_{K_2-1};Y_{K_2-1}\big|Q\big) + I\big(\mathbb{X}_{K_2};Y_{K_2}\big|Q\big)\Big) \tag{114}$$

*where* $\overset{*}{\mathcal{P}}_{sum}^{MAIN}$ *is given as in Proposition 6.*

*Proof of Theorem 3)* The achievability is derived using a simple treating interference as noise strategy: The messages $\mathbb{M}-\mathbb{M}^*$ are withdrawn from the transmission scheme where $\mathbb{M}^*$ are those elements of $\mathbb{M}$ which are not a satellite for any other message. The messages $\mathbb{M}^*$ are encoded based on the MACCM plan. The receiver $Y_j, j=1,\dots,K_2$, jointly decodes its respective messages $\mathbb{M}_{Y_j}^*$ and treats all the other signals as noise; therefore, based on Proposition 2 the following rate is achievable:

$$\max_{\{\text{PDFs in (106)}\}} \left( I\big(\mathbb{M}_{Y_1}^*;Y_1\big|Q\big) + \cdots + I\left(\mathbb{M}_{Y_{K_2-1}}^*;Y_{K_2-1}\middle|Q\right) + I\left(\mathbb{M}_{Y_{K_2}}^*;Y_{K_2}\middle|Q\right)\right) \tag{115}$$

Similar to the procedure in (111), one can simply show that (115) is equivalent to (114). For the converse part, note that the degradedness condition in (113) implies the Markov relations (107). Therefore, (108) constitutes an outer bound on the sum-rate capacity of the network. Furthermore, for the generalized Z-interference network (112) shown in Fig. 25, the expression (108) is readily reduced to (114). The proof is thus complete. ■

***Remark 4:*** The result of Theorem 3 actually holds for all generalized Z-interference networks (112) where the marginal distributions of their transition probability function are given similar to the networks (113).

### IV.A.3) Many-to-One Interference Networks

We now discuss a special case of the generalized Z-interference networks in (112). Specifically, consider the case where the transition probability function is factorized as follows:

$$\mathbb{P}_{Y_1\dots Y_{K_2}|\mathbb{X}_1\dots\mathbb{X}_{K_2}} = \mathbb{P}_{Y_1|\mathbb{X}_1}\mathbb{P}_{Y_2|\mathbb{X}_2}\mathbb{P}_{Y_3|\mathbb{X}_3}\dots\mathbb{P}_{Y_{K_2-1}|\mathbb{X}_{K_2-1}}\mathbb{P}_{Y_{K_2}|\mathbb{X}_1\mathbb{X}_2\mathbb{X}_3\dots\mathbb{X}_{K_2}} \tag{116}$$

In other words, only one receiver experiences interference. Fig. 26 depicts the network model where each receiver has been linked to its connected transmitters by a dashed line.

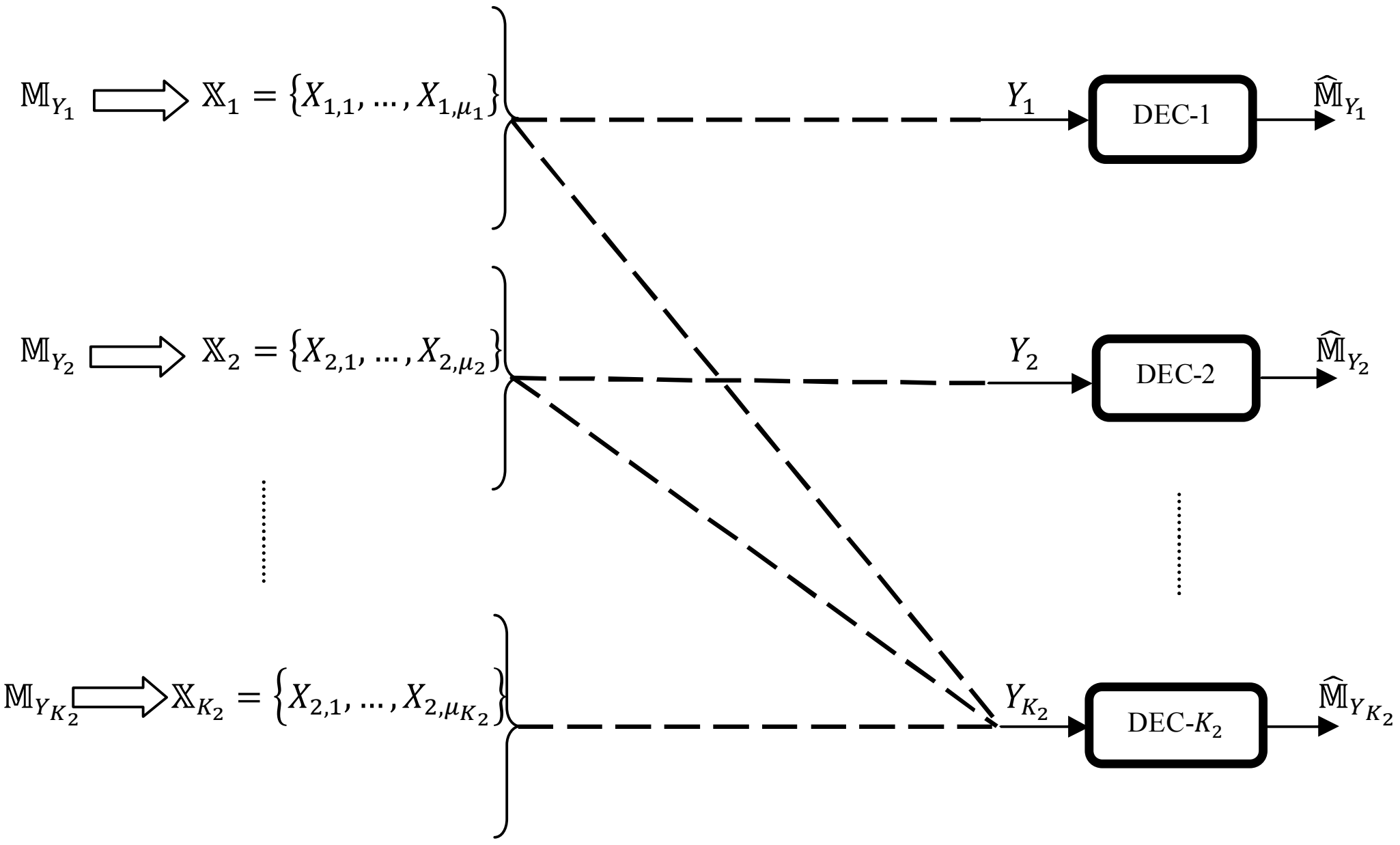

Figure 26. The Many-to-One Interference Network. For $i = 1, \ldots, K_2$, $\mathbb{X}_i$ denotes a set of arbitrary transmitters which send (in an arbitrary order) the messages $\mathbb{M}_{Y_j}$ to the receiver $Y_j$.

Note that the arrangement of messages among transmitters and receivers are similar to the generalized Z-interference networks in Fig. 25. These networks are a natural generalization of the so-called *Many-to-One Interference Channel* studied in [18] and [19] for the multi-message case. Hence, they are coined as the "Many-to-One Interference Networks". In these networks many *groups* of transmitters send information to their respective receivers (each group is concerned to one receiver) via a common media such that all receivers, except one, receive interference-free signals. In other words, only one of the receivers experiences interference, i.e., its received signal contains information regarding both desired and non-desired messages. For the Gaussian Many-to-One CIC, a noisy interference regime was identified in [20] using the outer bounds derived based on the genie-aided techniques. Here, we identify such a regime for the many-to-one interference networks in Fig. 26, for both discrete and Gaussian networks. Specifically, consider a network with the following constraint:

$$\mathbb{P}_{Y_1 \ldots Y_{K_2}|\mathbb{X}_1 \ldots \mathbb{X}_{K_2}} = \mathbb{P}_{Y_1|\mathbb{X}_1}\mathbb{P}_{Y_2|\mathbb{X}_2}\mathbb{P}_{Y_3|\mathbb{X}_3} \ldots \mathbb{P}_{Y_{K_2-1}|\mathbb{X}_{K_2-1}}\mathbb{P}_{Y_{K_2}|Y_1 \ldots Y_{K_2-1}\mathbb{X}_{K_2}} \tag{117}$$

For the networks in (117), conditioned on $\mathbb{X}_{K_2}$, the output $Y_{K_2}$ is a noisy (degraded) version of $Y_1, \ldots, Y_{K_2-1}$. In the next theorem, we prove that for such networks, treating interference as noise is sum-rate optimal.

***Theorem 4)*** *Consider the many-to-one interference network in Fig. 26. If the transition probability function satisfies the degradedness condition in (117), then the sum-rate capacity is given below:*

$$\mathcal{C}_{sum}^{MtO-IN_{deg}} = \max_{\overset{*}{\mathcal{P}}_{sum}^{MAIN}} \left(I(\mathbb{X}_1; Y_1|Q) + I(\mathbb{X}_2; Y_2|Q) + \cdots + I(\mathbb{X}_{K_2-1}; Y_{K_2-1}|Q) + I(\mathbb{X}_{K_2}; Y_{K_2}|Q)\right) \tag{118}$$

*where $\overset{*}{\mathcal{P}}_{sum}^{MAIN}$ is given as in Proposition 6.*

*Proof of Theorem 4)* To achieve the bound (118), similar to the scenario of Theorem 3, each receiver decodes its own messages and treats all the other signals as noise. For the converse part, note that the degradedness condition (117) implies the Markov relations given in (107). Therefore, (108) constitutes an outer bound on the sum-rate capacity of the network. The proof is completed by the fact that for the many-to-one interference network (116) shown in Fig. 26, the expression (108) is reduced to (118). ■

***Remark 5:*** It should be remarked that the degradedness condition (117) is in general different from (113), and the result of Theorem 4 cannot be deduced from Theorem 3. In fact, for all generalized Z-interference networks (112), where their transition probability function implies the Markov chains (107), treating interference as noise is sum-rate optimal.

Let us consider the special case of many-to-one CIC which is derived from the network in Fig. 26 by setting $K_1 = K_2 = K$, i.e., with respect to each receiver, there exists only one transmitter. The Gaussian channel is formulated as follows:

$$\begin{cases} Y_i = X_i + Z_i, \qquad i = 1, \dots, K-1 \\ Y_K = a_1X_1 + a_2X_2 + \dots + a_KX_K + Z_K \end{cases} \tag{119}$$

where $Z_1, \dots, Z_K$ are independent Gaussian noise with zero means and unit variances. Also, the input signals are subject to power constraints: $\mathbb{E}[X_i^2] \leq P_i, i = 1, \dots, K$. In [20], the authors showed that if the channel gains satisfy:

$$\sum_{i=1}^{K-1} a_i^2 \leq 1 \tag{120}$$

then, treating interference as noise is sum-rate optimal. One can show that the channel with (120) is (stochastically) degraded in the sense of (117). Therefore, its sum-rate capacity is also derived from Theorem 4.

### IV.A.4) Incorporation of Successive Decoding and Treating Interference as Noise

In Section III, we proved that for all degraded interference networks (37), the successive decoding scheme is sum-rate optimal. Also, in previous subsections, we introduced classes of interference networks for which treating interference as noise at the receivers achieves the sum-rate capacity. Now we intend to present interesting scenarios for which a combination of these two schemes is optimal. We identify classes of interference networks which are composed of two sets of receivers; to achieve the sum-rate capacity for these networks the receivers of one set apply the successive decoding strategy while the receivers of the other set treat interference as noise and decode only their own messages.

Consider a class of MAINs shown in Fig. 24 with $K_2 = \eta_1 + \eta_2$, where $\eta_1$ and $\eta_2$ are two arbitrary natural numbers. Let the transition probability function of the network satisfies the following factorization:

$$\begin{aligned} \mathbb{P}_{Y_1 \dots Y_{\eta_1} Y_{\eta_1+1} \dots Y_{\eta_1+\eta_2} | \mathbb{X}_1 \dots \mathbb{X}_{\eta_1} \mathbb{X}_{\eta_1+1} \dots \mathbb{X}_{\eta_1+\eta_2}} &= \mathbb{P}_{Y_1|\mathbb{X}_1} \mathbb{P}_{Y_2|\mathbb{X}_2} \dots \mathbb{P}_{Y_{\eta_1-1}|\mathbb{X}_{\eta_1-1}} \mathbb{P}_{Y_{\eta_1}|Y_1 \dots Y_{\eta_1-1}\mathbb{X}_{\eta_1}} \\ &\times \mathbb{P}_{Y_{\eta_1+1}|Y_{\eta_1}\mathbb{X}_{\eta_1+1}\dots\mathbb{X}_{\eta_1+\eta_2}} \mathbb{P}_{Y_{\eta_1+2}|Y_{\eta_1+1}} \dots \mathbb{P}_{Y_{\eta_1+\eta_2-1}|Y_{\eta_1+\eta_2-2}} \mathbb{P}_{Y_{\eta_1+\eta_2}|Y_{\eta_1+\eta_2-1}} \end{aligned} \tag{121}$$

Note that the networks in (121) is neither degraded in the sense of (37) nor satisfies the Z-networks' factorization (112). Therefore, they belong to none of the scenarios considered before. In fact, such networks can be considered as a combination of the degraded networks in (37) and the many-to-one networks in (117). In the next theorem, we establish the sum-rate capacity for these networks where we show that treating interference as noise at the receivers $Y_1, \dots, Y_{\eta_1}$ and the successive decoding scheme at the receivers $Y_{\eta_1+1}, \dots, Y_{\eta_1+\eta_2}$ is optimal.

***Theorem 5)*** *For the MAIN shown in Fig. 24 with $K_2 = \eta_1 + \eta_2$ if the transition probability function of the network satisfies the degradedness condition in* (121)*, the sum-rate capacity is given by:*

$$\max_{\overset{*}{\mathcal{P}}_{sum}^{MAIN}} \begin{pmatrix} I(\mathbb{X}_1; Y_1|Q) + I(\mathbb{X}_2; Y_2|Q) + \dots + I(\mathbb{X}_{\eta_1-1}; Y_{\eta_1-1}|Q) + I(\mathbb{X}_{\eta_1}; Y_{\eta_1}|Q) \\ +I(\mathbb{X}_{\eta_1+1}; Y_{\eta_1+1}|\mathbb{X}_{\eta_1+2}, \dots, \mathbb{X}_{\eta_1+\eta_2-1}, \mathbb{X}_{\eta_1+\eta_2}, Q) + \dots + I(\mathbb{X}_{\eta_1+\eta_2-1}; Y_{\eta_1+\eta_2-1}|\mathbb{X}_{\eta_1+\eta_2}, Q) + I(\mathbb{X}_{\eta_1+\eta_2}; Y_{\eta_1+\eta_2}|Q) \end{pmatrix} \tag{122}$$

*where $\overset{*}{\mathcal{P}}_{sum}^{MAIN}$ is given as in Proposition 6.*

*Proof of Theorem 5)* The achievability is derived using a combination of treating interference as noise and successive decoding strategy: The messages $\mathbb{M} - \mathbb{M}^*$ are not sent, where $\mathbb{M}^*$ are those elements of $\mathbb{M}$ which are not a satellite for any other message. The messages $\mathbb{M}^*$ are encoded based on the MACCM plan. The receiver $Y_j, j = 1, \dots, \eta_1$ jointly decodes its respective messages $\mathbb{M}_{Y_j}^*$ and treats all the other signals as noise. The decoding scheme at the receivers $Y_{\eta_1+1}, \dots, Y_{\eta_1+\eta_2}$ is the successive decoding scheme devised for the degraded networks given in (37). The receiver $Y_{\eta_1+\eta_2}$ decodes its respective messages $\mathbb{M}_{Y_{\eta_1+\eta_2}}^*$ using the joint decoding technique. The partial sum-rate due to this step is

$$I\left(\mathbb{M}_{Y_{\eta_1+\eta_2}}^*; Y_{\eta_1+\eta_2}\middle|Q\right)$$

At the receiver $Y_{\eta_1+\eta_2-1}$, first the messages $\mathbb{M}_{Y_{\eta_1+\eta_2}}^*$ are jointly decoded; this step does not introduce any new rate cost because $Y_{\eta_1+\eta_2}$ is a degraded version of $Y_{\eta_1+\eta_2-1}$. Then, it jointly decodes its respective messages $\mathbb{M}_{Y_{\eta_1+\eta_2-1}}^*$ using the received sequence as well as the previously decoded codewords. We know that there is no cloud center in $\mathbb{M}^* - \mathbb{M}_{Y_{\eta_1+\eta_2}}^*$ for the messages $\mathbb{M}_{Y_{\eta_1+\eta_2-1}}^*$ ; also, their cloud centers which possibly exist in $\mathbb{M}_{Y_{\eta_1+\eta_2}}^*$ have been already decoded. Thereby, successful decoding with the following partial sum-rate is attainable:

$$I\left(\mathbb{M}_{Y_{\eta_1+\eta_2-1}}^*; Y_{\eta_1+\eta_2-1}\middle|\mathbb{M}_{Y_{\eta_1+\eta_2}}^*, Q\right)$$

This successive decoding strategy is repeated at other receivers step by step from the weaker receivers towards the stronger ones up to the receiver $Y_{\eta_1+1}$. This receiver also first decodes the messages $\mathbb{M}_{Y_{\eta_1+\eta_2}}^*, \mathbb{M}_{Y_{\eta_1+\eta_2-1}}^*, \dots, \mathbb{M}_{Y_{\eta_1+2}}^*$ successively and then decodes its own messages $\mathbb{M}_{Y_{\eta_1+1}}^*$. The rate cost due to this step is

$$I\left(\mathbb{M}_{Y_{\eta_1+1}}^*; Y_{\eta_1+1}\middle|\mathbb{M}_{Y_{\eta_1+2}}^*, \dots, \mathbb{M}_{Y_{\eta_1+\eta_2-1}}^*, \mathbb{M}_{Y_{\eta_1+\eta_2}}^*, Q\right)$$

Therefore, by this scheme the following sum-rate is achieved:

$$\max_{\{\text{PDFs in (106)}\}} \begin{pmatrix} I(\mathbb{M}_{Y_1}^*; Y_1|Q) + I(\mathbb{M}_{Y_2}^*; Y_2|Q) + \cdots + I\left(\mathbb{M}_{Y_{\eta_1-1}}^*; Y_{\eta_1-1}\middle|Q\right) + I\left(\mathbb{M}_{Y_{\eta_1}}^*; Y_{\eta_1}\middle|Q\right) \\ I\left(\mathbb{M}_{Y_{\eta_1+1}}^*; Y_{\eta_1+1}\middle|\mathbb{M}_{Y_{\eta_1+2}}^*, \dots, \mathbb{M}_{Y_{\eta_1+\eta_2-1}}^*, \mathbb{M}_{Y_{\eta_1+\eta_2}}^*, Q\right) + \cdots + I\left(\mathbb{M}_{Y_{\eta_1+\eta_2-1}}^*; Y_{\eta_1+\eta_2-1}\middle|\mathbb{M}_{Y_{\eta_1+\eta_2}}^*, Q\right) + I\left(\mathbb{M}_{Y_{\eta_1+\eta_2}}^*; Y_{\eta_1+\eta_2}\middle|Q\right) \end{pmatrix} \tag{123}$$

Similar to the procedure in (111), one can show that (123) is equivalent to (122). For the converse part, note that the degradedness condition in (121) implies the Markov relations (107). Therefore, (108) constitutes an outer bound on the sum-rate capacity. Moreover, for the factorization (121), the expression in (108) is reduced to (122). The proof is thus complete. ■

In fact, many other scenarios could be identified for which the outer bound in (101) is sum-rate optimal. Essentially, this outer bound is optimal for the degraded networks in (37) for which the successive decoding strategy achieves the sum-rate capacity. The degradedness condition in (37) implies that a weaker receiver, given a stronger receiver, is statistically independent of all the input signals. Nonetheless, for certain networks with receivers statistically independent of some of the transmitters such as those introduced in this subsection, it is always possible to somewhat relax the crucial degradedness constraint (37) and derive situations where the outer bound (101) is still sum-rate optimal.

### IV.A.5) Discussion

Let us once more revisit the capacity region of the MACCM discussed in Subsection II.A. Consider the procedure described in (2) and (3) where we proved the converse theorem. In this procedure, we derived four constraints on the communication rates and their linear combinations. These constraints were actually sufficient to prove the capacity region. Nevertheless, by following exactly the same lines as (2), one can derive the constraints below as well:

$$\begin{aligned} R_0 &\leq I(M_0; Y_1|M_1, M_2, Q) \\ R_0 + R_1 &\leq I(M_0, M_1; Y_1|M_2, Q) \\ R_0 + R_2 &\leq I(M_0, M_2; Y_1|M_1, Q) \end{aligned} \tag{124}$$

Adding these constraints to (4), we derive the following outer bound on the capacity region;

$$\bigcup_{\substack{P_Q P_{M_0} P_{M_1} P_{M_2} \\ X_i = f_i(M_0, M_i, Q), i=1,2}} \left\{ \begin{array}{l} (R_0, R_1, R_2) \in \mathbb{R}_+^3 : \\ R_0 \leq I(M_0; Y|M_1, M_2, Q) \\ R_1 \leq I(M_1; Y|M_0, M_2, Q) \\ R_2 \leq I(M_2; Y|M_0, M_1, Q) \\ R_0 + R_1 \leq I(M_0, M_1; Y|M_2, Q) \\ R_0 + R_2 \leq I(M_0, M_2; Y|M_1, Q) \\ R_1 + R_2 \leq I(M_1, M_2; Y|M_0, Q) \\ R_0 + R_1 + R_2 \leq I(M_0, M_1, M_2; Y|Q) \end{array} \right\} \tag{125}$$

It is clear that the outer bound (125) is a subset of (4). On the one side, since (4) is the capacity region, (125) also coincides with the capacity region. This new description of the capacity region for the MACCM, which is due to Han [21], involves a more complicated characterization; however, it represents a very simple achievability scheme. It states that to achieve the capacity region of the MACCM, it is sufficient to encode the messages separately using independent codewords (without any binning or superposition coding); in other words, the message $M_i$ is encoded using a codeword $M_i^n$ generated based on $P_{M_i}(m_i), i = 0,1,2$. A time-sharing codeword $Q^n$ is also generated independently based on $P_Q(q)$ which is revealed to all parties. The transmitter $X_i, i = 1,2$ then generates its codeword $X_i^n$ as $X_i^n = f_i(M_0^n, M_i^n, Q^n)$[4], where $f_i(.)$ is an arbitrary deterministic function, and sends it over the channel. At the receiver side, all the codewords $M_0^n, M_1^n, M_2^n$ are jointly decoded. A simple analysis of this scheme leads to the rate region (125). Inspired by such an achievability scheme for the MACCM, we can present a rather simple scheme to achieve the sum-rate capacity of the degraded interference networks. Consider once more the outer bound we derived in Lemma 2 for the degraded networks. We now provide a simple scheme to directly achieve this bound. Similar to the MACCM, all the messages are encoded separately using independent codewords. Typically, the message $M_\Delta^\nabla \in \mathbb{M}$ is encoded using a codeword $(M_\Delta^\nabla)^n$ generated based on $P_{M_\Delta^\nabla}(m_\Delta^\nabla)$. Also, a time-sharing codeword $Q^n$ is generated independently based on $P_Q(q)$ and is revealed to all parties. Then, each transmitter $X_i \in \{X_1, \ldots, X_{K_1}\}$ generates its codeword $X_i^n$ as $X_i^n = f_i(\{(M_\Delta^\nabla)^n : M_\Delta^\nabla \in \mathbb{M}_{X_i}\}, Q^n)$, where $f_i(.)$ is an arbitrary deterministic function, and transmits it. At the receivers the respective codewords are decoded, successively. At the weakest receiver $Y_{K_2}$, the codewords $\{(M_\Delta^\nabla)^n : M_\Delta^\nabla \in \mathbb{M}_{Y_{K_2}}\}$ are successively decoded (in an arbitrary order, (see Proposition 2)). At the receiver $Y_{K_2-1}$, the codewords $\{(M_\Delta^\nabla)^n : M_\Delta^\nabla \in \mathbb{M}_{Y_{K_2}}\}$ are successively decoded first similar to the receiver $Y_{K_2}$; this step does not include any new rate cost because $Y_{K_2}$ is a degraded version of $Y_{K_2-1}$. Then, the codewords $\{(M_\Delta^\nabla)^n : M_\Delta^\nabla \in \mathbb{M}_{Y_{K_2-1}} - \mathbb{M}_{Y_{K_2}}\}$ are decoded. This process is followed at the other receivers step by step from the weaker receivers towards the stronger ones. One can easily check that by this scheme the sum-rate (40) is achieved.

It should be noted that the achievability scheme described above, however, is simpler than the one presented in Subsection III based on superposition coding, but instead that scheme results to a substantially simpler characterization of the sum-rate capacity. In fact, by using the algorithm given in Subsection III, many of the messages (actually auxiliary random variables) are removed from the general expression in (40). Accordingly, the maximization problem which is contained in the characterization of the sum-rate capacity is considerably resolved. The simplicity of our characterization in Theorem 1 is crucial when we consider the Gaussian networks because the complexity in the explicit characterization of the sum-rate capacity for these networks grows very rapidly with the number of messages. This simplicity is also very helpful when considering the outer bound derived in Theorem 2 for the sum-rate capacity of general interference networks. For example, by our approach we derived the outer bound (104) for the large multi-message network in Fig. 16; this bound does not contain any auxiliary random variable and hence it is easily computable. Another significant point regarding our scheme described in Section III is that it provides us a deep insight regarding the nature of information flow in degraded networks because it explicitly determines that which messages are actually required to be transmitted to achieve the sum-rate capacity for these networks.

## IV.B) Unified Outer Bound on the Capacity Region

In this subsection, we derive unified outer bounds on the entire capacity region of the general interference networks in Fig. 14. These outer bounds, which are derived by taking insight from our results for the degraded networks, are given in the following theorems. To present our results, we need to some new notations.

---

[4]Here, the notation $X_i^n = f_i(M_0^n, M_i^n, Q^n)$ actually indicates $X_{i,t} = f_i(M_{0,t}, M_{i,t}, Q_t), t = 1, \ldots, n$.

***Definition 7:*** *Let $\{R_1,\ldots,R_K\}$ be a K-tuple of non-negative real numbers, where K is a natural number. Let also $\mathbb{M}=\{M_1,\ldots,M_K\}$ be a set of K indexed elements. Assume that $\Omega$ is an arbitrary subset of $\mathbb{M}$. The partial-sum $\boldsymbol{R}_{\sum\Omega}$ with respect to $\Omega$ is defined as follows:*

$$\boldsymbol{R}_{\sum\Omega}\triangleq\sum_{l\in\underline{id}_\Omega}R_l \tag{126}$$

Note that the identification of the set $\Omega$, i.e., $\underline{id}_\Omega$, was defined in Part I [1, Def. II.1].

***Theorem 6)*** *Consider the general interference network in Fig. 14. Define $\mathfrak{R}_{o:(1)}^{GIN}$ as follows:*

$$\mathfrak{R}_{o:(1)}^{GIN}\triangleq\bigcup_{\mathcal{P}_{o:(1)}^{GIN}}\left\{\begin{array}{l}(R_1,R_2,\ldots,R_K)\in\mathbb{R}_+^K:\\ \forall\,\Omega\in\mathbb{M},\qquad\forall\,\lambda(.):\{1,\ldots,K_2\}\to\{1,\ldots,K_2\}\\ \overline{\overline{Y}}_{\lambda(j)}\triangleq\left(Y_{\lambda(j)},Y_{\lambda(j+1)},\ldots,Y_{\lambda(K_2)}\right),\qquad j=1,\ldots,K_2\\ \boldsymbol{R}_{\sum\Omega}\le\left(\begin{array}{c}I\left(\Omega\cap\mathbb{M}_{Y_{\lambda(1)}};\overline{\overline{Y}}_{\lambda(1)}\middle|\Omega\cap\mathbb{M}_{Y_{\lambda(2)}},\ldots,\Omega\cap\mathbb{M}_{Y_{\lambda(K_2-1)}},\Omega\cap\mathbb{M}_{Y_{\lambda(K_2)}},\mathbb{M}-\Omega,Q\right)\\ +\cdots+I\left(\Omega\cap\mathbb{M}_{Y_{\lambda(K_2-1)}};\overline{\overline{Y}}_{\lambda(K_2-1)}\middle|\Omega\cap\mathbb{M}_{Y_{\lambda(K_2)}},\mathbb{M}-\Omega,Q\right)+I\left(\Omega\cap\mathbb{M}_{Y_{\lambda(K_2)}};\overline{\overline{Y}}_{\lambda(K_2)}\middle|\mathbb{M}-\Omega,Q\right)\end{array}\right)\end{array}\right\} \tag{127}$$

*where $\mathcal{P}_{o:(1)}^{GIN}$ denotes the set of all joint PDFs $P_{QM_1\ldots M_KX_1\ldots X_{K_1}}(q,m_1,\ldots,m_K,x_1,\ldots,x_{K_1})$ satisfying:*

$$P_{QM_1\ldots M_KX_1\ldots X_{K_1}}=P_Q\times P_{M_1}\times\ldots\times P_{M_K}\times P_{X_1|\mathbb{M}_{X_1},Q}\times\ldots\times P_{X_{K_1}|\mathbb{M}_{X_{K_1}},Q} \tag{128}$$

*Also, the PDFs $P_{M_l}, l=1,\ldots,K$, are uniformly distributed, and $P_{X_i|\mathbb{M}_{X_i},Q}\in\{0,1\}$ for $i=1,\ldots,K_1$. The set $\mathfrak{R}_{o:(1)}^{GIN}$ constitutes an outer bound on the capacity region.*

***Remarks 6:***

1. The functions $\lambda(.)$ in (127) are permutations of the elements of the set $\{1,\ldots,K_2\}$.
2. The auxiliary random variables in the outer bound $\mathfrak{R}_{o:(1)}^{GIN}$ in (127) are the messages $M_1,\ldots,M_K$ and the time-sharing parameter $Q$.
3. As the capacity region of the interference networks depends on marginal distributions of their transition probability function, a tighter outer bound is derived by the intersection of rate regions as (127) over all conditional PDFs $\widetilde{\mathbb{P}}_{Y_1\ldots Y_{K_2}|X_1\ldots X_{K_1}}$ which have the same marginal distributions as the original network $\mathbb{P}_{Y_1\ldots Y_{K_2}|X_1\ldots X_{K_1}}$.

*Proof of Theorem 6)* This outer bound is actually derived similar to the outer bound (40) given in Lemma 2 on the sum-rate capacity. Note that the virtual network with the outputs $\overline{\overline{Y}}_{\lambda(1)},\ldots,\overline{\overline{Y}}_{\lambda(K_2)}$ is degraded in the sense that $X_1,\ldots,X_{K_1}\to\overline{\overline{Y}}_{\lambda(1)}\to\overline{\overline{Y}}_{\lambda(2)}\to\cdots\to\overline{\overline{Y}}_{\lambda(K_2)}$ forms a Markov chain. Also, the constraint in (127) on the partial sum-rate with respect the messages $\Omega$ is essentially similar to expression (40) except that:

- The signal $Y_j$ (40) is replaced everywhere by $\overline{\overline{Y}}_{\lambda(j)}, j=1,\ldots,K_2$.
- The messages $\mathbb{M}-\Omega$ are given as virtual side information to all signals $\overline{\overline{Y}}_{\lambda(j)}, j=1,\ldots,K_2$.
- The set $\mathbb{M}_{Y_j}$ in (40) is replaced everywhere by $\Omega\cap\mathbb{M}_{Y_{\lambda(j)}}, j=1,\ldots,K_2$.

By the above re-formulation, one can verify that all the derivations (A~1)-(A~10) still hold. Thus, we derive the outer bound (127). ■

It is clear that the outer bound (127) is sum-rate optimal for all degraded networks considered in previous subsections. It is indeed a useful capacity outer bound which is tighter than the trivial cut-set bound [10] for almost all interference networks. However, it is somewhat complex from computational aspects, specifically for large networks. In fact, describing capacity bounds for large multi-message networks in essence is complex so that even for the MACCM, where a full characterization of the capacity region is known, to describe the capacity (see (9)), corresponding to each common message an auxiliary random variable is required. Therefore, the complexity is a part of the essence of the problem and not due to our formulation in (127). Nonetheless, by the procedure we developed in Section III to simplify the sum-rate capacity for degraded networks, below we extract an outer bound from (127) which includes a fewer number of auxiliary random variables.

Consider the general interference network in Fig. 14 and the corresponding MACCM plan of messages depicted in Fig. 15. Let $\lambda(.)$ in (127) be a permutation of the elements of the set $\{1,\dots,K_2\}$. Given a subset $\nabla$ of the set $\{1,\dots,K_2\}$, define:

$$\lambda^{-1}(\nabla) \triangleq \{\lambda^{-1}(j) \,:\, j\in\nabla\} \tag{129}$$

Also, let $\Omega$ be a subset of the messages $\Omega\subseteq\mathbb{M}$. For every $\Delta\subseteq\{1,\dots,K_1\}$, define:

$$\Theta_\Delta^{\Omega,\lambda} \triangleq \begin{cases}\min\left\{\max\lambda^{-1}(\nabla) \,:\, \nabla\subseteq\{1,\dots,K_2\},\quad M_\Delta^\nabla\in\Omega\cap\mathbb{M}_\Delta\right\} & \text{if }\ \mathbb{M}_\Delta\cap\Omega\neq\emptyset\\ K_2+1 & \text{if }\ \mathbb{M}_\Delta\cap\Omega=\emptyset\end{cases} \tag{130}$$

Note that the subsets $\mathbb{M}_\Delta$ are given by (42). Now, consider the constraint given in (127) on the partial sum-rate with respect to $\Omega$. For every $\mathbb{M}_\Delta$, one of the following cases occurs:

*Case 1)* $\mathbb{M}_\Delta\cap\Omega\neq\emptyset$: In this case, using the constraint in (127) and similar to the derivations (46)-(52), it is not difficult to show that:

$$\boldsymbol{R}_{\Sigma\Omega}\leq\left(\begin{array}{c} I\left(\Omega\cap\mathbb{M}_{Y_{\lambda(1)}}-\mathbb{M}_\Delta;\bar{\bar{Y}}_{\lambda(1)}\middle|\Omega\cap\mathbb{M}_{Y_{\lambda(2)}}-\mathbb{M}_\Delta,\dots,\Omega\cap\mathbb{M}_{Y_{\lambda(K_2)}}-\mathbb{M}_\Delta,\mathbb{M}-\Omega-\mathbb{M}_\Delta,\mathbb{M}_\Delta,Q\right)+\dots+\\ I\left(\Omega\cap\mathbb{M}_{Y_{\lambda\left(\Theta_\Delta^{\Omega,\lambda}-1\right)}}-\mathbb{M}_\Delta;\bar{\bar{Y}}_{\lambda\left(\Theta_\Delta^{\Omega,\lambda}-1\right)}\middle|\Omega\cap\mathbb{M}_{Y_{\lambda\left(\Theta_\Delta^{\Omega,\lambda}\right)}}-\mathbb{M}_\Delta,\dots,\Omega\cap\mathbb{M}_{Y_{\lambda(K_2)}}-\mathbb{M}_\Delta,\mathbb{M}-\Omega-\mathbb{M}_\Delta,\mathbb{M}_\Delta,Q\right)+\\ +I\left(\Omega\cap\mathbb{M}_{Y_{\lambda\left(\Theta_\Delta^{\Omega,\lambda}\right)}}-\mathbb{M}_\Delta,\mathbb{M}_\Delta;\bar{\bar{Y}}_{\lambda\left(\Theta_\Delta^{\Omega,\lambda}\right)}\middle|\Omega\cap\mathbb{M}_{Y_{\lambda\left(\Theta_\Delta^{\Omega,\lambda}+1\right)}}-\mathbb{M}_\Delta,\dots,\Omega\cap\mathbb{M}_{Y_{\lambda(K_2)}}-\mathbb{M}_\Delta,\mathbb{M}-\Omega-\mathbb{M}_\Delta,Q\right)\\ +\dots+I\left(\Omega\cap\mathbb{M}_{Y_{\lambda(K_2)}}-\mathbb{M}_\Delta;\bar{\bar{Y}}_{\lambda(K_2)}\middle|\mathbb{M}-\Omega-\mathbb{M}_\Delta,Q\right)\end{array}\right) \tag{131}$$

*Case 2)* $\mathbb{M}_\Delta\cap\Omega=\emptyset$: In this case, $\mathbb{M}_\Delta$ is a subset of $\mathbb{M}-\Omega$. Thus, all the messages in $\mathbb{M}_\Delta$ appear after the conditioning operator in the mutual information functions of the constraint (127). In other words, we have:

$$\boldsymbol{R}_{\Sigma\Omega}\leq\left(\begin{array}{c} I\left(\Omega\cap\mathbb{M}_{Y_{\lambda(1)}};\bar{\bar{Y}}_{\lambda(1)}\middle|\Omega\cap\mathbb{M}_{Y_{\lambda(2)}},\dots,\Omega\cap\mathbb{M}_{Y_{\lambda(K_2-1)}},\Omega\cap\mathbb{M}_{Y_{\lambda(K_2)}},\mathbb{M}-\Omega-\mathbb{M}_\Delta,\mathbb{M}_\Delta,Q\right)\\ +\dots+I\left(\Omega\cap\mathbb{M}_{Y_{\lambda(K_2-1)}};\bar{\bar{Y}}_{\lambda(K_2-1)}\middle|\Omega\cap\mathbb{M}_{Y_{\lambda(K_2)}},\mathbb{M}-\Omega-\mathbb{M}_\Delta,\mathbb{M}_\Delta,Q\right)+I\left(\Omega\cap\mathbb{M}_{Y_{\lambda(K_2)}};\bar{\bar{Y}}_{\lambda(K_2)}\middle|\mathbb{M}-\Omega-\mathbb{M}_\Delta,\mathbb{M}_\Delta,Q\right)\end{array}\right) \tag{132}$$

Next, considering (131) and (132), we can derive the following result.

***Theorem 7)*** *Consider the general interference network in Fig. 14 with the associated messages $\mathbb{M}$ and its subsets $\mathbb{M}_\Delta$, $\Delta\subseteq\{1,\dots,K_1\}$ as defined in (42). Let $\{W_\Delta \,:\, \Delta\subseteq\{1,\dots,K_1\},\ \mathbb{M}_\Delta\neq\emptyset\}$ be a set of auxiliary random variables. Also, let $\mathcal{P}_{o:(2)}^{GIN}$ denotes the set of all joint PDFs with:*

$$P_Q\prod_\Delta P_{W_\Delta}\prod_{i=1}^{K_1}P_{X_i|\{W_\Delta\}_{\Delta:\,i\in\Delta},Q},\qquad P_{X_i|\{W_\Delta\}_{\Delta:\,i\in\Delta},Q}\in\{0,1\} \tag{133}$$

*Define the rate region $\mathfrak{R}_{o:(2)}^{GIN}$ as follows:*

$$\mathfrak{R}_{o:(2)}^{GIN} \triangleq \bigcup_{\mathcal{P}_{o:(2)}^{GIN}} \left\{ \begin{array}{l} (R_1, R_2, \dots, R_K) \in \mathbb{R}_+^K : \\ \forall\, \Omega \in \mathbb{M}, \qquad \forall\, \lambda(.) : \{1, \dots, K_2\} \to \{1, \dots, K_2\} \\ \bar{\bar{Y}}_{\lambda(j)} \triangleq \left(Y_{\lambda(j)}, Y_{\lambda(j+1)}, \dots, Y_{\lambda(K_2)}\right), \qquad j = 1, \dots, K_2 \\ \boldsymbol{R}_{\sum\Omega} \leq \left( \begin{array}{c} I\left(\{W_\Delta\}_{\Delta:\, \Theta_\Delta^{\Omega,\lambda}=1} ; \bar{\bar{Y}}_{\lambda(1)} \middle| \{W_\Delta\}_{\Delta:\, \Theta_\Delta^{\Omega,\lambda}\geq 2}, Q\right) + I\left(\{W_\Delta\}_{\Delta:\, \Theta_\Delta^{\Omega,\lambda}=2} ; \bar{\bar{Y}}_{\lambda(2)} \middle| \{W_\Delta\}_{\Delta:\, \Theta_\Delta^{\Omega,\lambda}\geq 3}, Q\right) \\ + \cdots + I\left(\{W_\Delta\}_{\Delta:\, \Theta_\Delta^{\Omega,\lambda}=K_2} ; \bar{\bar{Y}}_{\lambda(K_2)} \middle| \{W_\Delta\}_{\Delta:\, \Theta_\Delta^{\Omega,\lambda}\geq K_2+1}, Q\right) \end{array} \right) \end{array} \right\} \tag{134}$$

*The set $\mathfrak{R}_{o:(2)}^{GIN}$ consider an outer bound on the capacity region.*

*Proof of Theorem 7)* Let $\Omega$ be a subset of messages $\mathbb{M}$. Consider the constraint given in the outer bound (127) on the partial sum-rate with respect to $\Omega$. As discussed before, for every $\Delta \subseteq \{1, \dots, K_1\}$, one can enlarge the constraint by collecting the messages $\mathbb{M}_\Delta$ next to each other as given in (131) and (132). Also, due to the definition (42), if $i \in \Delta$, then $\mathbb{M}_\Delta \subseteq \mathbb{M}_{X_i}$ and we have:

$$P_{X_i|\mathbb{M}_{X_i}Q} = P_{X_i|\mathbb{M}_\Delta, \mathbb{M}_{X_i} - \mathbb{M}_\Delta, Q} \tag{135}$$

Otherwise, if $i \notin \Delta$, then $\mathbb{M}_\Delta \cap \mathbb{M}_{X_i} = \emptyset$. This means that in the distributions $P_{X_i|\mathbb{M}_{X_i}Q}, i = 1, \dots, K_1$ in (128) all the messages belonging to $\mathbb{M}_\Delta$ appear next to each other. Now, the constraint in (134) is derived by representing all the messages $\mathbb{M}_\Delta$ using a unique auxiliary random variable $W_\Delta$ (for all $\Delta \subseteq \{1, \dots, K_1\}$). ■

***Remark 7)***

1. The outer bound $\mathfrak{R}_{o:(2)}^{GIN}$ in (134) may be weaker than that in (127). However, for a broad range of network topologies, specifically for the MAINs in Fig. 24, it is still tighter than the trivial cut-set bound [10].
2. For those networks in which each of the subsets $\mathbb{M}_\Delta, \Delta \subseteq \{1, \dots, K_1\}$ includes at most one message (for example the MAINs in Fig. 24) both the outer bounds (127) and (134) are identical. For the MACCM in Fig. 4, both coincide with the capacity region.

Evaluation of the outer bound (134) is substantially simpler than that one in (127) because in (134) with respect to each subset $\mathbb{M}_\Delta$ of $\mathbb{M}$, there is only one auxiliary random variable, i.e., $W_\Delta$, while in (127) with respect to each message there is an auxiliary random variable. In fact, the outer bound $\mathfrak{R}_{o:(2)}^{GIN}$ in (134) can be efficiently evaluated for many networks. Nonetheless, for large multi-message networks the number of its constraints very rapidly grows.

Let us concentrate on the MAINs shown in Fig. 24. For these networks, inspired by MACCM capacity region in (9), one can deduce that it is more efficient to restrict $\Omega$ in either (127) or (134) to be a right-sided subset of messages (see Definition 3). In this case, the computational complexity of the outer bound is of the order of that for a MACCM capacity region (9). Moreover, we can consider the union in (134) over all joint PDFs which are given exactly similar to the MACCM encoding graph probability low in (11). In other words, characterizing these outer bounds follows the same arguments as the MACCM capacity region. We conclude this subsection by providing an example in this regard. Consider a special class of the MAINs in Fig. 24 where there is no common message among transmitters. Fig. 27 depicts such a network. In this scenario, the transmitters $X_{j,1}, \dots, X_{j,\mu_j}$ send the messages $M_{j,1}, \dots, M_{j,\mu_j}$, respectively, to the receiver $Y_j, j = 1, \dots, K_2$.

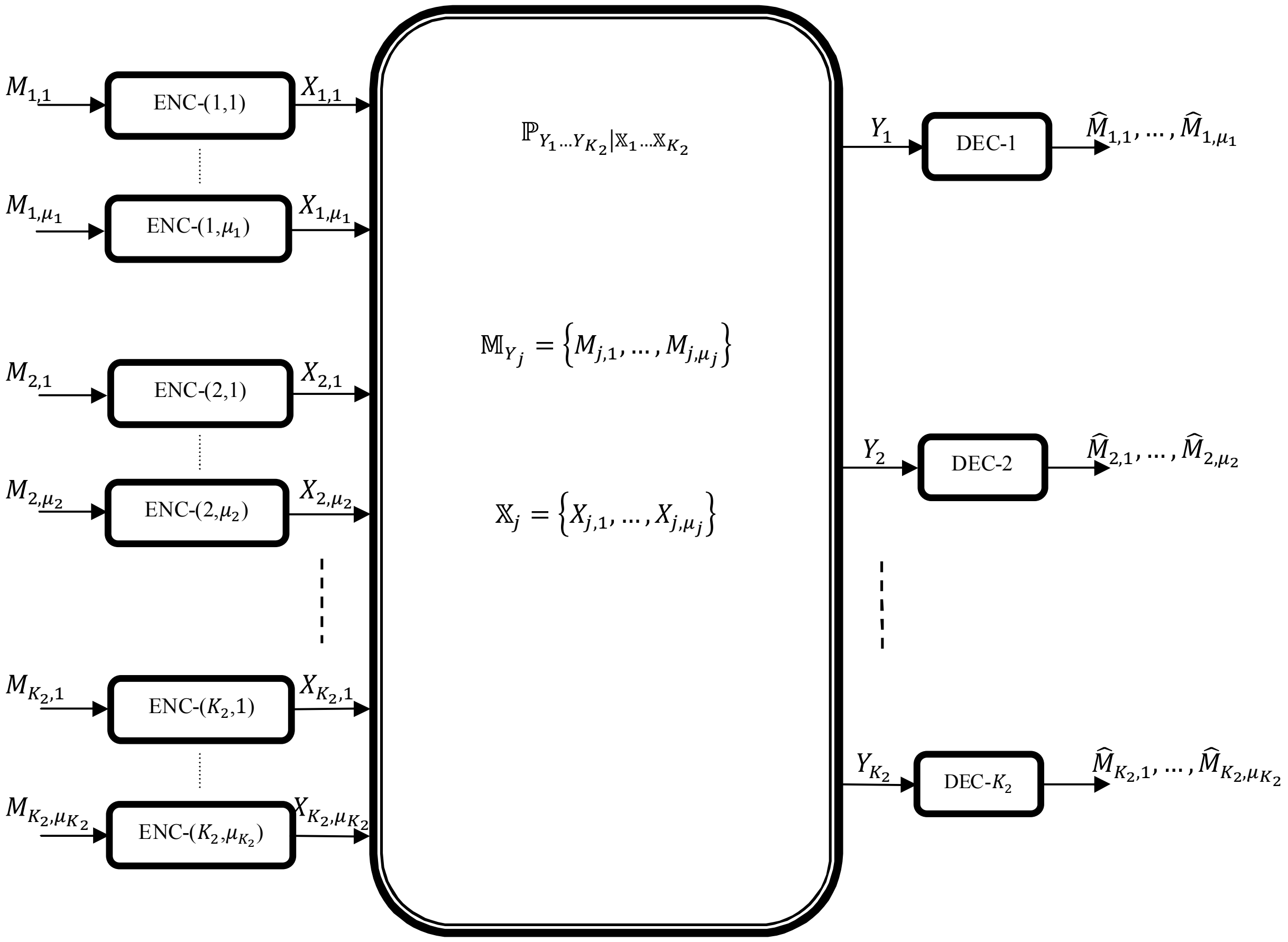

Figure 27. The MAIN with private messages. In this scenario, the transmitters $X_{j,1},\dots,X_{j,\mu_j}$ send the messages $M_{j,1},\dots,M_{j,\mu_j}$, respectively, to the receiver $Y_j$, $j=1,\dots,K_2$.

For this network, using the MACCM MG\EG translation, one can re-write the outer bound (127) as follows:

$$\bigcup_{\mathcal{P}^{IN\to Fig.27}} \left\{\begin{array}{l} \left(R_{1,1},\dots,R_{1,\mu_1},R_{2,1},\dots,R_{2,\mu_2},\dots,R_{K_2,1},\dots,R_{K_2,\mu_{K_2}}\right)\in\mathbb{R}_+^{\sum_{j=1}^{K_2}\mu_j}: \\ \forall\ \ \mathbb{Y}_j\subseteq\{1,\dots,\mu_j\} \qquad \tilde{\mathbb{Y}}_j\triangleq\{1,\dots,\mu_j\},\qquad j=1,\dots,K_2 \\ \forall\ \ \lambda(.):\{1,\dots,K_2\}\to\{1,\dots,K_2\}\qquad \bar{\bar{Y}}_{\lambda(j)}\triangleq\left(Y_{\lambda(j)},Y_{\lambda(j+1)},\dots,Y_{\lambda(K_2)}\right),\qquad j=1,\dots,K_2 \\ \\ \displaystyle\sum_{j=1}^{K_2}\sum_{i\in\mathbb{Y}_j}R_{j,i}\le\left(\begin{array}{c} I\left(\left\{X_{\lambda(1),l}\right\}_{l\in\mathbb{Y}_{\lambda(1)}};\bar{\bar{Y}}_{\lambda(1)}\middle|\left\{X_{\lambda(2),l}\right\}_{l\in\mathbb{Y}_{\lambda(2)}},\dots,\left\{X_{\lambda(K_2-1),l}\right\}_{l\in\mathbb{Y}_{\lambda(K_2-1)}},\left\{X_{\lambda(K_2),l}\right\}_{l\in\mathbb{Y}_{\lambda(K_2)}},\cup_{j=1}^{K_2}\left\{X_{\lambda(j),l}\right\}_{l\in\tilde{\mathbb{Y}}_{\lambda(j)}},Q\right) \\ +\cdots+I\left(\left\{X_{\lambda(K_2-1),l}\right\}_{l\in\mathbb{Y}_{\lambda(K_2-1)}};\bar{\bar{Y}}_{\lambda(K_2-1)}\middle|\left\{X_{\lambda(K_2),l}\right\}_{l\in\mathbb{Y}_{\lambda(K_2)}},\cup_{j=1}^{K_2}\left\{X_{\lambda(j),l}\right\}_{l\in\tilde{\mathbb{Y}}_{\lambda(j)}},Q\right) \\ +I\left(\left\{X_{\lambda(K_2),l}\right\}_{l\in\mathbb{Y}_{\lambda(K_2)}};\bar{\bar{Y}}_{\lambda(K_2)}\middle|\cup_{j=1}^{K_2}\left\{X_{\lambda(j),l}\right\}_{l\in\tilde{\mathbb{Y}}_{\lambda(j)}},Q\right) \end{array}\right) \end{array}\right\} \tag{136}$$

where $\mathcal{P}^{IN\to Fig.27}$ denotes the set of all joint PDFs $P_Q\prod_{j=1}^{K_2}\prod_{i=1}^{\mu_j}P_{X_{j,i}|Q}$.

Thus, the rate region (136) constitutes an outer bound on the capacity region of the MAIN with private messages in Fig. 27. Note that (136) does not include any auxiliary random variable except the time-sharing parameter $Q$. It is trivial that this is tighter than the cut-set bound.

# CONCLUSION

In the present part of our multi-part papers, we started the study of capacity limits for network topologies of arbitrary large sizes by considering *degraded networks*. We derived an explicit characterization of the sum-rate capacity for the degraded interference networks with any number of transmitters, any number of receivers, and any distribution of messages among transmitters and receivers. We showed that a *successive decoding strategy* is sum-rate optimal for these networks. Also, we proved that to achieve the sum-rate capacity, only a certain subset of messages is required to be considered for the transmission scheme. By developing graphical illustrations coined as "*MACCM Plan of Messages*", we presented algorithms to explicitly determine these desired messages. In addition, using the closed-form expression derived for the sum-rate capacity of the degraded networks, we obtained unified outer bounds for the sum-rate capacity of the general (non-degraded) interference networks. We made use of these outer bounds to identify noisy interference regimes for certain classes of interference networks such as the *generalized Z-interference networks* and the *many-to-one interference networks*. Also, by using the derived outer bounds, for the first time we identified network scenarios for which incorporation of successive decoding and treating interference as noise achieves the sum-rate capacity. Finally, by taking insight from our results for the degraded networks, we established unified outer bounds on the entire capacity region of the general non-degraded networks. For a broad range of network topologies, these outer bounds are tighter than the existing trivial cut-set bound [10].

Our systematic study regarding capacity bounds for the interference networks will be continued by investigating the behavior of information flow in strong and mixed interference e regimes in Part III of our multi-part papers [3].

# APPENDIX

- ***Proof of Lemma 2***

Consider a code of length $n$ with the communication rates $R_1, \dots, R_K$ for the messages $M_1, \dots, M_K$, respectively; assume that the code has vanishing average error probability. Define the sets $\mathcal{L}_j, j = 1, \dots, K_2$, as follows:

$$\mathcal{L}_j \triangleq \mathbb{M}_{Y_j} - \left(\mathbb{M}_{Y_{j+1}} \cup \dots \cup \mathbb{M}_{Y_{K_2}}\right), \qquad j = 1, \dots, K_2 \tag{A~1}$$

where $\mathbb{M}_{Y_{K_2+1}} \triangleq \emptyset$. Let recall that in the following analysis, for a given subset of messages $\Omega$, the notation $\underline{id}_\Omega$ denotes the identification of the set $\Omega$, as defined in Part I [1, Def. II.1]. Now using Fano's inequality, we can write:

$$\begin{aligned}
\sum_{l \in \underline{id}_{\mathcal{L}_{K_2}}} R_l &\leq \frac{1}{n} I\left(\mathcal{L}_{K_2}; Y_{K_2}^n\right) + \epsilon_{K_2,n} \\
&= \frac{1}{n} \sum_{t=1}^{n} I\left(\mathcal{L}_{K_2}; Y_{K_2,t} \middle| Y_{K_2}^{t-1}\right) + \epsilon_{K_2,n} \\
&\leq \frac{1}{n} \sum_{t=1}^{n} I\left(\mathcal{L}_{K_2}, Y_{K_2}^{t-1}; Y_{K_2,t}\right) + \epsilon_{K_2,n} \\
&\overset{(a)}{=} \frac{1}{n} \sum_{t=1}^{n} I\left(\mathbb{M}_{Y_{K_2}}, Y_{K_2}^{t-1}; Y_{K_2,t}\right) + \epsilon_{K_2,n}
\end{aligned} \tag{A~2}$$

where $\epsilon_{K_2,n} \to 0$ as $n \to \infty$, and equality (a) holds because $\mathcal{L}_{K_2} = \mathbb{M}_{Y_{K_2}}$. Also, we have:

$$\begin{aligned}
\sum_{l \in \underline{id}_{\mathcal{L}_{K_2-1}}} R_l &\leq \frac{1}{n} I\left(\mathcal{L}_{K_2-1}; Y_{K_2-1}^n\right) + \epsilon_{K_2-1,n} \\
&\leq \frac{1}{n} I\left(\mathcal{L}_{K_2-1}; Y_{K_2-1}^n, Y_{K_2}^n, \mathbb{M}_{Y_{K_2}}\right) + \epsilon_{K_2-1,n}
\end{aligned}$$

$$\overset{(a)}{=}\frac{1}{n}I\left(\mathcal{L}_{K_2-1};Y_{K_2-1}^{n},Y_{K_2}^{n}\middle|\mathbb{M}_{Y_{K_2}}\right)+\epsilon_{K_2-1,n}$$
$$=\frac{1}{n}\sum_{t=1}^{n}I\left(\mathcal{L}_{K_2-1};Y_{K_2-1,t},Y_{K_2,t}\middle|\mathbb{M}_{Y_{K_2}},Y_{K_2-1}^{t-1},Y_{K_2}^{t-1}\right)+\epsilon_{K_2-1,n}$$
$$\overset{(b)}{=}\frac{1}{n}\sum_{t=1}^{n}I\left(\mathcal{L}_{K_2-1};Y_{K_2-1,t}\middle|\mathbb{M}_{Y_{K_2}},Y_{K_2-1}^{t-1},Y_{K_2}^{t-1}\right)+\epsilon_{K_2-1,n}$$
$$\leq\frac{1}{n}\sum_{t=1}^{n}I\left(\mathcal{L}_{K_2-1},Y_{K_2-1}^{t-1};Y_{K_2-1,t}\middle|\mathbb{M}_{Y_{K_2}},Y_{K_2}^{t-1}\right)+\epsilon_{K_2-1,n}$$
$$\overset{(c)}{=}\frac{1}{n}\sum_{t=1}^{n}I\left(\mathbb{M}_{Y_{K_2-1}},Y_{K_2-1}^{t-1};Y_{K_2-1,t}\middle|\mathbb{M}_{Y_{K_2}},Y_{K_2}^{t-1}\right)+\epsilon_{K_2-1,n}$$

(A~3)

where $\epsilon_{K_2-1,n}\to 0$ as $n\to\infty$; equality (a) holds because the elements in the set $\mathcal{L}_{K_2-1}$ are independent of those in $\mathbb{M}_{Y_{K_2}}$, equality (b) is due to degradedness property of the channel and equality (c) holds because $\left(\mathbb{M}_{Y_{K_2-1}}-\mathcal{L}_{K_2-1}\right)\subseteq\mathbb{M}_{Y_{K_2}}$. Similarly, one can derive:

$$\sum_{l\in \underline{id}_{\mathcal{L}_j}}R_l\leq\frac{1}{n}\sum_{t=1}^{n}I\left(\mathbb{M}_{Y_j},Y_j^{t-1};Y_{j,t}\middle|\mathbb{M}_{Y_{j+1}},\dots,\mathbb{M}_{Y_{K_2}},Y_{j+1}^{t-1},\dots,Y_{K_2}^{t-1}\right)+\epsilon_{j,n},\qquad j=1,\dots,K_2$$

(A~4)

where $\epsilon_{j,n}\to 0$ as $n\to\infty$. Define:

$$\varSigma_j\triangleq\frac{1}{n}\sum_{t=1}^{n}I\left(\mathbb{M}_{Y_j},Y_j^{t-1};Y_{j,t}\middle|\mathbb{M}_{Y_{j+1}},\dots,\mathbb{M}_{Y_{K_2}},Y_{j+1}^{t-1},\dots,Y_{K_2}^{t-1}\right)$$

(A~5)

Now we have:

$$\varSigma_{K_2}+\varSigma_{K_2-1}=\frac{1}{n}\sum_{t=1}^{n}I\left(\mathbb{M}_{Y_{K_2}},Y_{K_2}^{t-1};Y_{K_2,t}\right)+\frac{1}{n}\sum_{t=1}^{n}I\left(\mathbb{M}_{Y_{K_2-1}},Y_{K_2-1}^{t-1};Y_{K_2-1,t}\middle|\mathbb{M}_{Y_{K_2}},Y_{K_2}^{t-1}\right)$$
$$=\frac{1}{n}\sum_{t=1}^{n}I\left(\mathbb{M}_{Y_{K_2}};Y_{K_2,t}\right)+\frac{1}{n}\sum_{t=1}^{n}I\left(Y_{K_2}^{t-1};Y_{K_2,t}\middle|\mathbb{M}_{Y_{K_2}}\right)+\frac{1}{n}\sum_{t=1}^{n}I\left(\mathbb{M}_{Y_{K_2-1}},Y_{K_2-1}^{t-1};Y_{K_2-1,t}\middle|\mathbb{M}_{Y_{K_2}},Y_{K_2}^{t-1}\right)$$
$$\overset{(a)}{\leq}\frac{1}{n}\sum_{t=1}^{n}I\left(\mathbb{M}_{Y_{K_2}};Y_{K_2,t}\right)+\frac{1}{n}\sum_{t=1}^{n}I\left(Y_{K_2}^{t-1};Y_{K_2-1,t}\middle|\mathbb{M}_{Y_{K_2}}\right)+\frac{1}{n}\sum_{t=1}^{n}I\left(\mathbb{M}_{Y_{K_2-1}},Y_{K_2-1}^{t-1};Y_{K_2-1,t}\middle|\mathbb{M}_{Y_{K_2}},Y_{K_2}^{t-1}\right)$$
$$=\frac{1}{n}\sum_{t=1}^{n}I\left(\mathbb{M}_{Y_{K_2}};Y_{K_2,t}\right)+\frac{1}{n}\sum_{t=1}^{n}I\left(\mathbb{M}_{Y_{K_2-1}},Y_{K_2-1}^{t-1},Y_{K_2}^{t-1};Y_{K_2-1,t}\middle|\mathbb{M}_{Y_{K_2}}\right)$$

(A~6)

where inequality (a) is due to the degradedness property of the channel. Then, using (A~4) we can write:

$$\varSigma_{K_2}+\varSigma_{K_2-1}+\varSigma_{K_2-2}$$
$$\overset{(a)}{\leq}\frac{1}{n}\sum_{t=1}^{n}I\left(\mathbb{M}_{Y_{K_2}};Y_{K_2,t}\right)+\frac{1}{n}\sum_{t=1}^{n}I\left(\mathbb{M}_{Y_{K_2-1}},Y_{K_2-1}^{t-1},Y_{K_2}^{t-1};Y_{K_2-1,t}\middle|\mathbb{M}_{Y_{K_2}}\right)$$
$$+\frac{1}{n}\sum_{t=1}^{n}I\left(\mathbb{M}_{Y_{K_2-2}},Y_{K_2-2}^{t-1};Y_{K_2-2,t}\middle|\mathbb{M}_{Y_{K_2-1}},\mathbb{M}_{Y_{K_2}},Y_{K_2-1}^{t-1},Y_{K_2}^{t-1}\right)$$

$$=\frac{1}{n}\sum_{t=1}^{n} I\left(\mathbb{M}_{Y_{K_2}};Y_{K_2,t}\right)+\frac{1}{n}\sum_{t=1}^{n} I\left(\mathbb{M}_{Y_{K_2-1}};Y_{K_2-1,t}\middle|\mathbb{M}_{Y_{K_2}}\right)+\frac{1}{n}\sum_{t=1}^{n} I\left(Y_{K_2-1}^{t-1},Y_{K_2}^{t-1};Y_{K_2-1,t}\middle|\mathbb{M}_{Y_{K_2-1}},\mathbb{M}_{Y_{K_2}}\right)$$

$$+\frac{1}{n}\sum_{t=1}^{n} I\left(\mathbb{M}_{Y_{K_2-2}},Y_{K_2-2}^{t-1};Y_{K_2-2,t}\middle|\mathbb{M}_{Y_{K_2-1}},\mathbb{M}_{Y_{K_2}},Y_{K_2-1}^{t-1},Y_{K_2}^{t-1}\right)$$

$$\overset{(b)}{\leq}\frac{1}{n}\sum_{t=1}^{n} I\left(\mathbb{M}_{Y_{K_2}};Y_{K_2,t}\right)+\frac{1}{n}\sum_{t=1}^{n} I\left(\mathbb{M}_{Y_{K_2-1}};Y_{K_2-1,t}\middle|\mathbb{M}_{Y_{K_2}}\right)+\frac{1}{n}\sum_{t=1}^{n} I\left(Y_{K_2-1}^{t-1},Y_{K_2}^{t-1};Y_{K_2-2,t}\middle|\mathbb{M}_{Y_{K_2-1}},\mathbb{M}_{Y_{K_2}}\right)$$

$$+\frac{1}{n}\sum_{t=1}^{n} I\left(\mathbb{M}_{Y_{K_2-2}},Y_{K_2-2}^{t-1};Y_{K_2-2,t}\middle|\mathbb{M}_{Y_{K_2-1}},\mathbb{M}_{Y_{K_2}},Y_{K_2-1}^{t-1},Y_{K_2}^{t-1}\right)$$

$$=\frac{1}{n}\sum_{t=1}^{n} I\left(\mathbb{M}_{Y_{K_2}};Y_{K_2,t}\right)+\frac{1}{n}\sum_{t=1}^{n} I\left(\mathbb{M}_{Y_{K_2-1}};Y_{K_2-1,t}\middle|\mathbb{M}_{Y_{K_2}}\right)+\frac{1}{n}\sum_{t=1}^{n} I\left(\mathbb{M}_{Y_{K_2-2}},Y_{K_2-2}^{t-1},Y_{K_2-1}^{t-1},Y_{K_2}^{t-1};Y_{K_2-2,t}\middle|\mathbb{M}_{Y_{K_2-1}},\mathbb{M}_{Y_{K_2}}\right)$$

(A~7)

where inequality (a) is obtained from (A~6) and (b) is due to the degradedness property of the channel. By continuing the steps (A~6) and (A~7), one can derive:

$$\Sigma_{K_2}+\Sigma_{K_2-1}+\Sigma_{K_2-2}+\cdots+\Sigma_1$$

$$\leq\frac{1}{n}\sum_{t=1}^{n} I\left(\mathbb{M}_{Y_{K_2}};Y_{K_2,t}\right)+\frac{1}{n}\sum_{t=1}^{n} I\left(\mathbb{M}_{Y_{K_2-1}};Y_{K_2-1,t}\middle|\mathbb{M}_{Y_{K_2}}\right)+\frac{1}{n}\sum_{t=1}^{n} I\left(\mathbb{M}_{Y_{K_2-2}};Y_{K_2-2,t}\middle|\mathbb{M}_{Y_{K_2-1}},\mathbb{M}_{Y_{K_2}}\right)+\cdots+$$

$$+\frac{1}{n}\sum_{t=1}^{n} I\left(\mathbb{M}_{Y_2};Y_{2,t}\middle|\mathbb{M}_{Y_3},\ldots,\mathbb{M}_{Y_{K_2-1}},\mathbb{M}_{Y_{K_2}}\right)+\frac{1}{n}\sum_{t=1}^{n} I\left(\mathbb{M}_{Y_1},Y_1^{t-1},\ldots,Y_{K_2-1}^{t-1},Y_{K_2}^{t-1};Y_{1,t}\middle|\mathbb{M}_{Y_2},\ldots,\mathbb{M}_{Y_{K_2-1}},\mathbb{M}_{Y_{K_2}}\right)$$

$$\overset{(a)}{=}\frac{1}{n}\sum_{t=1}^{n} I\left(\mathbb{M}_{Y_{K_2}};Y_{K_2,t}\right)+\frac{1}{n}\sum_{t=1}^{n} I\left(\mathbb{M}_{Y_{K_2-1}};Y_{K_2-1,t}\middle|\mathbb{M}_{Y_{K_2}}\right)+\frac{1}{n}\sum_{t=1}^{n} I\left(\mathbb{M}_{Y_{K_2-2}};Y_{K_2-2,t}\middle|\mathbb{M}_{Y_{K_2-1}},\mathbb{M}_{Y_{K_2}}\right)+\cdots+$$

$$+\frac{1}{n}\sum_{t=1}^{n} I\left(\mathbb{M}_{Y_2};Y_{2,t}\middle|\mathbb{M}_{Y_3},\ldots,\mathbb{M}_{Y_{K_2-1}},\mathbb{M}_{Y_{K_2}}\right)+\frac{1}{n}\sum_{t=1}^{n} I\left(\mathbb{M}_{Y_1};Y_{1,t}\middle|\mathbb{M}_{Y_2},\ldots,\mathbb{M}_{Y_{K_2-1}},\mathbb{M}_{Y_{K_2}}\right)$$

(A~8)

where equality (a) holds because:

$$\frac{1}{n}\sum_{t=1}^{n} I\left(Y_1^{t-1},\ldots,Y_{K_2-1}^{t-1},Y_{K_2}^{t-1};Y_{1,t}\middle|\mathbb{M}_{Y_1},\mathbb{M}_{Y_2},\ldots,\mathbb{M}_{Y_{K_2-1}},\mathbb{M}_{Y_{K_2}}\right)=\frac{1}{n}\sum_{t=1}^{n} I\left(Y_1^{t-1},\ldots,Y_{K_2-1}^{t-1},Y_{K_2}^{t-1};Y_{1,t}\middle|\mathbb{M}\right)=0$$

(A~9)

Thus, using (A~8) we can write:

$$\mathcal{C}_{\Sigma_{\mathbb{M}}}^{GIN-deg}=\sum_{j=1}^{K_2}\left(\sum_{l\in \underline{id}_{\mathcal{L}_j}} R_l\right)\leq\sum_{j=1}^{K_2}\Sigma_j+\sum_{j=1}^{K_2}\epsilon_{j,n}$$

$$\leq\frac{1}{n}\sum_{t=1}^{n} I\left(\mathbb{M}_{Y_1};Y_{1,t}\middle|\mathbb{M}_{Y_2},\ldots,\mathbb{M}_{Y_{K_2-1}},\mathbb{M}_{Y_{K_2}}\right)+\cdots+\frac{1}{n}\sum_{t=1}^{n} I\left(\mathbb{M}_{Y_{K_2-1}};Y_{K_2-1,t}\middle|\mathbb{M}_{Y_{K_2}}\right)+\frac{1}{n}\sum_{t=1}^{n} I\left(\mathbb{M}_{Y_{K_2}};Y_{K_2,t}\right)+\sum_{j=1}^{K_2}\epsilon_{j,n}$$

(A~10)

Finally, by applying a standard time-sharing argument and also by letting $n$ tends to infinity, we derive the outer bound (40). The proof is thus complete. ■

➢ ***Converse Proof for Proposition 4***

We show that the Gaussian distributions are optimal for the sum-rate (86). There are several ways to prove this result. Here, we make use of the EPI. Consider the argument of the maximization (86). Fix a joint PDF $P_Q P_{X_1X_2|Q} P_{X_3|Q} P_{X_4|X_1X_2X_3Q}$. We have:

$$\begin{aligned}
&I(X_4;Y_1|X_1,X_2,X_3,Q)+I(X_3;Y_2|X_1,X_2,Q)+I(X_1,X_2;Y_3|Q)\\
&\quad=H(Y_1|X_1,X_2,X_3,Q)-H(Y_2|X_1,X_2,X_3,Q)+H(Y_2|X_1,X_2,Q)-H(Y_3|X_1,X_2,Q)+H(Y_3|Q)-H(Y_1|X_1,X_2,X_3,X_4,Q)\\
&\quad=H(a_4X_4+Z_1|X_1,X_2,X_3,Q)-H\left(\tfrac{1}{b_2}(a_4X_4+Z_1)+\sqrt{1-\tfrac{1}{b_2^2}}\,\tilde{Z}_2\middle|X_1,X_2,X_3,Q\right)\\
&\qquad+H\left(\tfrac{a_3}{b_2}X_3+\tfrac{a_4}{b_2}X_4+Z_2\middle|X_1,X_2,Q\right)-H\left(\tfrac{1}{b_3}\left(\tfrac{a_3}{b_2}X_3+\tfrac{a_4}{b_2}X_4+Z_2\right)+\sqrt{1-\tfrac{1}{b_3^2}}\,\tilde{Z}_3\middle|X_1,X_2,Q\right)\\
&\qquad+H\left(\tfrac{a_1}{b_2b_3}X_1+\tfrac{a_2}{b_2b_3}X_2+\tfrac{a_3}{b_2b_3}X_3+\tfrac{a_4}{b_2b_3}X_4+Z_3\middle|Q\right)-H(Z_1)
\end{aligned}$$

(A~11)

Next let $X_1^G,X_2^G,X_3^G,X_4^G$ be jointly Gaussian RVs with a covariance matrix identical to that of $X_1,X_2,X_3,X_4$. Thus, we can decompose $X_4^G$ as follows:

$$X_4^G=\alpha\sqrt{\frac{P_4}{P_1}}X_1^G+\beta\sqrt{\frac{P_4}{P_3}}X_3^G+\sqrt{(1-(\alpha^2+\beta^2))P_4}Z$$

(A~12)

where $\alpha,\beta$ belong to the interval $[-1,1]$ with $\alpha^2+\beta^2\leq 1$, and $Z$ is a Gaussian RV independent of $X_1^G,X_2^G,X_3^G$ with zero mean and unit variance. Consider the expressions in the right hand side of (A~11). For the first term, we have:

$$\begin{aligned}
\frac{1}{2}\log(2\pi e)&\leq H(Z_1)\leq H(a_4X_4+Z_1|X_1,X_2,X_3,Q)\leq H(a_4X_4+Z_1|X_1,X_3)\\
&\leq H(a_4X_4^G+Z_1|X_1^G,X_3^G)=\tfrac{1}{2}\log\left(2\pi e\left(a_4^2(1-(\alpha^2+\beta^2))P_4+1\right)\right)
\end{aligned}$$

(A~13)

Comparing the two sides of (A~13), we deduce that there exists a $\lambda_1\in[0,1]$ such that:

$$H(a_4X_4+Z_1|X_1,X_2,X_3,Q)=\frac{1}{2}\log\left(2\pi e\left(\lambda_1a_4^2(1-(\alpha^2+\beta^2))P_4+1\right)\right)$$

(A~14)

Then consider the second term in (A~11). We can write:

$$\begin{aligned}
&H\left(\tfrac{1}{b_2}(a_4X_4+Z_1)+\sqrt{1-\tfrac{1}{b_2^2}}\,\tilde{Z}_2\middle|X_1,X_2,X_3,Q\right)\\
&\quad\overset{(a)}{\geq}\tfrac{1}{2}\log\left(2^{2H\left(\frac{1}{b_2}(a_4X_4+Z_1)\middle|X_1,X_2,X_3,Q\right)}+2^{2H\left(\sqrt{1-\frac{1}{b_2^2}}\tilde{Z}_2\middle|X_1,X_2,X_3,Q\right)}\right)\\
&\quad\overset{(b)}{=}\tfrac{1}{2}\log\left(2\pi e\tfrac{1}{b_2^2}\left(\lambda_1a_4^2(1-(\alpha^2+\beta^2))P_4+1\right)+2\pi e\left(1-\tfrac{1}{b_2^2}\right)\right)\\
&\quad=\tfrac{1}{2}\log\left(2\pi e\left(\lambda_1\tfrac{a_4^2}{b_2^2}(1-(\alpha^2+\beta^2))P_4+1\right)\right)
\end{aligned}$$

(A~15)

where (a) is due to the EPI and (b) is due to (A~14). Thereby, from (A~14) and (A~15) we obtain:

$$H(a_4X_4+Z_1|X_1,X_2,X_3,Q)-H\left(\tfrac{1}{b_2}(a_4X_4+Z_1)+\sqrt{1-\tfrac{1}{b_2^2}}\,\tilde{Z}_2\middle|X_1,X_2,X_3,Q\right)$$

$$\leq \frac{1}{2}\log\left(\frac{\lambda_1 a_4^2\big(1-(\alpha^2+\beta^2)\big)P_4+1}{\lambda_1 \frac{a_4^2}{b_2^2}\big(1-(\alpha^2+\beta^2)\big)P_4+1}\right) \overset{(a)}{\leq} \frac{1}{2}\log\left(\frac{a_4^2\big(1-(\alpha^2+\beta^2)\big)P_4+1}{\frac{a_4^2}{b_2^2}\big(1-(\alpha^2+\beta^2)\big)P_4+1}\right)$$

(A~16)

where inequality (a) holds because the expression in its left hand side is monotonically increasing in terms of $\lambda_1$ (for the case where $b_2 \geq 1$).

Next we evaluate the third and the forth terms in (A~11). We have:

$$\frac{1}{2}\log(2\pi e) \leq H(Z_2) \leq H\left(\frac{a_3}{b_2}X_3+\frac{a_4}{b_2}X_4+Z_2\middle|X_1,X_2,Q\right) \leq H\left(\frac{a_3}{b_2}X_3+\frac{a_4}{b_2}X_4+Z_2\middle|X_1\right)$$

$$\leq H\left(\frac{a_3}{b_2}X_3^G+\frac{a_4}{b_2}X_4^G+Z_2\middle|X_1^G\right) \leq \frac{1}{2}\log\left(2\pi e\left(\frac{1}{b_2^2}\left(|a_3|+|a_4\beta|\sqrt{\frac{P_4}{P_3}}\right)^2+\frac{a_4^2}{b_2^2}\big(1-(\alpha^2+\beta^2)\big)P_4+1\right)\right)$$

(A~17)

Comparing the two sides of (A~17), we deduce that there exists a $\lambda_2 \in [0,1]$ such that:

$$H\left(\frac{a_3}{b_2}X_3+\frac{a_4}{b_2}X_4+Z_2\middle|X_1,X_2,Q\right) = \frac{1}{2}\log\left(2\pi e\left(\lambda_2\frac{1}{b_2^2}\left(\left(|a_3|+|a_4\beta|\sqrt{\frac{P_4}{P_3}}\right)^2+a_4^2\big(1-(\alpha^2+\beta^2)\big)P_4\right)+1\right)\right)$$

(A~18)

Considering (A~18) and using the EPI, one can derive:

$$H\left(\frac{1}{b_3}\left(\frac{a_3}{b_2}X_3+\frac{a_4}{b_2}X_4+Z_2\right)+\sqrt{1-\frac{1}{b_3^2}}\,\tilde{Z}_3\middle|X_1,X_2,Q\right) \geq \frac{1}{2}\log\left(2\pi e\left(\lambda_2\frac{1}{b_2^2 b_3^2}\left(\left(|a_3|+|a_4\beta|\sqrt{\frac{P_4}{P_3}}\right)^2+a_4^2\big(1-(\alpha^2+\beta^2)\big)P_4\right)+1\right)\right)$$

(A~19)

Now from (A~18) and (A~19), we obtain:

$$H\left(\frac{a_3}{b_2}X_3+\frac{a_4}{b_2}X_4+Z_2\middle|X_1,X_2,Q\right) - H\left(\frac{1}{b_3}\left(\frac{a_3}{b_2}X_3+\frac{a_4}{b_2}X_4+Z_2\right)+\sqrt{1-\frac{1}{b_3^2}}\,\tilde{Z}_3\middle|X_1,X_2,Q\right)$$

$$\leq \frac{1}{2}\log\left(\frac{\lambda_2\frac{1}{b_2^2}\left(\left(|a_3|+|a_4\beta|\sqrt{\frac{P_4}{P_3}}\right)^2+a_4^2\big(1-(\alpha^2+\beta^2)\big)P_4\right)+1}{\lambda_2\frac{1}{b_2^2 b_3^2}\left(\left(|a_3|+|a_4\beta|\sqrt{\frac{P_4}{P_3}}\right)^2+a_4^2\big(1-(\alpha^2+\beta^2)\big)P_4\right)+1}\right)$$

$$\overset{(a)}{\leq} \frac{1}{2}\log\left(\frac{\frac{1}{b_2^2}\left(\left(|a_3|+|a_4\beta|\sqrt{\frac{P_4}{P_3}}\right)^2+a_4^2\big(1-(\alpha^2+\beta^2)\big)P_4\right)+1}{\frac{1}{b_2^2 b_3^2}\left(\left(|a_3|+|a_4\beta|\sqrt{\frac{P_4}{P_3}}\right)^2+a_4^2\big(1-(\alpha^2+\beta^2)\big)P_4\right)+1}\right)$$

(A~20)

where inequality (a) holds because the expression in its left hand side is monotonically increasing in terms of $\lambda_2$ (for the case where $b_2, b_3 \geq 1$). Finally, consider the fifth term of (A~11). We have:

$$H\left(\frac{a_1}{b_2 b_3}X_1+\frac{a_2}{b_2 b_3}X_2+\frac{a_3}{b_2 b_3}X_3+\frac{a_4}{b_2 b_3}X_4+Z_3\middle|Q\right)$$

$$\leq H\left(\frac{a_1}{b_2 b_3}X_1^G+\frac{a_2}{b_2 b_3}X_2^G+\frac{a_3}{b_2 b_3}X_3^G+\frac{a_4}{b_2 b_3}X_4^G+Z_3\right)$$

$$\leq \frac{1}{2}\log\left(2\pi e\left(\frac{1}{b_2^2b_3^2}\begin{pmatrix} a_1^2P_1+a_2^2P_2+a_3^2P_3+a_4^2P_4 \\ +2|a_1a_2\mathbb{E}[X_1^GX_2^G]|+2|a_1a_4\mathbb{E}[X_1^GX_4^G]|+2|a_2a_4\mathbb{E}[X_2^GX_4^G]|+2|a_3a_4\mathbb{E}[X_3^GX_4^G]| \end{pmatrix}+1\right)\right)$$

$$\overset{(a)}{=} \frac{1}{2}\log\left(2\pi e\left(\frac{1}{b_2^2b_3^2}\begin{pmatrix} a_1^2P_1+a_2^2P_2+a_3^2P_3+a_4^2P_4 \\ +2|a_1a_2\mathbb{E}[X_1^GX_2^G]|+2|a_1a_4\alpha|\sqrt{P_1P_4}+2|a_2a_4\alpha\mathbb{E}[X_1^GX_2^G]|\sqrt{\frac{P_4}{P_1}}+2|a_3a_4\beta|\sqrt{P_3P_4} \end{pmatrix}+1\right)\right)$$

$$\overset{(b)}{\leq} \frac{1}{2}\log\left(2\pi e\left(\frac{1}{b_2^2b_3^2}\begin{pmatrix} a_1^2P_1+a_2^2P_2+a_3^2P_3+a_4^2P_4 \\ +2|a_1a_2|\sqrt{P_1P_2}+2|a_1a_4\alpha|\sqrt{P_1P_4}+2|a_2a_4\alpha|\sqrt{P_2P_4}+2|a_3a_4\beta|\sqrt{P_3P_4} \end{pmatrix}+1\right)\right)$$

(A~21)

where equality (a) is due to (A~12) and inequality (b) holds because $\mathbb{E}[X_1^GX_2^G] \leq \sqrt{P_1P_2}$. By substituting (A~16), (A~20), and (A~21) in (A~11), we derive the desired result. ■

➢ ***Converse Proof for Proposition 5***

We prove that the Gaussian inputs are optimal for (96). Consider the argument of the maximization (96); we have:

$$I(X_1,X_2;Y_1|W)+I(W;Y_2) = H(X_1+aX_2+Z_1|W) - H\left(b(X_1+aX_2+Z_1)+\sqrt{1-b^2}\,\tilde{Z}_2\middle|W\right) + H(bX_1+X_2+Z_2) - H(Z_1)$$

(A~22)

Define $\alpha$ and $\beta$ as follows:

$$\alpha \triangleq \operatorname{sign}(b)\sqrt{\frac{\mathbb{E}[(\mathbb{E}[X_1|W])^2]}{\mathbb{E}[X_1^2]}}, \qquad \beta \triangleq \sqrt{\frac{\mathbb{E}[(\mathbb{E}[X_2|W])^2]}{\mathbb{E}[X_2^2]}}$$

(A~23)

where $\operatorname{sign}(b)$ is equal to 1 if $b>0$, and is equal to $-1$ if $b<0$. Note that we have:

$$\mathbb{E}[X_i^2]-\mathbb{E}[(\mathbb{E}[X_i|W])^2] = \mathbb{E}[\mathbb{E}[X_i^2|W]-(\mathbb{E}[X_i|W])^2] = \mathbb{E}\big[\mathbb{E}^2[X_i|W]\big] \geq 0, \;\; i=1,2$$

(A~24)

where $\mathbb{E}^2[X_i|W]$ is derived from the Definition II.2 of Part I [1]. Therefore, $\alpha^2$ and $\beta^2$ both belong to the interval $[0,1]$. Then consider the expressions in the right side of (A~22). We have:

$$H(X_1+aX_2+Z_1|W) \geq H(X_1+aX_2+Z_1|X_1,X_2,W) = \frac{1}{2}\log(2\pi e)$$

(A~25)

Also, using the "Gaussian maximizes entropy" principle, we can write:

$$\begin{aligned} H(X_1+aX_2+Z_1|W) &\leq \mathbb{E}\left[\frac{1}{2}\log\big(2\pi e(\mathbb{E}^2[X_1+aX_2+Z_1|W])\big)\right] \\ &\overset{(a)}{=} \mathbb{E}\left[\frac{1}{2}\log\big(2\pi e(\mathbb{E}^2[X_1|W]+\mathbb{E}^2[aX_2|W]+1)\big)\right] \\ &\overset{(b)}{\leq} \frac{1}{2}\log\Big(2\pi e\big(\mathbb{E}\big[\mathbb{E}^2[X_1|W]\big]+a^2\mathbb{E}\big[\mathbb{E}^2[X_2|W]\big]+1\big)\Big) \\ &= \frac{1}{2}\log\big(2\pi e(\mathbb{E}[\mathbb{E}[X_1^2|W]-(\mathbb{E}[X_1|W])^2]+a^2\mathbb{E}[\mathbb{E}[X_2^2|W]-(\mathbb{E}[X_2|W])^2]+1)\big) \\ &\overset{(c)}{=} \frac{1}{2}\log\Big(2\pi e\big((1-\alpha^2)P_1+a^2(1-\beta^2)P_2+1\big)\Big) \end{aligned}$$

(A~26)

where (a) is due to Lemma II.1 of Part I [1], (b) is due to Jensen inequality, and (c) is derived by (A~23). Based on (A~25) and (A~26), we deduce that there exist $\mu_\alpha$ and $\mu_\beta$ with:

$$H(X_1 + aX_2 + Z_1|W) = \frac{1}{2}\log\left(2\pi e\left((1-\mu_\alpha)P_1 + a^2(1-\mu_\beta)P_2 + 1\right)\right), \qquad \alpha^2 \le \mu_\alpha \le 1, \qquad \beta^2 \le \mu_\beta \le 1$$

(A~27)

Now for the second expression of (A~22), using the EPI, we have:

$$\begin{aligned} H\left(b(X_1 + aX_2 + Z_1) + \sqrt{1-b^2}\,\tilde{Z}_2\middle|W\right) &\ge \frac{1}{2}\log\left(2^{2H(b(X_1+aX_2+Z_1)|W)} + 2^{2H\left(\sqrt{1-b^2}\,\tilde{Z}_2|W\right)}\right) \\ &= \frac{1}{2}\log\left(2\pi e\left(b^2\left((1-\mu_\alpha)P_1 + a^2(1-\mu_\beta)P_2\right) + 1\right)\right) \end{aligned}$$

(A~28)

By combining (A~27) and (A~28), we obtain:

$$H(X_1 + aX_2 + Z_1|W) - H\left(b(X_1 + aX_2 + Z_1) + \sqrt{1-b^2}\,\tilde{Z}_2\middle|W\right) \le \frac{1}{2}\log\left(\frac{(1-\mu_\alpha)P_1 + a^2(1-\mu_\beta)P_2 + 1}{b^2\left((1-\mu_\alpha)P_1 + a^2(1-\mu_\beta)P_2\right) + 1}\right)$$

(A~29)

Consider the following deterministic function:

$$f(\mu_\alpha, \mu_\beta) \triangleq \frac{(1-\mu_\alpha)P_1 + a^2(1-\mu_\beta)P_2 + 1}{b^2\left((1-\mu_\alpha)P_1 + a^2(1-\mu_\beta)P_2\right) + 1}$$

(A~30)

It is easy to show that for the case of $ab = 1, |a| \ge 1$, the function $f(\mu_\alpha, \mu_\beta)$ is monotonically decreasing in terms of both $\mu_\alpha$ and $\mu_\beta$. Thereby, since $\alpha^2 \le \mu_\alpha$ and $\beta^2 \le \mu_\beta$, we derive:

$$H(X_1 + aX_2 + Z_1|W) - H\left(b(X_1 + aX_2 + Z_1) + \sqrt{1-b^2}\,\tilde{Z}_2\middle|W\right) \le \frac{1}{2}\log\left(\frac{(1-\alpha^2)P_1 + a^2(1-\beta^2)P_2 + 1}{b^2((1-\alpha^2)P_1 + a^2(1-\beta^2)P_2) + 1}\right)$$

(A~31)

Finally, for the third expression in the right side of (A~22), we have:

$$\begin{aligned} H(bX_1 + X_2 + Z_2) &\le \frac{1}{2}\log(2\pi e(b^2P_1 + P_2 + 2b\mathbb{E}[X_1X_2] + 1)) \\ &\overset{(a)}{\le} \frac{1}{2}\log\left(2\pi e\left(b^2P_1 + P_2 + 2|b|\sqrt{\mathbb{E}[(\mathbb{E}[X_1|W])^2]}\sqrt{\mathbb{E}[(\mathbb{E}[X_1|W])^2]} + 1\right)\right) \\ &\overset{(b)}{=} \frac{1}{2}\log\left(2\pi e\left(b^2P_1 + P_2 + 2b\alpha\beta\sqrt{P_1P_2} + 1\right)\right) \end{aligned}$$

(A~32)

where inequality (a) is due to Lemma II.1 of Part I in [1] and equality (b) is due to (A~23). Now by substituting (A~31) and (A~32) in (A~22), we derive the desired result. ■

## ACKNOWLEDGEMENT

The author would like to appreciate Dr. H. Estekey, head of School of Cognitive Sciences, IPM, Tehran, Iran, and Dr. R. Ebrahimpour, faculty member at School of Cognitive Sciences, for their kindly support of the first author during this research. He also greatly thanks his mother whose love made this research possible for him. Lastly, F. Marvasti is acknowledged whose editing comments improved the language of this work.

# REFERENCES

[1] R. K. Farsani, "Fundamental limits of communications in interference networks-Part I: Basic structures," *IEEE Trans. Information Theory, Submitted for Publication, 2012, available at* http://arxiv.org/abs/1207.3018.

[2] __________, "Fundamental limits of communications in interference networks-Part II: Information flow in degraded networks," *IEEE Trans. Information Theory, Submitted for Publication, 2012, available at* http://arxiv.org/abs/1207.3027.

[3] __________, "Fundamental limits of communications in interference networks-Part III: Information flow in strong interference regime," *IEEE Trans. Information Theory, Submitted for Publication, 2012, available at* http://arxiv.org/abs/1207.3035.

[4] __________, "Fundamental limits of communications in interference networks-Part IV: Networks with a sequence of less-noisy receivers," *IEEE Trans. Information Theory, Submitted for Publication, 2012, available at* http://arxiv.org/abs/1207.3040.

[5] __________, "Fundamental limits of communications in interference networks-Part V: A random coding scheme for transmission of general message sets," *IEEE Trans. Information Theory, To be submitted, preprint available at* http://arxiv.org/abs/1107.1839.

[6] __________, " Capacity theorems for the cognitive radio channel with confidential messages," 2012, *available at* http://arxiv.org/abs/1207.5040.

[7] R. K. Farsani and F. Marvasti, "Interference networks with general message sets: A random coding scheme," [Online] *available at*: http://arxiv.org/abs/1107.1839.

[8] ____________________, "Interference networks with general message sets: A random coding scheme", 49th *Annual Allerton Conference on Communication, Control, and Computing.*, Monticello, IL, Sep. 2011.

[9] D. Slepian and J. K. Wolf, "A coding theorem for multiple access channels with correlated sources," *Bell Syst. Tech. J.*, vol. 52, pp. 1037–1076, 1973.

[10] T. M. Cover and J. A. Thomas, *Elements of Information Theory*. 2nd ed, New York: Wiley, 2006.

[11] A. El Gamal and Y.-H. Kim, "*Lecture notes on network information theory*," *arXiv:1001.3404*, 2010.

[12] K. Marton, "A coding theorem for the discrete memoryless broadcast channel," *IEEE Trans. Inf. Theory*, vol. IT-25, no. 3, pp. 306–311, May 1979.

[13] A. El Gamal, "The capacity of a class of broadcast channels," *IEEE Trans. Inf. Theory*, vol. IT-25, no. 2, pp. 166–169, Mar. 1979.

[14] T. M. Cover, "Broadcast channels," *IEEE Trans. Inf. Theory*, vol. IT-18, no. 1, pp. 2–14, Jan. 1972.

[15] P. P. Bergmans, "Random coding theorem for broadcast channels with degraded components," *IEEE Trans. Inf. Theory*, vol. 19, no. 2, pp. 197–207, 1973.

[16] R. G. Gallager, "Capacity and coding for degraded broadcast channels," *Probl. Inf. Transm.*, vol. 10, no. 3, pp. 3–14, 1974.

[17] J. Jose and S. Vishwanath, "Sum capacity of k-user Gaussian degraded interference channels," *submitted to IEEE Transactions on Information Theory*. arXiv:1004.2104.

[18] A. Jovicic, H.Wang, and P. Viswanath, "On network interference management," in *IEEE Trans. Inform. Theory*, vol. 56, pp. 4941– 4955, Oct. 2010.

[19] G. Bresler, A. Parekh, and D. Tse, "The approximate capacity of the many-to-one and one-to-many Gaussian interference channels," *IEEE Trans. Inf. Theory*, vol. 56, no. 9, pp. 4566-4592, Sep. 2010.

[20] V. S. Annapureddy and V. V. Veeravalli, "Gaussian interference networks: Sum capacity in the low-interference regime and new outer bounds on the capacity region," in *IEEE Trans. Inform. Theory*, vol. 55(7), July 2009, pp. 3032 – 3050.

[21] T. Han, "The capacity region of general multiple-access channel with certain correlated sources," *Information and Control*, vol. 40, no. 1, pp. 37–60, 1979.

[22] R. K. Farsani, "The K-user interference channel: strong interference regime," *2012, available at* http://arxiv.org/abs/1207.3045.

[23] __________, "How much rate splitting is required for a random coding scheme? A new achievable rate region for the broadcast channel with cognitive relays," 50th *Annual Allerton Conference on Communication, Control, and Computing.*, Monticello, IL, Oct. 2012.